\documentstyle[aps,epsf]{revtex}

\newcommand{\eq}[1]{(\ref{#1})}
\newcommand{\kb}{k_{\rm B}}
\newcommand{\tw}{t_{\rm w}}
\newcommand{\twz}{t_{{\rm w}1}}
\newcommand{\two}{t_{{\rm w}2}}
\newcommand{\twt}{t_{{\rm w}3}}
\renewcommand{\vec}[1]{\bbox{#1}}

\newcommand{\LOVLP}{\xi_{\rm \Delta T}}
\newcommand{\trec}{\tau_{\rm rec}}
\newcommand{\twtotal}{t_{\rm w}^{\rm total}}

\begin{document}
\epsfclipon

\title{Multiple Domain Growth and Memory in the Droplet Model for Spin-Glasses}

\author{Hajime Yoshino\thanks{E-mail: yoshino@.ess.sci.osaka-u.ac.jp; Present address: Department 
of Earth and Space Science Faculty of Science, Osaka University,
Machikaneyama Toyonaka 560-0043 Osaka Japan
}, Ana\"el
 Lema\^{\i}tre\thanks{E-mail: anael@lob.physics.ucsb.edu; Present address: Department of Physics, University of California, 
 Santa Barbara, CA 93106 USA}, Jean-Philippe
 Bouchaud\thanks{E-mail: bouchau@spec.saclay.cea.fr}}
\address{Service de Physique de l'Etat Condens\'e, CEA Saclay,\\ 91191 Gif sur Yvette Cedex, France}
\maketitle

\begin{abstract} 
We study domain growth dynamics when the target state is suddenly 
changed on all length scales. This procedure mimics the `chaos' 
effect postulated by the droplet theory of spin-glasses, and allows 
us to investigate in details its various dynamical consequences. 
We study the problem by a variety of methods, including scaling arguments,
analytical solution of the spherical Mattis model, and Monte Carlo simulations 
of a 2-dimensional Ising Mattis model.
We show that successive coarsening with respect to different equilibrium 
states imprints multiple domain structures on top of each other, plus extra  
noise due to random interferences. We demonstrate that the domain structures 
can be retrieved by an additional series of coarsening in the reversed order
which removes the noises.
We discuss the rejuvenation (chaos) and memory effects observed in 
temperature-cycling experiments in glassy systems from the 
present point of view, and discuss some open problems and alternative
descriptions.
\end{abstract}


\section{Introduction}
\label{sec.intro}

Marginal stability of the glassy equilibrium states to weak perturbations 
such as a small shift of temperatures has been a fundamental interest in 
the studies of spin-glasses and related systems including vortex lines systems 
in dirty type-II super-conductors. In particular, the droplet picture based 
on scaling arguments and Migdal-Kadanoff type real-space renormalization-group 
calculations \cite{BM,FH88a,FH91,NH} claims that any small but finite 
perturbation, such as changes of temperature by an amount $\Delta T$,
is enough to change the equilibrium states in such glassy systems completely 
at length scales larger than the so called overlap length $\LOVLP$. 
Such a dramatic effect has been coined {\it temperature chaos}.
It was anticipated that such a change of the equilibrium states,
if it exists, should have significant consequences on dynamical observable
such as the dynamical linear-susceptibility. \cite{FH88b,KH}

From the experimental side, a series of interesting experiments 
have been done with different spin-glasses measuring relaxations
of thermo-remanent magnetization (TRM), zero-field cool magnetization (ZFC)
and AC-magnetic susceptibilities. They show strikingly 
rich dynamical aspects of spin-glasses subjected to small temperature 
cycles (within the spin-glass phase)
\cite{saclay-exp-rev-1,saclay-exp-rev-2,RVHO87,VBHL,LHMV,Nordblad,uppsala-1,uppsala-2,JVHBN,interference}. The main outcome of the experiments is 
the coexistence of two seemingly contradictory aspects, namely 
`rejuvenation' upon cooling and `memory' upon heating back. 
These experiments in spin-glasses have motivated similar
experimental studies in other glassy systems including
polymer glass \cite{polymer-glass},
frozen ferrofluid \cite{magneticparticle},
random ferromagnetic system \cite{pinned-domainwall}, 
random ferroelectric system \cite{dipole-glass,dipole-glass-2}
and structural glass \cite{structural-glass}.

The rejuvenation effect can be interpreted as a signature of the 
chaotic change of the underlying equilibrium states as anticipated 
by the droplet picture. However the simultaneous memory effect is not 
obvious to account for within the droplet picture, and the previous attempts
have remained unsatisfactory \cite{FH88b,KH,uppsala-1}. 
The need for some mechanism which allows conservation of large scale
spatial structures to preserve memory is now clearly realized \cite{uppsala-2,JVHBN,JP}.

On the other hand, there has been recent remarkable progress 
in the dynamical mean-field 
theory for glassy systems \cite{CK,FM,CKD96,BCKM,DMFT}. 
It was in particular shown within 
the mean-field theory \cite{DMFT} that temperature-cycling 
processes amount to a cycling of break-point $q_{\rm EA}$ which 
separates the stationary and aging part of the correlation 
and response functions. The cycling of $q_{\rm EA}$ can push 
a part of the stationary signal into the non-stationary regime 
(thereby leading to rejuvenation), while preserving the rest of 
the non-stationary part (memory) for large enough time scales.

A somewhat similar picture was advocated within a hierarchical 
phase space picture \cite{saclay-exp-rev-1}, where each level of 
the hierarchical tree has its own glass temperature. Therefore, 
a small temperature drop drives a certain level out of 
equilibrium (rejuvenation) while higher level of the tree are 
frozen (memory) \cite{BD}. An important motivation for this picture 
is the Parisi's replica-symmetry breaking solution 
for the static properties of mean-field spin-glass models.\cite{MFT}
This scenario has been recently substantiated by interesting 
numerical simulations \cite{SN}. 
Its real-space transcription in terms of a hierarchy of time scales associated with different 
length scales was developed in \cite{JP}, in particular in the context 
of pinned domain walls \cite{pinned-domainwall}.

In the present paper, we want to go back to the original droplet picture and work out 
in details the dynamical consequences of chaotic changes of the underlying equilibrium state, 
and in particular address the question of memory-conservation. To simplify the approach, 
we restrict ourselves to the simplest scenario for the relaxational dynamics as in \cite{FH88b}.
Namely we assume that relaxational dynamics at any temperature is a coarsening process 
of the domain walls between an equilibrium pure state and its time-reversal state. 
As we noted above, changes of temperature amount to complete changes of the equilibrium states
beyond $\LOVLP$ in the droplet picture. In order to investigate the temperature cycling 
procedures based on the droplet picture, we have studied coarsening subjected to cycling of 
the underlying equilibrium state which we impose by hand. In the present paper, we disregard 
possible transient short-time behaviors associated with length scales smaller than $\LOVLP$ 
and concentrate on the large time phenomena. Somewhat unexpectedly, we found 
that the domain structure corresponding to the different equilibrium states that are 
encountered can indeed be preserved and retrieved dynamically. We will show that the droplet 
picture itself can provide a suggestive and interesting scenario for the rejuvenation and 
memory effects observed in experiments of spin-glasses.

It is known that changes of temperature in a class of frustrated
systems can change the effective coupling between certain `block spins'
due subtle entropy effects, so that interesting re-entrant phase transitions 
can occur \cite{frustration-reentrance}. This has motivated a recent work in which
somewhat similar ideas for the mechanism of temperature-cycling experiments 
are presented \cite{MV}. 

The outline of the present paper is as the following.
In section \ref{sec:droplet} we briefly review the droplet picture which underlies
the present study. In section \ref{sec:scaling} we introduce our models and
discuss the generic features of coarsening systems subjected to equilibrium 
states cycling, based on the standard phenomenology of coarsening systems. 
In section \ref{sec.spherical} we study the dynamics of the $O(n)$ Mattis model 
in the spherical limit which is described by a Time Dependent Ginzburg Landau (TDGL) equation. 
We solve the TDGL equation exactly under cycling of equilibrium states
and examine the physical picture discussed in section \ref{sec:scaling}.
In section \ref{sec.ising} we study 2-dimensional Ising Mattis model by Monte Carlo
simulations to further check our picture.
In section \ref{sec:scenario} we compare our results with 
the rejuvenation (chaos) and memory effects observed in temperature-cycling 
experiments in a spin-glass system.
Finally, in section \ref{sec.discussion} we summarize our result and 
underline important open questions.
In the appendices, we present some technical details of the calculations
of the $O(n)$ Mattis model.

\section{The Droplet Picture}
\label{sec:droplet}

Here we briefly review the droplet picture which is the background 
of the present study. For simplicity let us consider 
spin-glasses which have $Z_{2}$ symmetry like Ising spin-glasses.
In the droplet picture, it is assumed that there exist only one 
equilibrium states and its time-reversal state at each temperature below 
the spin-glass transition temperature $T_{\rm c}$. 
In equilibrium, the most important contributions to physical 
observables such as the magnetic susceptibility comes from thermally
activated excitations of compact clusters of spins, called droplets.

Let us consider for simplicity the equilibrium state at zero temperature,
i.e. the ground state. The total number of spins at the surface of a 
droplet of size $L$ is postulated to scale typically as  
$L_{0}(L/L_{0})^{d_{s}}$ where $d_{s}$ is
the fractal dimension of the surface and $L_{0}$ is a microscopic
length scale. By definition, a droplet of size $L$ should have 
a non-zero excitation energy gap. The excitation energy of the droplet 
$E_{\rm gap}$  typically scales with $L$ as 
$E_{\rm gap}\sim \Upsilon(L/L_{0})^{\theta}$.
Here $\Upsilon$ is the stiffness constant and $\theta$ 
is the stiffness exponent.

The dynamics of droplets is considered to be a thermally activated process. 
The energy barrier $E_{\rm barrier}$ to create a droplet is supposed to scales with $L$ as 
$E_{\rm barrier}\sim \Upsilon(L/L_{0})^{\psi}$ with $\psi \geq \theta$.
The relaxation time is given by the Arrhenius law,
\begin{equation}
t_{L} \sim \tau_{0} \exp (\Upsilon(L/L_{0})^{\psi}/\kb T),
\label{eq:arrhenius}
\end{equation}
where $\tau_{0}$ is the attempt time for the activated process.

\subsection{Effect of Temperature Change on Equilibrium States}

In the droplet picture, small temperature changes cause substantial changes
of the equilibrium state. The argument goes as follows: the entropy associated 
to a droplet is the sum of contributions which are random in sign over the surface of 
the droplet. The latter implies the entropy associated with a droplet of size $L$ is 
random in sign, and of magnitude $\sim \kb \sqrt{(L/L_{0})^{d_{s}}}$. 
A subtle conjecture is that the (free-)energy exponent $\theta$ satisfies the inequality 
$\theta < d_{s}/2$. Therefore, a small change of temperature can ruin the balance between 
energy and entropy. In particular, the ground state becomes unstable at finite temperatures 
due to the gain in entropy, and is transformed into a `new' equilibrium state that it is 
completely uncorrelated with the ground state beyond the {\it overlap length},
\begin{equation}
\frac{\LOVLP}{L_{0}} \propto \left ( \frac{\kb \Delta T}{\Upsilon}\right)^{-\zeta},
\label{eq:lovlp}
\end{equation}
where $\zeta=1/(d_{s}/2-\theta)$. It is conjectured that this kind of first-order like 
phase transitions occur continuously within the whole temperature range below $T_{c}$.

\subsection{Domain Growth}
\label{subsec:droplet-domaingrowth}

Within the droplet picture\cite{FH88b}, the aging of spin-glasses
that starts from an out-of-equilibrium initial condition is thought of
as a coarsening process, where the domain walls between the two equilibrium states
progressively disappear. The coarsening is driven by successive nucleation
and annihilation of droplets. From \eq{eq:arrhenius}, the typical size of 
droplet which can be thermally activated within a
given time scale $t$ is expected to scale as,
\begin{equation}
L_{T}( t)=L_{0} \left [ \left (\frac{\kb T}{\Upsilon} 
\log \left (\frac{t}{\tau_{0}} \right)
 \right) \right]^{1/\psi}.
\label{eq:growth-law}
\end{equation}
Thus the mean separation of the domain walls after time
$t$ starting from a random initial condition is also expected to be given 
by \eq{eq:growth-law}.

While there is no experimental way to observe directly such a domain growth 
in spin-glasses, AC magnetic susceptibility can be a useful probe.
In the droplet picture, the AC susceptibility 
at frequency $\omega$ is considered to be proportional to the inverse 
of the stiffness $\Upsilon$ of droplet excitations whose size is  
$L_{T}(\omega^{-1})$. During aging, the excitation energy gap and hence the 
effective stiffness is smaller than in complete equilibrium
because some droplets of size $L_{T}(\omega^{-1})$ can happen to share their
surface with the ``frozen-in'' droplet of size $L_{T}(t)$, and lower their 
energy. Using scaling arguments, the resultant reduction of the stiffness 
is obtained as $\Delta \Upsilon (L_{T}(\omega^{-1}),L_{T}(t))
\sim (L_{T}(\omega^{-1})/L_{T}(t))^{d-\theta}\Upsilon$. From the latter, the
relaxation of the out-of-phase AC susceptibility is obtained as,
\begin{equation}
\chi''(\omega,\tw) 
\sim \chi''(\omega,\infty) 
\left [1-\left (\frac{L_{T}(\omega^{-1})}{L_{T}(\tw)} \right)^{d-\theta} \right]^{-1},
\label{eq:droplet-scaling-chi}
\end{equation}
where $\chi''(\omega,\infty)$ is the equilibrium susceptibility.
For an experimental analysis of the AC-susceptibility based on 
this scaling ansatz, see \cite{SMW}. Some related analysis was performed
also in numerical simulations of the Edward-Anderson spin-glass model
\cite{3DEA-KYT,4DEA-HYT}. However, the following analysis will not depend on the
detailed shape of \eq{eq:droplet-scaling-chi}, but rather on the
existence of some general (inverse) relation between the AC-susceptibility and the
typical size of the non-equilibrium droplets.

\subsection{Separation of Time and Length Scales}

An important consequence of thermally activated dynamics is that it
induces a natural hierarchy of time scales (at a given temperature) and 
a strong separation of time scales (between different temperatures)\cite{JP}.
The latter is very useful to understand the temperature cycling experiments 
in spin-glasses. Due to the Arrhenius law, the time needed to cross 
a certain energy barrier can be extremely different at two different 
temperatures, say $T$ and $T+\Delta T$. 
The time $t_{T}$ needed at temperature $T$ to jump over a barrier 
crossed at time $t_{T+\Delta T}$ at temperature $T+\Delta T$ is given by:
\begin{equation}
t_{T}= \tau_{0} \left (\frac{t_{T+\Delta T}}{\tau_{0}} \right)^{1+\Delta T/T},
\label{eq:enhanced-separation}
\end{equation}
or:
\begin{equation}
\log \left ( \frac{t_{T}}{t_{T+\Delta T}} \right) =
  \frac{\Delta T}{T} \log \left (\frac{t_{T+\Delta T}}{\tau_{0}}\right).
\label{eq:enhanced-separation-2}
\end{equation}
The number of decades separating $t_{T}$ and $t_{T+\Delta T}$ is thus equal to the 
number of decades separating $t_{T+\Delta T}$ and $\tau_0$ times $\Delta T/T$. 
In experiments, the latter is typically $15$ or so, so that a $10 \%$ temperature change
multiplies the time scales by $30$. Note that this separation is much weaker in 
numerical simulations, where the 
number of decades separating $t_{T+\Delta T}$ and $\tau_0$ is $\sim 5$. 

It is also useful to consider the separation of length scales. 
Let us consider two temperatures $T$ and $T+\Delta T$. 
The ratio of the length scale explored at the two temperatures
within the same time, say $\tw$, is obtained from \eq{eq:growth-law} as,
\begin{equation}
\frac{L_{T+\Delta T}(\tw)}{L_{T}(\tw)}=
 \left ( 1+\frac{\Delta T}{T}  \right)^{-1/\psi}.
\label{eq:separation-length-1}
\end{equation}
One should note that the latter formula does not imply strong separation 
of length scales: in order to have appreciable separation of length scale,
$\Delta T$ must be comparable to $T$ itself.

Finally it should be remarked that the time/length separation is even more sharp in reality
because the typical energy barriers grow when the temperature is decreased \cite{LOHOV}.
This can be interpreted as a growth of the stiffness  \cite{FH88b,usinprep,4DEA-DF,4DEA-HYT} as, 
\begin{equation}
\Upsilon \sim J |1-T/T_{c}|^{\psi\nu}.
\label{eq:stiffness-temp}
\end{equation}

\section{Coarsening towards different Equilibrium States}
\label{sec:scaling}

We now start to consider the possible dynamical consequences of 
the chaos effect within the droplet picture. 
Because we have in mind the temperature-cycling experiments 
in spin-glasses, we consider coarsening dynamics under cycling of the 
underlying equilibrium state.
In this section we discuss intuitively how and when a succession of coarsening 
with 
respect to different states can create and store in memory the domain
structures of all of them.  Some essential aspects of the picture will 
be verified quantitatively in the following sections, based on some
analytical and numerical study of the Mattis model.

To be specific, let us consider an Ising spin model on a lattice in which 
a spin at site $i$ is $S_i$. We denote an equilibrium state as $\alpha$ and 
consider that it consists of  a spin-configuration $\sigma^{\alpha}_{i}$ 
where $\sigma^{\alpha}_{i}$ takes $\pm 1$ randomly with zero mean.  It is convenient 
to introduce a projection of a spin configuration ${S}_{i}$
to the equilibrium state $\sigma^{\alpha}_{i}$  as,
\begin{equation}
\tilde{S}_i^{\alpha}=\sigma^{\alpha}_{i} {S}_{i}.
\label{eq:real-projection}
\end{equation}
Projections to two different ground states
say $\{\sigma_{i}^{\alpha}\}$ and $\{\sigma^{\beta}_{i}\}$ 
are related as
\begin{equation}
\tilde{S}^{\alpha}_{i}
=\sigma^{\alpha}_{i}\sigma^{\beta}_{i}\tilde{S}^{\beta}_{i}.
\label{eq:projection-projection}
\end{equation}

If a uniform external magnetic field $h_{\rm uni}$ 
is applied to the system, the Zeeman energy becomes,
\begin{equation}
h_{\rm uni} \sum_{i}{S}_{i}
= \sum_{i}\tilde{h}_i^\alpha  \tilde{S}_i^\alpha,
\end{equation}
where we introduce a random field $\tilde{h}_i^\alpha$
defined as,
\begin{equation}
\tilde{h}_i^\alpha=h_{\rm uni}\sigma^\alpha_{i}.
\label{eq:uniform-staggered}
\end{equation}

In the following, we suppose that the equilibrium state at 
temperature $T_{A}$ is a certain configuration $\alpha=\pm A$, and 
at temperature $T_{B}$ is a different configuration $\alpha=\pm B$;
we suppose that the two states $A$ and $B$ are completely uncorrelated 
beyond the overlap length $\LOVLP$.

Concerning experiments, it should be noted that temperature is controlled 
within certain finite resolution $\delta T$. Thus the overlap length 
associated with the limited accuracy should be large enough compared with the 
dynamical length scales explored within some laboratory time scales. 
Otherwise, neither isothermal aging nor `cycling' can even be achieved.

\subsection{Mattis Model}
\label{subsec.mattis}

It will be useful to study a specific model which allows
coarsening towards various equilibrium states
in a transparent way.
In later sections (section \ref{sec.spherical} and \ref{sec.ising}),  
we analyze in detail the so called Mattis model \cite{mattis},
\begin{equation}H(\{S\})=
-J \sum_{i,j}\tilde{S}_i^\alpha\tilde{S}_{j}^\alpha- \sum_{i}\tilde{h}_{i}^\alpha\tilde{S}_i^\alpha.
\end{equation}
As one can see easily, this model clearly has the spin-configuration 
$\tilde{S}_{i}^\alpha \equiv \pm 1$ as ground states (for $h_{\rm uni}=0$). Since this model
is equivalent to ferromagnetic models, the relaxational dynamics at
low temperatures is nothing but the progressive coarsening of the equilibrium 
states.\cite{B94}  
In order to implement the droplet picture more precisely, 
one could introduce some disorder to the coupling parameter $J$ in 
order to have thermally activated dynamics due to pinning of domain 
walls \cite{HH85}. The latter does lead to slow growth of the domain 
as in \eq{eq:growth-law}. However, we will not perform specific analysis 
of the decorated model in the present paper.

In section \ref{sec.spherical}, we study coarsening of the Mattis model in the spherical limit approximation for general spatial dimension $d$, and obtain
a fully analytical solution, which we confirm in section \ref{sec.ising}
by a zero-temperature Monte Carlo simulation of the Mattis model
in two-dimension ($d=2$).

\subsection{A Cycle on a Symmetry Broken State}

We begin with a simple cycling procedure between $T_A$ and
$T_B$, which provides the basic intuition about the
coarsening process where the target equilibrium state is cycled.
We first grow the $B$ phase with $A$ as the initial configuration,
and then revert to $A$ as the target state.

\subsubsection{Noise Imprinting}
\label{subsec.noise}

The coarsening towards $B$ given $A$  
as the initial configuration is a standard coarsening 
process, because $A$ is simply a random configuration with 
respect to $B$: the projection $\tilde{S}^{B}_i(t=0)$ is random 
in sign with short range correlation only up to $\LOVLP$.
We focus on how the symmetry-broken state $A$ is 
affected by this process.

After time $t$, the spin-configuration $\{ S_{i}(t) \}$ 
has coarsened with respect to B: the 
spatial pattern of the projection $\{ \tilde{S}^{B}_{i}(t) \}$
consists of  domains of $B$ and $-B$ separated
by domain walls. The typical distance between the domain walls
is $L_{T_B}(t)$, and increases with time $t$. [The growth law $L_{T}(t)$
depend on temperature $T$ in spin-glasses (see \eq{eq:growth-law}).]
The correlation between the configuration at time $t \gg \tau_0$ and
the initial configuration is given by:
\begin{equation}
C(t,0)=(1/N)\sum_{i}S_{i}(t)S_{i}(0)\sim( L_{T}(t)/\LOVLP)^{-\lambda}.\label{eq:lambda}
\end{equation}
The last equality is a general property of coarsening systems, and
defines the non-equilibrium dynamical exponent $\lambda$.\cite{B94}
Note that we have included the effect of short-range spatial correlations 
given by $\LOVLP$ in the initial condition.

Now let us consider the projection of the spin-configuration
$\{S_{i}(t)\}$ onto the initial state $A$.  We  expect that
the projection $\{ \tilde{S}^{A}_{i}(t)\}$ are random numbers 
with only short-ranged spatial correlation. The mean value, however,
is nothing but the staggered magnetization $\rho_{A}$ with respect to $A$ which is non-zero. Indeed:
\begin{equation}
\rho_{A}(t) \equiv (1/N) \sum_{i} \tilde{S}^{A}_{i}(t=0) \tilde{S}^{A}_{i}(t)
=C(t,0)\sim (L_{T_{B}}(t)/\LOVLP)^{-\lambda}, \label{eq:rho}
\end{equation}
where we have used the initial condition $\{ \tilde{S}^{A}_{i}(0) \equiv 1 \}$, and the simple identity $S_{i}(t)S_{i}(0)=\tilde{S}^{A}_{i}(t)\tilde{S}^{A}_{i}(0)$.

To summarize, if one starts from a completely symmetry broken state $A$, the coarsening with respect to a different state $B$ adds some noise 
to $A$, and reduces the magnetization to $\rho_A(t)$ which decreases
with $t$. However it is very important to note that {\it for any
finite time $t$ the bias is non-zero}: the symmetry between $A$ 
and $-A$ remains broken.

\subsubsection{Noise Cleaning}
\label{subsec.recovery}

We now revert back to $A$ as the target state, and evolve the 
configuration $\{S_{i}(t)\}$ obtained above. The initial configuration is
a random configuration with small bias $\rho=\rho_{A}(t)$ given
in \eq{eq:rho}. Obviously the symmetry-broken state $A$ 
should be finally restored, because the bias is present.
An important question is the time is needed for the recovery.

If the bias $\rho$ has been made sufficiently small, 
the coarsening with respect to $A$ proceeds for a long time almost as if the initial condition was un-biased random configuration with short-range correlation of order $\LOVLP$: both the majority ($A$) and minority 
($-A$) phase coarsen. From the initial condition $\{S_{i}(t)\}$ the correlation
behaves as \eq{eq:lambda} for a long time. However, in the large time limit,
this correlation has to converge to $\rho$ since the state $A$ is finally recovered.  The matching between the two regimes allows one to obtain
the {\it recovery time} $\trec$ as: 
\begin{equation}
\left(\frac{L_{T_{A}}(\trec(\rho))}{\LOVLP}\right)^{-\lambda} \sim \rho. 
\label{eq:recover-noise}
\end{equation}
This is the characteristic time around which the symmetry-broken state is
{\it almost} recovered.

The fact that coarsening with a biased initial condition dies out
after a finite time scale has been analyzed analytically in the $O(n)$ 
model.\cite{FBP95} The mechanism is similar to the interruption of 
coarsening under finite magnetic field studied analytically in the 
relaxational dynamics of the spherical Sherrington-Kirkpatrik
mean field spin-glass model \cite{CD95}.

Now combining \eq{eq:rho} and \eq{eq:recover-noise}, we find a simple 
relation between the recovery time $\trec$ of a symmetry-broken state 
and coarsening time $t$ with respect to an unrelated phase,
\begin{equation}
L_{T_{A}}(\trec) \sim L_{T_{B}}(t).
\end{equation}
Note that the role of the overlap-length $\LOVLP$ does not appear
explicitly.

Here an important point in spin-glasses is that the
time $\trec$ can be extremely different from $t$
due to the strong separation of time scales \eq{eq:enhanced-separation}
discussed above:
\begin{equation}
\trec= \tau_{0} \left (\frac{t}{\tau_{0}} \right)^{T_{B}/T_{A}}.
\end{equation}
In the case of negative cycling $T_{A} \to T_{B} < T_A$, the
recovery time $\trec$ can be much shorter than the 
coarsening time $t$. Conversely, for positive cycling $T_{A} \to T_{B}
> T_A$, the recovery time $\trec$ can be much larger than $t$.

\subsection{Double Coarsening in One Step Cycling}
\label{subsec.double}

We now extend the two stage process discussed above to
the following three stage process: we first coarsen the system
towards $A$ for time $\twz$ starting from a totally random initial 
configuration, unrelated to both $A$ and $B$. Subsequently, we 
take $B$ as the target state for a time $\two$, and finally 
coarsen again towards $A$ for time $\twt$. In spirit, this corresponds to
the one-step temperature-cycling protocol used in experiments \cite{saclay-exp-rev-1,saclay-exp-rev-2,Nordblad}. The results of the previous section corresponds to the limit where $\twz \to \infty$.

\subsubsection{First and Second Stage}

\begin{figure}
\begin{center}
\leavevmode \epsfxsize=0.45\linewidth
\epsfbox{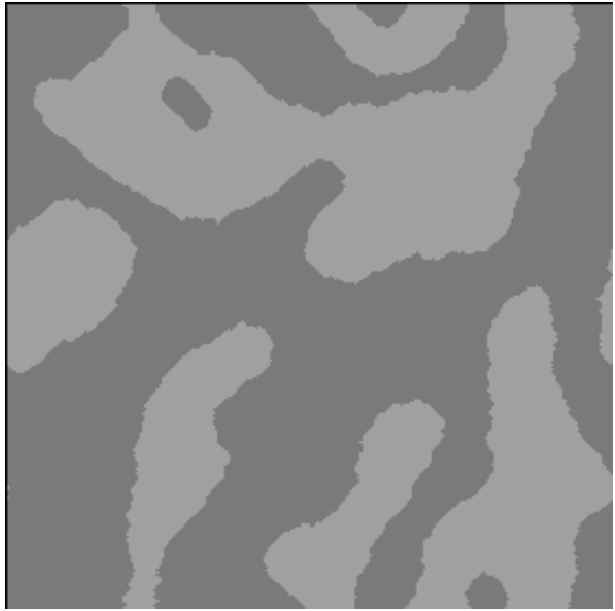}
\end{center}
\caption{Projection of the spin configuration
onto the ground state $A$ at end of the first stage ($t=\twz$). 
The picture is obtained by a zero-temperature Monte Carlo simulation 
of the 2-dimensional Ising Mattis Model. 
The initial condition is a random initial configuration
and the duration of the first stage is chosen to be $\twz=1000$ MCS.}
\label{image-0.fig}
\end{figure}

In the first stage, domains of ${A}$ and $-{A}$
grow in competition with each other. After time $\twz$, the mean distance 
between the domain walls is $L_{T_{A}}(\twz)$, that we suppose much larger 
than any microscopic length scale $L_0$ and the overlap-length $\LOVLP$. 
In Fig. \ref{image-0.fig} we show a picture of the domain structure
of a two-dimensional Ising Mattis model 
(see section \ref{sec.ising}) during coarsening.

\begin{figure}
\begin{center}
\leavevmode \epsfxsize=0.45\linewidth
\epsfbox{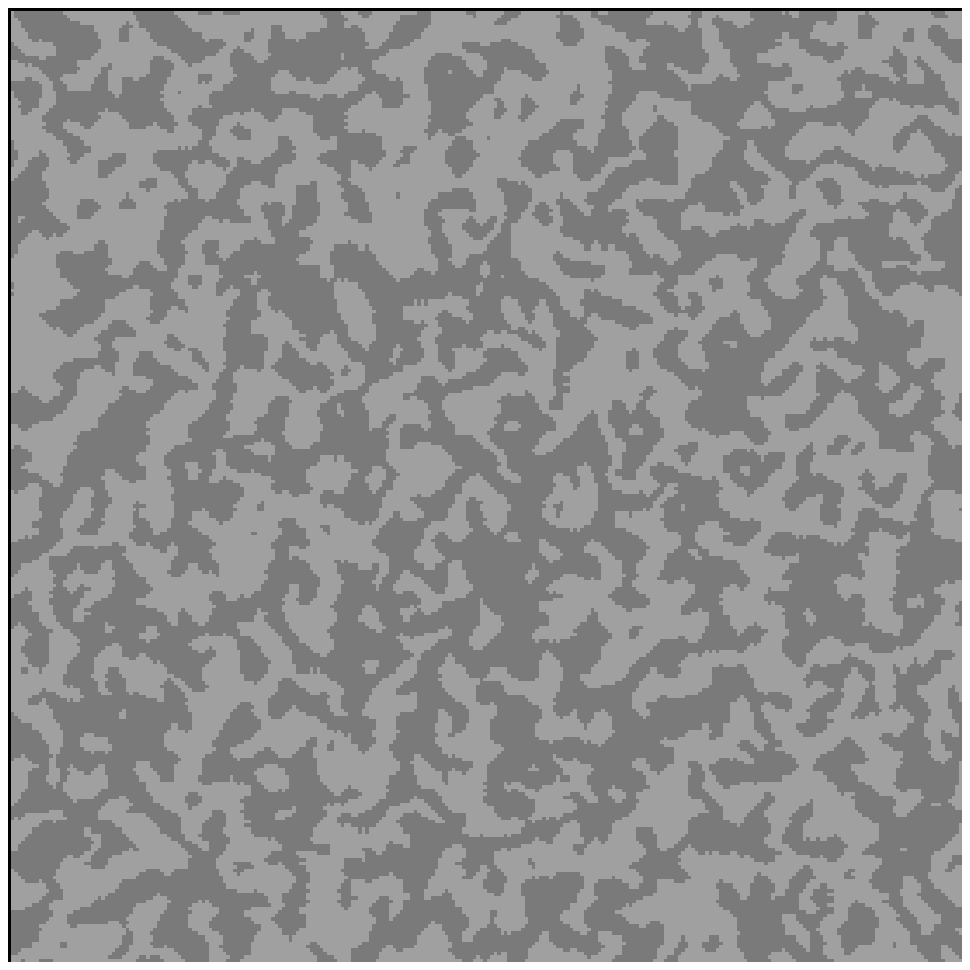}
\leavevmode \epsfxsize=0.45\linewidth
\epsfbox{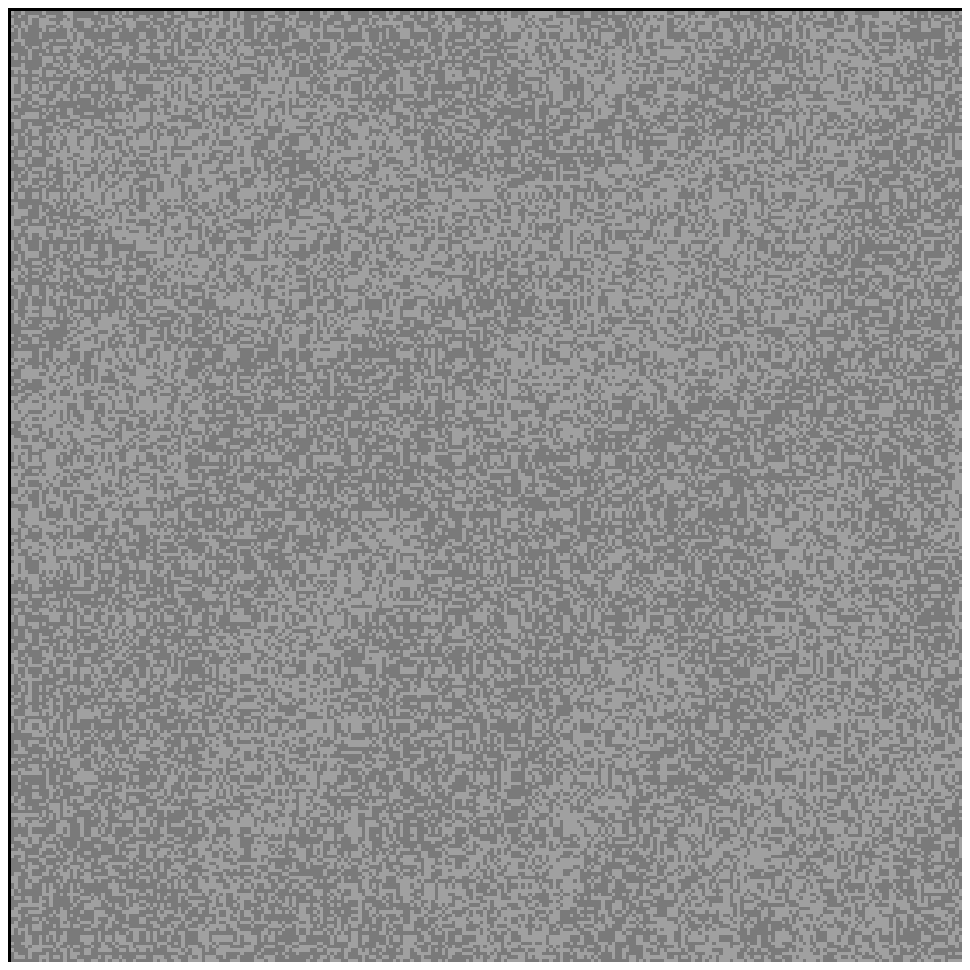}
\end{center}
\caption{The spin-configuration of the Ising Mattis Model
after the second stage of coarsening towards a state $B$,
completely uncorrelated with $A$ ($\LOVLP=1$).
The duration is chosen to be $\two=20$ MCS, much less than $\twz$.
The upper figure is the projection onto $B$ whereas the lower figure is 
the projection onto $A$.
One can compare the lower figure with Fig. \ref{image-0.fig}
and clearly distinguish the `ghost domain' structure.} 
\label{image-1-2.fig}
\end{figure}

Subsequently the system coarsen towards $B$. Since the spin 
configuration obtained by the first stage is a random initial
configuration with respect to $B$, coarsening of the domains of
${B}$ and $-{B}$ starts from the overlap length $\LOVLP$.
After time $\two$, the mean separation of the domain walls is $L_{T_{B}}(\two)$.
(see the upper figure of Fig \ref{image-1-2.fig}).

An interesting way to monitor the time evolution of the spin configuration 
during the second stage is to use its projection onto ${A}$.
(see the lower figure of Fig \ref{image-1-2.fig}).
Remember that at the end of the first stage, the spin configuration
is divided into domains of ${A}$ and $-{A}$
separated by domain walls at a distance $L_{T_{A}}(\twz)$
from each other. Let us consider such a domain as a {\it window cell} to
monitor the time evolution of the projection onto $A$ during the second 
stage. Within such a cell, the initial spin configuration
is completely polarized with respect either to ${A}$ or to 
$-{A}$. Then the subsequent time evolution of the
spin configuration in the bulk of such a cell 
should be the same as the case discussed in section \ref{subsec.noise}.
Thus we expect that the projection to ${A}$  within 
the cell becomes a random configuration beyond $\LOVLP$ but with a remnant mean bias
whose sign is the same as at time $\twz$. The amplitude of the bias, 
however, decreases in magnitude as in \eq{eq:rho}. 

To summarize, after time $\two$ of the second stage, the projection onto 
${A}$ (or $-{A}$) is a random configuration beyond $\LOVLP$ but with a remnant local bias,
\begin{equation}
\rho_2 \sim (L_{T_{B}}(\two)/\LOVLP)^{-\lambda}.
\label{eq:rho-one-step}
\end{equation} 
The remarkable point is that the {\it spatial structure} of the sign of
the bias, coarse-grained over the length $L_{T_A}$ is the same as at the end of the first stage. The `real' domain walls of size $L_{T_A}$ which separate 
${A}$ and $-{A}$ at the end of the first stage have been destroyed. 
However, the `sign' of the bias retains the very same spatial structure. For convenience, we call the latter `ghost' domains.  This is the mechanism
to install and conserve memory of the thermal history of the system 
in the present picture.

\subsubsection{Third Stage}
\label{sec.one-step-third}

In the third stage, the state $A$ is restored as the target state, 
given the final configuration of the second stage as the 
new initial condition. We continue to monitor the spin configuration 
using the window cell defined above, of size $L_{T_{A}}(\twz)$.
From the discussion in section \ref{subsec.recovery}, coarsening of
${A}$ and $-{A}$ re-starts within the cell. Let us call this regime the {\it inner-coarsening} regime.

How long does this inner-coarsening regime last ? 
Suppose that the size of the cell $L_{T_{A}}(\twz)$,
which is the typical length of the spatial structure of the bias field,
can be regarded as large enough so that the situation is essentially 
the same as with an infinitely large system with biased random initial 
condition (see \ref{subsec.noise}). The inner-coarsening finishes at a 
recovery time related to the strength of the bias as given as 
\eq{eq:recover-noise}. However if the size of the cell 
$L_{T_{A}}(\twz)$ is small the inner-coarsening will be interrupted 
when the size of the new domains reaches that of the cell. 
The condition separating these two regimes reads:
\begin{eqnarray}
&& \mbox{a)} \qquad   \rho_2^{-1/\lambda} \ll L_{T_{A}}(\twz)/\LOVLP. \qquad\nonumber \\ 
&& \mbox{b)} \qquad \rho_2^{-1/\lambda} \gg L_{T_{A}}(\twz)/\LOVLP \qquad \mbox{(`finite size effect')}
\label{eq:condition-a-b}
\end{eqnarray}
Thus we obtain the life time $\trec$ 
of the inner-coarsening regime as,
\begin{equation}
L_{T_{A}}(\trec)/\LOVLP 
\sim \mbox{min} (\rho_2^{-1/\lambda}, L_{T_{A}}(\twz)/\LOVLP)
 \qquad \mbox{or} \qquad
\trec= \mbox{min} \left (
\tau_{0} \left (\frac{\two}{\tau_{0}} \right)^{T_{B}/T_{A}},\twz
\right).
\label{eq:recover-noise-cycle}
\end{equation}

Let us consider the case a) more closely. In this case, 
a natural expectation is that after time $\trec$,
the magnitude of the polarization (bias) within the bulk 
of the cell is almost fully recovered.  The latter implies that 
the `ghost' domains of sizes $L_{T_{A}}(\twz)$, which are the trace 
of the real domain constructed in the first stage, become the 
`real' domains again. The retrieved domain will then re-start to grow 
just as the continuation of the 1st stage. We call this regime as 
{\it outer-coarsening regime}.

\begin{figure}
\begin{center}
\leavevmode \epsfxsize=0.45\linewidth
\epsfbox{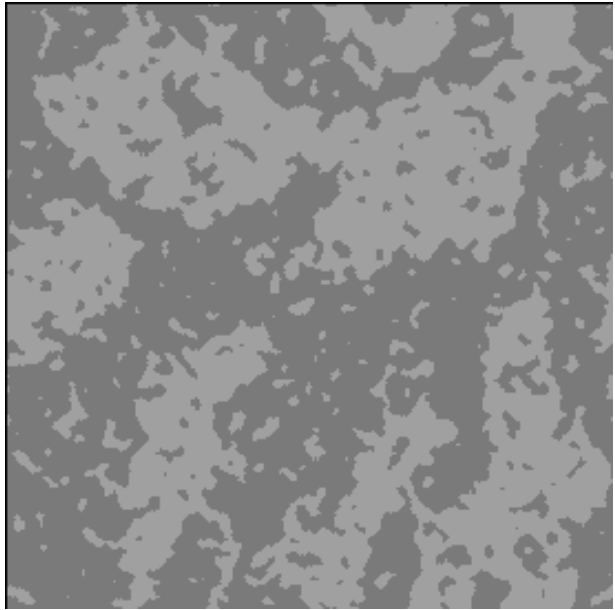}
\label{image-4.fig}
\end{center}
\caption{The projection of the spin configuration
onto the ground state $A$ in the third stage of coarsening. The snapshot is 
taken after $20$ MCS in the 3rd stage.
Comparing with Fig. \ref{image-0.fig} and  Fig. \ref{image-1-2.fig}, 
one can find that the `ghost domains' have become the `real domains' again.
The duration of the second stage ($20$ MCS)
is chosen to be much smaller than that of the first stage ($1000$ MCS) 
so that the {\it plateau regime} exists. 
}
\end{figure}

An important feature is that the domain pattern 
retrieved after the time
$\trec \ll \twz$ will remain almost frozen in the interval 
$[\trec,\twz]$ (See Fig. \ref{image-4.fig}). Thus there is 
a clear separation between the inner-coarsening regime
and outer-coarsening regime, when the retrieved domain structure expands
appreciably. We call this intermediate regime the {\it plateau regime}.
This is the mechanism which allows retrieval of the memory of 
the thermal history of the system in the present picture.

Next let us consider the case b). In this case, the noise on the
`ghost domain' is too large and the inner-coarsening finishes 
only at around $\trec \sim \twz$. The crossover from
inner-coarsening to outer-coarsening takes place very smoothly
and there is no plateau regime. In this case `memory' cannot be retrieved 
because it is impossible to recover the amplitude of 
the bias with its spatial structure frozen: the shape 
of the domain at around $\trec \sim \twz$ will be already different from installed one. 

To summarize, the separation of the inner-coarsening and outer-coarsening
regime is different in the case a) and b).
Combining \eq{eq:condition-a-b} and \eq{eq:recover-noise-cycle} we obtain,
\begin{eqnarray}
&& \mbox{a) Wide separation:} \qquad  L_{T_{A}}(\trec)  \ll  L_{T_{A}}(\twz) \nonumber \\ 
&& \mbox{b) No separation :} \qquad L_{T_{A}}(\trec)  \sim L_{T_{A}}(\twz). 
\label{eq:condition-a-b-2}
\end{eqnarray}
Remember that the amplitude $\rho_2$ of the bias is related 
to the duration of the second stage $\two$ 
through \eq{eq:rho-one-step}. Combining the latter with 
the classification \eq{eq:condition-a-b-2} we obtain a very
simple condition:
\begin{eqnarray}
&& \mbox{a) Wide separation:} \qquad L_{T_{B}}(\two) \ll  L_{T_{A}}(\twz)\nonumber \\ 
&& \mbox{b) No separation :} \qquad L_{T_{B}}(\two)  \sim L_{T_{A}}(\twz). 
\label{eq:condition-a-b-3}
\end{eqnarray}
Thus the separation between the inner- and outer-coarsening regimes
in the third regime depends on relative domain size in the
first and second stages.

Finally, let us note for completeness what is happening 
on the projection onto $B$ during the 3rd stage: 
the projection onto $B$ is becoming more and more noisy but 
still the 'ghost domains' of size $L_{T_{B}}(\two)$, 
which is the remnant of the 'real domains' of $B$
created in the second state, remain.

\subsection{AC susceptibility in One Step Cycling}
\label{subsec.ac-one-step}

It is useful to consider how the double coarsening
can be observed through the AC susceptibility.
Within the droplet picture, the relaxation of the
AC susceptibility is due to the decrease of the domain
wall density (see section \ref{subsec:droplet-domaingrowth}).
The specific scaling form \eq{eq:droplet-scaling-chi} is derived for standard 
isothermal aging where $L(\tw)$ is the size of the domain monotonically
increasing with time. In the case of the one step cycling, we only need
to replace $L(\tw)$ by the relevant domain size.

In the first stage $(0 < t < \twz)$ , the susceptibility simply decays as
standard aging \eq{eq:droplet-scaling-chi} where the relevant size
of the domain is that of the $A$ phase $L_{T_{A}}(t)$. 
In the second stage $(\twz < t < \twz+\two)$,
the relevant domains are that of the $B$ phase so the relevant 
length scale is now $L_{T_{B}}(t-\twz)$. Thus the relaxation re-starts and there is a discontinuity at $t=\twz$, as clearly observed 
in many experiments.\cite{saclay-exp-rev-1,saclay-exp-rev-2,Nordblad}

In the previous section, we argued  that 
the third stage  $(\twz+\two < t) $ can be divided into 
two asymptotic regimes, namely, an inner-coarsening 
regime for $(\twz+\two < t \ll \twz+\two+ \trec)$
and an outer-coarsening regime for $(\twz+\two+ \trec \ll  t)$.

In the inner-coarsening regime, the relevant size of domain is $L_{T_{A}}(t-\twz-\two)$. Thus the relaxation of the AC-susceptibility re-starts and there is again a discontinuity at $t=\twz+\two$. 
This feature can be observed only if the frequency $\omega$ of the AC field is large enough
compared with the inverse lifetime of the inner-coarsening regime,
\begin{equation}
\omega^{-1} \ll  \trec.
\label{eq:cond-inner-coarsening}
\end{equation}

On the other hand, in the outer-coarsening regime
the relevant domain is that of the revived ghost domains.
Thus in the latter regime, the relevant size of the domain is simply
$L_{T_{A}}(t-\two)$ and the relaxation of the AC-susceptibility
is the continuation of the first stage as if the second stage
were absent. This can be observed if the experiment is continued up
to large enough time $\twt$ compared with the lifetime of the 
inner-coarsening regime,
\begin{equation}
\twt \gg  \trec.
\label{eq:cond-outer-coarsening}
\end{equation}

\begin{figure}
\begin{center}
\leavevmode \epsfxsize=0.6\linewidth
\epsfbox{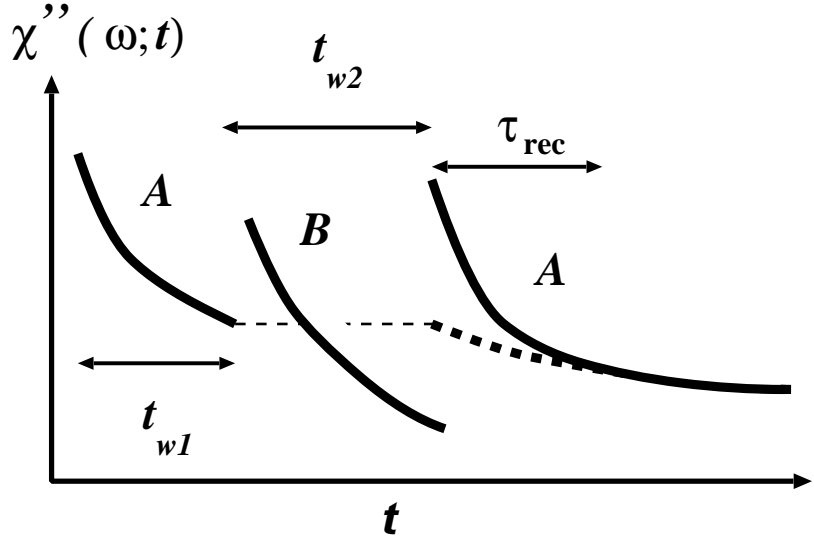}
\end{center}
\caption{Schematic behavior of the relaxation of the
out-of-phase AC-susceptibility in an one-step cycling procedure.
The thick dotted line is the reference curve which is the
direct continuation of the
first stage. `A' and `B' indicate which equilibrium state is coarsening.} 
\label{concept-one-step-ac.fig}
\end{figure}

We summarize the generic behavior of the relaxation of the
out-of-phase AC-susceptibility in the one-step cycling procedure
in Fig. \ref{concept-one-step-ac.fig}. 
Here a very important remark is that 
the strong separation of time scale \eq{eq:enhanced-separation}
in spin-glasses due to the activated dynamics can explain
the strong differences in the third regime between negative and 
positive cycling observed in experiment (this point was emphasized in
\cite{uppsala-1,KH}).
In the case of negative cycling, the outer-coarsening regime 
can be easily observed but the lifetime of the inner-coarsening regime
can be so short that \eq{eq:cond-inner-coarsening} is not satisfied.
On the contrary, positive cycling makes the inner-coarsening regime
easily observed (rejuvenation) but its effect is rapidly too large
to allow the observation of some `memory'. As we will discuss later in 
section \ref{sec:scenario}, the experimental data can be (at least qualitatively) interpreted along these lines.

\subsection{Relaxation of DC susceptibilities after One-Step Cycling}
\label{subsec.dc-one-step}

Another powerful experimental tool to study the temperature-cycling process
is the DC-magnetic susceptibilities in the third stage of the one-step cycling.
In a class of experiments \cite{saclay-exp-rev-1}, very small magnetic
field $h$ is applied during temperature-cycling 
$T_{1} (\twz) \rightarrow T_{2} (\two) \rightarrow T_{1} (\twt)$.
The magnetic field is then cut-off at time 
\begin{equation}
\twtotal \equiv \twt+\two+\twz
\end{equation}
and relaxation of the 
magnetization (thermo-remanent magnetization (TRM)) is measured
subsequently at time $\tau+\twtotal$ with increasing $\tau$
in the third stage where the temperature is kept to $T_{1}$. 

As far as linear-response holds, the magnetization
can be written as $h \chi_{\rm TRM}(\tau+\twtotal,\twtotal)$
where we introduced a dynamical DC-magnetic susceptibility.
In another class of experiments\cite{Nordblad}, 
the temperature cycling is done under zero-field. Then a small 
magnetic field is switched on 
and the growth  of the magnetization (zero-field cooled  magnetization (ZFC))
is measured.  Again the magnetization can be written as
$h \chi_{\rm ZFC}(\tau+\twtotal,\twtotal)$ where
we introduced another dynamical DC-magnetic susceptibility.
As far as linear response holds, the TRM and ZFC are simply 
related \cite{Nordblad} as 
\begin{equation}
\chi_{\rm ZFC}(\tau+\twtotal,\twtotal)+\chi_{\rm TRM}(\tau+\twtotal,\twtotal)
=\chi_{\rm ZFC}(\tau+\twtotal,0). \label{eq:zfc-trm}
\end{equation}
The rightside of the last equation
is the magnetization (divided by $h$) measured if the magnetic field
is applied from the beginning and afterwards. Experimentally
such a magnetization  almost saturates to 
a constant which is oftenly called as field cool (FC) magnetization.
If one assumes {\it naively} 
that the fluctuation dissipation theorem (FDT) holds, 
the DC-susceptibilities
(per spin) are related to auto correlation function 
$C(\tau+\twtotal,\twtotal) \equiv <m(\tau+\twtotal)m(\twtotal)>$
of magnetization $m(t)$ (per spin). 
Assuming that the correlation function is normalized as $C(t,t)=1$
one finds the ZFC susceptibility as,
\begin{equation}
\chi_{\rm ZFC}(\tau+\twtotal,\twtotal)=
(\kb T)^{-1} (1-C(\tau+\twtotal,\twtotal)).
\end{equation}
and the TRM susceptibility $\chi_{\rm TRM}(\tau+\twtotal,\twtotal)$
is related via \eq{eq:zfc-trm}.

Let us now consider the one-step cycling protocol.
Suppose that the magnetic field is applied during 
the 1st and 2nd stage and then cut-off right at the beginning
of the third stage, i.e. $\twt=0$ so that 
$\twtotal=\two+\twz$. Then TRM magnetization at time 
$\tau+\twtotal=\tau+(\two+\twz)$ should relax with increasing $\tau$
just like the auto-correlation function
between the magnetization right at time
$\two+\twz$ and that after time $\tau$ later.  
From the discussions in the previous
sections, we expect that auto-correlation 
function will be generically the following,
\begin{eqnarray}
C(\tau+(\two+\twz),\two+\twz)= & C_{0}(\tau,0) 
& \qquad L(\tau) \ll L(\trec(\rho))\qquad \mbox{inner-coarsening}  \\
& \rho C_{0}(\tau+\twz,\twz) 
& \qquad L(\tau) \gg L(\trec(\rho)) \qquad \mbox{outer-coarsening},
\label{eq:c-third-step}
\end{eqnarray}
where $C_{0}$ is the auto-correlation function
in the standard coarsening \eq{eq:lambda} and $\rho$ is the amplitude
of the ghost domain right after the 2nd stage, which decreases for
larger $\two$.
We have confirmed the above feature 
analytically  within the spherical Mattis model 
(see Fig.\ref{on_c_1.fig})  
in section  \ref{subsubsec.c-third-2} 
and numerically in two-dimensional Ising Mattis model
(see Fig.\ref{tw0-tw1=20.fig}) in section \ref{sec.ising}.
It is interesting to note that quite similar features have also been
obtained within the dynamical mean-field theory \cite{DMFT} in the sense
that the effect of the second stage amounts to a reduction of 
the plateau value $q_{\rm EA}$ at which the rejuvenation and memory
effects are separated.

Here we are assuming the case a) $L(\twz) \gg L(\two)$
which allows clear separation between the inner- and outer-coarsening regimes.
The initial decay is that due to the inner-coarsening regime,
($L(\tau) \ll L(\trec)$) where the correlation decays as if the memory of the 
first stage was completely lost. The remarkable feature
is the plateau regime $L(\trec) \ll L(\tau) \ll L(\twz)$, 
where the correlation 
function stays almost constant. The subsequent drop is due to the 
outer coarsening $L(\twz) \ll L(\tau) $ where the correlation 
function decays as if the second stage is absent. But here the
amplitude is reduced from 1 to $\rho$.
Note that in the limit $L(\twz) \rightarrow \infty$, the second
relaxation does not occur $C_{0}(\tau+\twz,\twz)=1$.
The latter is the same as the case of a cycling on a
symmetry broken state discussed in section \ref{subsec.recovery}.

In the previous TRM experiments \cite{saclay-exp-rev-1}, the field 
change is made not right at the beginning of the third stage but slightly
afterwards when some additional time $\twt$ is spent in the third stage.
The auto correlation corresponding to the DC-magnetic susceptibilities
is now $C(\tau+\twtotal,\twtotal(=\twt+\two+\twz))$ with non-zero $\twt > 0$. 
The behavior becomes more complicated because the noise is already removed to
a certain extent during the additional period $\twt$ 
thus the effect of rejuvenation tends to be obscured. 
Nonetheless we explicitly compute such 
an auto-correlation function
in section \ref{subsubsec.c-third-2} and later compare with experimental
curves in section \ref{sec:scenario}.

Finally let us note that standard FDT assumed above naively does not hold
in non-stationary dynamics as the one we are concerned here.
It is by now well known \cite{CK,CKD96,BCKM} that 
in spin-glass systems one should consider modified forms of FDT.
As compared with the AC-susceptibility discussed in the previous 
section, DC-susceptibilities contain
integral contributions of wider range of the non-stationary parts of the 
response function where the standard FDT is very likely violated.
Unfortunately, the conventional droplet picture \cite{FH88a,FH88b} 
is not able to take into account strongly non-stationary part of 
responses in spin-glasses.
Recent progress of the dynamical mean-field theories \cite{CK,FM,CKD96} 
suggests that it can be quite different  from usual coarsening systems.
Nonetheless, {\it qualitative} features of relaxation curves 
of auto-correlation functions and DC-magnetic susceptibilities, 
such as the waiting time effects, 
are known to appear very similar in the case of isothermal aging. 
The latter implies that qualitative feature discussed 
above concerning the relaxation after temperature-cycling also applies  
for the DC-magnetic susceptibilities.

\subsection{Multiple Coarsening in Multiple Step Cycling}
\label{subsec.multi}

One can naturally extend the one step cycling 
to multiple steps cycling of the equilibrium states trying to mimic
continuous temperature-cycling experiments \cite{Nordblad,JVHBN}.
For example, let us consider the coarsening of three different 
equilibrium states $A$, $B$ and $C$ which take place in turn as 
$A \rightarrow B \rightarrow C$
with durations $\twz$, $\two$ and $\twt$ respectively.
Since we are interested with large time behaviors, we disregard 
the differences between the corresponding overlap lengths 
and simply set them to the microscopic length $L_{0}$.
At the beginning of each stage, a new coarsening process is started.
After the first, second and third stages, the domain sizes of $A$, $B$ and $C$
are $L_{T_{A}}(\twz)$, $L_{T_{B}}(\two)$ and $L_{T_{C}}(\twt)$
respectively.

Let us consider the noise imprinted on the projected configuration
(onto the reference state) due to the
coarsening of unrelated phases.  The second stage reduces
intensity of the bias of the $A$ phase  from $1$ down to
$\rho_{A}(\twz+\two)\sim (L_{T_{B}}(\two)/L_{0})^{-\lambda}$.
Similarly, the third stage reduces the bias of the $B$ phase down to,
$\rho_{B}(\twz+\two+\twt)\sim (L_{T_{C}}(\twt)/L_{0})^{-\lambda}$.
An interesting question is how the third stage influences the 
projection onto the $A$ phase. A natural expectation is that the noise 
effect is {\it multiplicative}
\footnote{We explicitly verify this relation within the $O(n)$ Mattis model
in appendix \ref{sec:appendix-noise-correlation}.},
\begin{equation}
\rho_{A}(\twt+\two+\twz) \sim \rho_{A}(\twz+\two) \times (L_{T_{C}}(\twt)/L_{0})^{-\lambda} \sim (L_{T_{B}}(\two)/L_{0})^{-\lambda} 
 (L_{T_{C}}(\twt)/L_{0})^{-\lambda}.
\end{equation}
To summarize, we obtain a `real' domain of phase $C$ with
size $L_{T_{C}}(\twt)$ and  `ghost' domains of phases $A$ and $B$,
with sizes $L_{T_{A}}(\twz)$ and  $L_{T_{B}}(\two)$ respectively
but with reduced intensities. Thus the information of all the three
phases are now stored in the spin configuration but with noises due to
random interferences.

Now let us consider how we can retrieve memories 
installed above by removing the noise that blurred stored information. 
Here we have to remember that  memory can be retrieved by additional 
conjugate coarsening but only in the case a): when the noise 
is small enough (see \eq{eq:condition-a-b}-\eq{eq:condition-a-b-3}).

To be specific, let us suppose that the previous coarsening of $A$, $B$,
and $C$ is  done in a well separated manner in the sense that
$\twz \gg \tau_{0} (\two/\tau_{0})^{T_{B}/T_{A}}$ and
$ \two \gg \tau_{0} (\twt/\tau_{0})^{T_{C}/T_{B}}$.
Then let us consider reversed order coarsening of the above 
procedure. First, we perform coarsening of $B$ 
with durations  $\two'$ given the final spin configuration obtained above.
According to the result of the previous section, it is sufficient to choose
the duration $\two' \sim \trec= \tau_{0} (\twt/\tau_{0})^{T_{C}/T_{B}}$
to remove the noise on $B$ due to $C$ and recover the spin configuration 
just before the coarsening of $C$. We will not choose a larger $\two'$ 
in order to avoid the outer-coarsening of $B$ which adds some additional 
noise to $A$, which will be treated later. Second we perform 
coarsening of $A$. By the same argument, it will be again enough to choose 
its duration as 
$\twz' \sim \trec= \tau_{0} (\two/\tau_{0})^{T_{B}/T_{A}}$.

If one skips the 2nd coarsening of $B$ and try to remove
the noise by doing only the 2nd coarsening of  $A$, the time need
for the recovery $\trec$ will become astronomically large,
\begin{equation}
(L(\trec)/L_{0})^{-\lambda} = \rho_{A}\times \rho_{B}
\qquad \mbox{or} \qquad \trec=\tau_{0} 
(\two/\tau_{0})^{(T_{B}/T_{A})(T_{B}/T_{A})\log(\twt/\tau_{0})}
\end{equation}
Thus it is very important that the multiplicative noise is cured in
two steps and not by a single stroke. 

\begin{figure}
\begin{center}
\leavevmode \epsfxsize=0.6\linewidth
\epsfbox{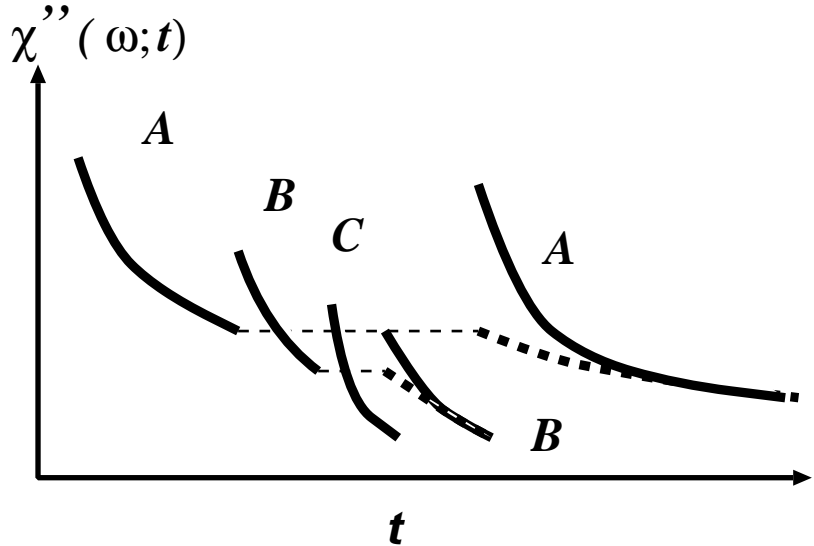}
\end{center}
\caption{Schematic behavior of the relaxation of the
out-of-phase AC-susceptibility in a two-step cycling procedure.
Thick dotted lines are the reference curves which show direct continuation of the
second and first stages.  `A', `B' and `C'  indicate 
the equilibrium states which are coarsening.
} 
\label{concept-two-step-ac.fig}
\end{figure}

More generally one can perform successive coarsening of arbitrary 
number of phases $A_{1} \rightarrow A_{2} 
\ldots  \rightarrow A_{n}$ with durations $\two$, $\twt$, $\ldots$, $t_{n}$
followed by the reversed processes $A_{n-1} \rightarrow A_{n-2} 
\ldots  \rightarrow A_{1}$ with durations $t'_{n-1}$, $t'_{n-2}$, $\ldots$, $t'_{1}$. The retrieval of the memory of each phase is ensured by the condition,
\begin{equation}
t_i \gg \tau_{0} (t_{i+1}/\tau_{0})^{T_{i+1}/T_{i}} \qquad t'_{i}\sim t_{i}.
\label{eq:retrieval-condition}
\end{equation} 
Not that is can be easily realized in negative cycling.
The domain structures of all the phases will be retrieved one after another.
\footnote{The phases of large $i$ which are recovered earlier
become noisy again as phases of smaller $i$ are treated later.}
The behavior of the AC-susceptibility during such a multi-step
(continuous) cycling will be interesting.  The generic behavior will be
`hierarchical' as depicted schematically in
Fig. \ref{concept-two-step-ac.fig}.
In section \ref{sec:scenario}, we discuss recent continuous
temperature-cycling experiments from this point of view.

\section{The $O(n)$ Mattis Model}
\label{sec.spherical}

In the previous section we discussed the peculiar properties of coarsening 
under a cycling of the equilibrium states, based on
scaling and heuristic arguments. 
In this section we study the Mattis model introduced in 
section \ref{subsec.mattis} in the spherical limit. 
This allows us to explicitly solve the dynamical equation under one-step cycling of 
the equilibrium states and check in details the physical picture presented
in the above sections.  
This section is rather technical and can be skipped at
first reading (the previous section is indeed the summary of the study 
in the present section and section \ref{sec.ising}). 
The reader interested by a more physical discussion can go directly to
section \ref{sec:scenario}.

\subsection{Model and Definitions}

\subsubsection{Dynamics of Projection Field}

Here we generalize the projection field $\tilde{S}$ to be 
a $n$-dimensional vector field
$\vec{\phi}(x,t)$ continuously varying in a $d$-dimensional space.  
The time evolution of the projection field $\vec{\phi}(x,t)$ is given by 
the time dependent Ginzburg-Landau (TDGL) equation, 
\begin{equation}
\frac{\partial \vec{\phi}}{\partial t}
=-\frac{\delta F_{[\vec{\phi}]}}{\delta \vec{\phi}},
\end{equation}
with the free-energy functional defined as,
\begin{equation}
F_{[\vec{\phi}]}=\int d^{d} x 
\left[\frac{1}{2} (\nabla \vec{\phi})^{2} 
+  \alpha \frac{(n-|\vec{\phi}|^{2})^{2}}{4n} \right].
\end{equation}
Here we disregard the Langevin force  due the to thermal noise
on the dynamics because the latter is irrelevant in the present model \cite{B94}.

In the spherical limit $n \rightarrow \infty$, under the assumption
of self-averageness, any one of the component satisfies,
\begin{equation}
\frac{\partial \phi}{\partial t}
= \nabla^{2} \phi -z(t)\phi,
\label{eq:TDGL}
\end{equation}
with
\begin{equation}
z(t)=\mu (1- \langle\phi^{2}(t)\rangle), \label{eq:self-con}
\end{equation}
where $\langle...\rangle$ means expectation value.  Thus we only need to 
consider a single component, i.e. a scalar field in the following.

\subsubsection{Random Equilibrium States}

We suppose that the equilibrium configuration 
is represented by a random scalar field $\sigma(x)$.
The spin configuration $\psi(x)$ is related to the projection field as,
\begin{equation}
\psi(x)=\sigma(x)\phi(x),
\label{eq:psi-phi}
\end{equation}
which is equivalent to the relation  \eq{eq:real-projection} 
on the lattice. 
We assume that $\sigma(x)$ is a Gaussian random scalar field with zero mean,
\begin{equation}
\langle\sigma(x)\rangle_{\sigma}=0 
\end{equation}
and short-ranged spatial correlations,
\begin{equation}
\langle\sigma(x)\sigma(x')\rangle_{\sigma}=\Delta \delta^{d}(x-x'). 
\label{eq:sig-sig}
\end{equation}
The latter means that we essentially disregard the finiteness 
of the overlap length \eq{eq:lovlp}. One could include this effect by
introducing a short-range correlated Gaussian field. However since we
are focusing on large time behavior, we do not go into such details in the present paper.

As for the lattice case \eq{eq:projection-projection}, the projections to different 
equilibrium states, say $\alpha$ and $\beta$, are related as,
\begin{equation}
\phi^{\alpha}(x)=\sigma^{\alpha \beta}(x)\phi^{\beta}(x).
\label{eq:intro-transform-1}
\end{equation}
where we defined a `transformation field',
\begin{equation}
\sigma^{\alpha \beta}(x)\equiv\sigma^{\alpha}(x)\sigma^{\beta}(x).
\label{eq:intro-transform-2}
\end{equation}

The transformation field should be also a random field with zero mean,
\begin{equation}
\langle\sigma^{\alpha \beta}(x)\rangle_{\sigma}=0 \label{eq:av-sig}
\end{equation}
and short-ranged correlation,
\begin{equation}
\langle\sigma^{\alpha \beta}(x)\sigma^{\alpha \beta}(x')\rangle_{\sigma}
=\Delta \delta^{d}(x-x').
\label{eq:corr-sig}
\end{equation}
More generally we have,
\begin{eqnarray}
 \langle\sigma^{\alpha_1 \alpha_2}(x_1)\sigma^{\alpha_2 \alpha_3}(x_2)
&\ldots &\sigma^{\alpha_{n-1} \alpha_{n}}(x_{n-1}) \sigma^{\alpha_n \alpha_1}(x_n)\rangle_{\sigma} \nonumber \\
&=&\Delta^{n} \delta^{d}(x_{1}-x_{2}) \delta^{d}(x_{2}-x_{3})\ldots
\delta^{d}(x_{n-1}-x_{n}).
\label{eq:corr-sig-general}
\end{eqnarray}
Some useful statistical properties of the 
transformation field are presented in appendix \ref{sec:transformation-field}.

\subsubsection{Formal Solution}

Taking Fourier transform $\hat{\phi}_{k}=\int d^{d}x \phi(x) e^{ikx}$,
one finds the formal solution to the TDGL equation \eq{eq:TDGL} as,
\begin{equation}
\hat{\phi}_{k}(t)=\hat{\phi}_{k}(t')
\frac{e^{-k^{2}(t-t')}}{\sqrt{\Gamma(t,t')}}, \label{eq:formal}
\end{equation}
where we introduced
\begin{equation}
\Gamma(t,t')=\exp \left(2\int_{t'}^{t}dt'z(t') \right).
\label{eq:def-gamma}
\end{equation}
Note that details of the solution are absorbed in $\Gamma(t,t')$.
By definition one must have the identity,
\begin{equation}
\Gamma(t,t)=1.\label{eq:gamma-0}
\end{equation}

To simplify our calculation, we consider 
$\mu \rightarrow \infty$ to enforce
the normalization of the magnitude of spin at any position in space and 
time: $\phi^{2}(x,t)\equiv 1$.\cite{normalization}
In Fourier space, the latter implies that the formal solution 
can be written as,
\begin{equation}
\hat{\phi}_{k}(t)\hat{\phi}_{l}(t)
=(2\pi)^{2d} \delta^{d}(k+l) W_{k}(t), \label{eq:norm-prop-1}
\end{equation}
Here $W_{k}(t)$  should satisfy,
\begin{equation}
\int d^{d}k \ W_{k}(t)=1. \label{eq:norm-prop-2}
\end{equation}
Note that $W_{k}(t)$ is the structure-factor of the projection field.
The factor $\Gamma(t,t')$ (or equivalently $z(t)$ )
can be determined self-consistently so as to satisfy
the normalization condition $\phi(x,t)^{2}=1$,
\begin{equation}
\Gamma(t,t')=\int \frac{d^{d}k}{(2\pi)^{d}}\frac{d^{d}l}{(2\pi)^{d}} 
e^{-k^{2}(t-t')}e^{-l^{2}(t-t')}
\langle\hat{\phi}_{k}(t')\hat{\phi}_{l}(t')\rangle
=\int d^{d}k e^{-2k^{2}(t-t')} W_{k}(t'). \label{eq:gamma-cal}
\end{equation}
Then the structure-factor \eq{eq:norm-prop-1} at time $t$ is obtained formally as
\begin{equation}
W_{k}(t)=W_{k}(t')\frac{e^{-2k^{2}(t-t')}}{\Gamma(t,t')}. 
\label{eq:evolution-structure-factor}
\end{equation}

\subsubsection{Physical Observables}

The properties of coarsening systems can be
well characterized using correlation functions. Given 
a structure-factor \eq{eq:norm-prop-1} at a certain time $s$, the 
auto-correlation function for two times $t > t' > s$ can 
be formally computed as,
\begin{equation}
C(r,t,t') = \langle\phi(r,t)\phi(0,t')\rangle=\int d^{d}k 
\frac{e^{-k^{2}((t-s)+(t'-s))}e^{ikr}
W_{k}(s)}{\sqrt{\Gamma(t,s)}\sqrt{\Gamma(t',s)}}.
\label{eq:how-to-gene-c}
\end{equation}
In particular, the equal-time $t'=t$ spatial correlation function is obtained 
simple as the inverse Fourier transform in space of the structure factor $W_{k}(t)$,
\begin{equation}
C(r,t,t)=\int d^{d}k e^{ikr} W_{k}(t).
\label{eq:how-to-c-space}
\end{equation}
Another important quantity is the local auto-correlation function,
\begin{equation}
C(r=0,t,t') 
=  \frac{\Gamma((t+t')/2,s)}{\sqrt{\Gamma(t,s)}\sqrt{\Gamma(t',s)}}.
\label{eq:how-to-c}
\end{equation}
In the following we denote the (local) auto-correlation function
$C(r=0,t,t')$ as $C(t,t')$ for simplicity.
Since coarsening is a non-stationary dynamics, the correlation 
functions depend not only on the time difference $t-\tw$ but 
explicitly on the two times $t$ and $\tw$.
This feature is called `waiting time effect' or 
`violation of time translational invariance'.

In addition to the correlation functions, linear-response functions
are also very interesting to study. In the spin-glass experiments, the 
measurement of the linear-response such as AC magnetic susceptibility
is one of the only detailed probe for the dynamics.
In appendix \ref{sec:response} we present the formal solution for
the linear-response function of the $O(n)$ Mattis model 
to uniform external field.

\subsection{Standard Coarsening}

Before studying coarsening under cycling of equilibrium states, let us review
essential results in the case of standard coarsening, i. e.
coarsening with un-biased random configuration with short-range correlation. 
The solution is well known and studied in detail. \cite{B94} 

Let us choose the random initial condition as,
\begin{equation}
\langle\hat{\phi}_{k}\hat{\phi}_{l}\rangle_{\rm ini}
=\Delta (2\pi)^{d} \delta^{d}(k+l),
\label{eq:random-ini-k}
\end{equation}
which is equivalent to
\begin{equation}
\langle\phi(x)\phi(x')\rangle_{\rm ini}=\Delta \delta^{d}(x-x'). 
\label{eq:random-ini}
\end{equation}
The latter means the structure-factor \eq{eq:norm-prop-1} is flat (white noise)
at the beginning,
\begin{equation}
W_{k}(0)=\frac{\Delta}{(2\pi)^{d}}. \label{eq:w-random}
\end{equation}
Then the solution is obtained using \eq{eq:gamma-cal} as,
\begin{equation}
\Gamma_{0}(t,0)=\frac{\Delta}{(2\pi)^{d}}  \int d^{d}k  e^{-2k^{2}t}
=\left (\frac{t}{\tau_{0}} \right)^{-d/2}
\sim (L(t)/L_{0})^{-2\lambda}, \label{eq:random-solution}
\end{equation}
where we defined a microscopic time scale $\tau_{0}=\Delta ^{2/d}/(8\pi)$. 
Note that this expression is valid for large enough time separation $t$ 
compared with $\tau_{0}$. In the limit $t=0$ we must have the identity
\eq{eq:gamma-0}. For the definition of $L(t)$ and the exponent
$\lambda$ in the last equation see \eq{eq:lt} and the following. 

The correlation functions can be obtained using \eq{eq:how-to-c-space},
\eq{eq:how-to-c} and \eq{eq:w-random}.
The spatial correlation function at equal-time becomes,
\begin{eqnarray}
C_{0}(r,t,t)=\langle\phi(r,t)\phi(0,t)\rangle &= &
\frac{\Delta}{(2\pi)^{d}}  \int d^{d}k \frac{e^{-2k^{2}t} e^{ikr}}{\Gamma_0(t,0)}
\nonumber \\
&=& \exp \left (-\frac{r^{2}}{8t} \right)
\equiv \exp \left [-\left (\frac{r}{L(t)} \right)^{2} \right],
\label{eq:domain-growth-random}
\end{eqnarray}
In the last equation we introduced a
characteristic length scale,
\begin{equation}
L(t) \propto L_{0}\sqrt{t}, \label{eq:lt}
\end{equation}
where $L_{0}$ is some microscopic unit of length.
Although there are no topological defects like domain walls in the
spherical limit $n \rightarrow \infty$, 
the latter characteristic length scale plays 
the role of scaling variable as played by the mean separation of domain 
walls in the systems with domain walls \cite{B94}.

The correlation between the random
initial configuration and the temporary configuration at time $t$
is obtained as
\begin{equation}
C_{0}(t,0)=\langle\phi^{\beta}(t)\phi^{\beta}(0)\rangle
=\frac{\Gamma(t/2,0)}{\sqrt{\Gamma(t,0)}}
=2^{d/2}(t/\tau_{0})^{-d/4} \sim (L(t)/L_{0})^{-\lambda}. \label{eq:c}
\end{equation}
The non-equilibrium dynamical exponent $\lambda$ (see \eq{eq:lambda}) 
of the $O(n)$ model is known to be,
\begin{equation}
\lambda=d/2 \qquad \mbox{$O(n \rightarrow \infty)$ model},
\label{eq:lambda-on}
\end{equation}
as one can see easily comparing with \eq{eq:lt}. 
More generally the two-time auto-correlation functions is obtained as,
\begin{equation}
C_{0}(t,\tw) =  \langle\phi^{\beta}(t)\phi^{\beta}(\tw)\rangle
=\frac{\Gamma((t+\tw)/2,0)}{\sqrt{\Gamma(t,0)}\sqrt{\Gamma(\tw,0)}}
 \sim_{t \gg \tw} (L(t)/L(\tw))^{-\lambda}. \label{eq:c-two-time}
\end{equation}
Here the waiting time effect follows the $L(t)/L(\tw)$ type scaling
behavior as in many other coarsening systems.\cite{B94}

\subsection{A Cycle on a Symmetry Broken State}
\label{subsec.rec-fsbs}

Let us now begin to analyze the effect of cycling the equilibrium states
on coarsening with the simplest version described in section
\ref{subsec.noise}. Here we look at how the projection of the temporary spin-configuration
onto an equilibrium state is progressively affected by the coarsening 
of a completely unrelated phase. Subsequently we study in detail how the noise 
induced by this process can be removed progressively  by performing the `conjugate' coarsening.

\subsubsection{Noise Imprinting}

Let us take a  random ground state 
$\{ \sigma^{\alpha}_{i} \}$ as the initial condition
so that the symmetry is fully broken with respect to $\alpha$ at the
beginning, 
\begin{equation}
\hat{\vec{\phi}}_{k}^{\alpha}(0)=(2\pi)^{d}\delta^{d}(k) \label{eq:fully-broken-ini}.
\end{equation}
We then perform coarsening with respect to 
a completely unrelated ground state $\{\sigma^{\beta}_{i} \}$.
The initial condition should look as a completely random configuration
in the projection onto $\beta$. Due to \eq{eq:transform} 
which is the Fourier transform of \eq{eq:psi-phi} we find, 
\begin{equation}
\hat{\vec{\phi}}_{k}^{\beta}(0)=\int \frac{d^{d}k'}{(2\pi)^{d}}
(\hat{\sigma}^{\alpha \beta})_{k'}(2\pi)^{d}\delta^{d}(k+k')
=(\hat{\sigma}^{\alpha \beta})_{k} ,
\end{equation}
Here $(\hat{\sigma}^{\alpha \beta})_{k}$ is 
the transformation field defined in \eq{eq:trans-field}
which is a Gaussian random field (see appendix \ref{sec:transformation-field}).
By \eq{eq:map-noise-mean} and \eq{eq:map-noise-correlation}, one can check
that the initial condition for the coarsening with respect to $\beta$
is indeed a random initial condition with zero mean 
and short range correlation as it should. The solution of the equation
of motion with such random initial condition is known as shown in
\eq{eq:random-solution}.

Let us  monitor the time evolution of the spin-configuration
by projecting onto $\alpha$ through \eq{eq:transform},
\begin{equation}
\hat{\phi}^{\alpha}_{k}(t)=
\int \frac{d^{d}k'}{(2\pi)^{d}}
(\hat{\sigma}^{\alpha \beta})_{k'}\hat{\phi}^{\beta}_{k-k'}(t).
\end{equation}
Using the 2-body correlation function of 
$(\hat{\sigma}^{\alpha \beta})_{k}$ shown in \eq{eq:map-noise-correlation}, one finds 
that the resultant configuration have the following properties.

First, it is easy to see that the $k=0$ component of the projection onto $\alpha$ has non-zero mean
while the others have zero mean,
\begin{equation}
\langle\hat{\phi}_{k}^{\alpha}(t)\rangle_{\sigma}=(2\pi)^{d} \delta^{d}(k)\ \rho
\label{eq:reduced-bias}
\end{equation}
or 
\begin{equation}
\langle\hat{\phi}^{\alpha}(x,t)\rangle_{\sigma}=\rho
\end{equation}
with
\begin{equation}
\rho=C_{0}(t,0) \sim (L(t)/L_{0})^{-\lambda} \label{eq:rho-c}.
\end{equation}
The result means that the symmetry remains broken with a weaker and
weaker bias $\rho$ as the coarsening time $t$ with respect 
the unrelated phase $\beta$ increase. This feature has been discussed in
section \ref{subsec.noise} (see \eq{eq:rho}).

Second, the spatial correlation of the projection to $\alpha$
can also be obtained
as described in appendix \ref{sec:appendix-noise-correlation}. 
The initial condition
\eq{eq:fully-broken-ini} implies that initial structure-factor is
$W^{\alpha}_{k}(0)=\delta^{d}(k)$ in \eq{eq:norm-prop-1}.
Then from \eq{eq:coarsening-noise-corr-2}, we obtain the correlation function
as,
\begin{equation}
\langle\hat{\phi}_{k}^{\alpha}(t)\hat{\phi}_{l}^{\alpha}(t)\rangle_{\sigma}
= (2\pi)^{2d} \delta^{d}(k+l) W^{\alpha}_{k}(t), \label{eq:noisy-field}
\end{equation}
with the structure-factor,
\begin{equation}
W^{\alpha}_{k}(t)=\rho^{2}\delta^{d}(k)+\frac{\Delta}{(2\pi)^{d}}
(1-2\rho^{2}+e^{-k^{2}t/2}). \label{eq:structure-factor-randomized}
\end{equation}
By taking the inverse Fourier transform one finds,
\footnote{In the derivation of the last equation 
we used $\Gamma(t/4,0)=C^{2}(t,0)=\rho^{2}$ 
as one can check from \eq{eq:c} and \eq{eq:random-solution}.}
\begin{eqnarray}
 \langle\phi^{\alpha}(x,t)\phi^{\alpha}(x',t)\rangle_{\sigma}
&-&\langle\phi^{\alpha}(x,t)\rangle_{\sigma}\langle\phi^{\alpha}(x',t)\rangle_{\sigma}  \nonumber \\
&=& \Delta d^{d}(x-x') [1-2\rho^{2}]+ e^{-(x-x')/2t}\rho^{2}.
\end{eqnarray}
To summarize, coarsening of unrelated phase produces 
essentially short-ranged correlated random field with 
weak bias as we expected in section \ref{subsec.noise}.

\subsubsection{Noise Cleaning}
\label{subsubsec.noise-removal-On}

\newcommand{\mb}{m_{\rm mem}}
\newcommand{\mf}{m_{\rm rej}}

\newcommand{\gcycle}{\Gamma_{\rm cycle}}

Let us stop the coarsening of the unrelated phase $\beta$ at time $t=\two$
and see closely how the noise imprinted on the projection field to $\alpha$ are removed 
by coarsening with respect to $\alpha$ for some additional time $\twt$.
Given $\phi^{\alpha}(x,\two)$ obtained in the last stage as the initial condition for this stage,
the solution at time $\twt+\two$ is obtained as
\begin{equation}
\hat{\phi}^{\alpha}_{k}(\twt+\two)=\hat{\phi}_{k}^{\alpha}(\two)
\frac{e^{-k^{2}\twt}}{\sqrt{\gcycle(\twt+\two,\two)}}. 
\end{equation}
Here the factor $\gcycle(\twt+\two,\two)$ is obtained
using \eq{eq:gamma-cal}, \eq{eq:structure-factor-randomized}
 and \eq{eq:random-solution} as,
\footnote{One can check that $\Gamma(\two,\two)=1$ is satisfied since
$\Gamma(\two/4)=C^{2}(\two,0)=\rho^{2}$ as one can check
and $\Gamma_{0}(0,0)=1$ because of the identity \eq{eq:gamma-0}.}
\begin{eqnarray}
\gcycle(\twt+\two,\two) 
 =  \rho^{2} + (1-2 \rho^{2})\Gamma_{0}(\twt,0)
+ \Gamma_{0}(\twt+\two/4,0),
\label{eq:gamma-complex}
\end{eqnarray}
with
\begin{equation}
\rho=C_{0}(\two,0) \sim (L(\two)/L_{0})^{-\lambda}.
\end{equation}
Here $\Gamma_{0}$ is the one obtained in the solution for
standard coarsening \eq{eq:random-solution} with un-biased 
random initial condition.

For the following analysis, it is useful to 
introduce the relative ratios of the three terms in \eq{eq:gamma-complex}, which will find a natural 
interpretation later:
\begin{eqnarray}
&& \mb^{2}(\twt,\two)  =   \rho^{2}/\gcycle(\twt+\two,\two),
\label{eq:m-para-retrieval} \\
&& \mf^{2}(\twt,\two)  =  (1 -  2 \rho^{2}) \Gamma_{0}(\twt,0)/
\gcycle(\twt+\two,\two),
\label{eq:r-para-retrieval} \\
&& \tilde{m}^{2}(\twt,\two)  =  
\Gamma_{0}(\twt+\two/4,0)/\gcycle(\twt+\two,\two).
\label{eq:tilde-m-para-retrieval}
\end{eqnarray}
By definition the sum of the three is always $1$.

Suppose that $\two$ has been chosen to sufficiently large.
Then for small enough $\twt$ compared with $\two$, $\mf$ is dominant,
\begin{equation}
\mf^{2} \simeq 1 \qquad \mbox{or} \qquad \gcycle(\twt+\two,\two) \simeq 
\Gamma_{0}(\twt,0) \sim (L(\twt)/L_{0})^{-2\lambda}.
\qquad \twt \ll \trec 
\label{eq:gamma-complex-short}
\end{equation}
here we used \eq{eq:random-solution}. On the other hand, at large
enough time $\twt$, $\mf$ and $\tilde{m}^{2}$ go to zero,
\begin{equation}
\mb^{2} \simeq 1 \qquad \mbox{or} \qquad \gcycle(\twt+\two,\two) 
\simeq \rho^{2}
\qquad \twt \gg \trec. 
\label{eq:gamma-complex-long}
\end{equation}
The crossover between the two limits
takes place when the ratio $\mf^{2}$ and $\mb^{2}$
becomes of the same order. (we are assuming $\tilde{m}^{2} \ll \mf^{2}$).
Then using \eq{eq:random-solution} one finds 
the crossover time $\trec(\rho)$ as,
\begin{equation}
\rho \sim  (L(\trec(\rho))/L_{0})^{-\lambda} \label{eq:taueq-rho}.
\end{equation} 
The above feature has direct consequences on the
physical observables.

First, let us consider the density of staggered magnetization.
Using \eq{eq:reduced-bias}, it is obtained simply as,
\begin{eqnarray}
\langle\phi^{\alpha}(\twt+\two)\rangle_{\sigma}
&=&\int \frac{d^{d}k}{(2\pi)^{d}}
\frac{e^{-k^{2}\twt}}{\sqrt{\gcycle(\twt+\two,\two)}} \langle\phi_{k}^{\alpha}(\two)\rangle_{\sigma}
\nonumber \\
& = &  \frac{\rho}{\sqrt{\gcycle(\twt+\two,\two)}} = \mb(\twt,\two).
\label{eq:m-recover}
\end{eqnarray}
In the last equation we used
the definition of  $\mb^{2}(\twt,\two)$ given in \eq{eq:m-para-retrieval}.
Thus $\mb(\twt,\two)$ is the staggered 
magnetization in the bulk, that will contribute to the memory effect.
The staggered magnetization starts from $\rho$ and saturates to the 
full moment $1$ at time scales large enough compared with
$\trec(\rho)$ defined  in \eq{eq:taueq-rho}. It means that
the density of the minority phase shrinks and
the fully symmetry broken state is almost recovered within 
a finite time scale.

Second, let us consider the equal time spatial correlation 
function of fluctuation of the projected field around the 
mean $\langle\phi^{\alpha}(\twt+\two)\rangle_{\sigma}$,
\begin{equation}
\delta \phi^{\alpha}(r,\twt+\two)=\phi^{\alpha}(r,\twt+\two)-\langle\phi^{\alpha}(\twt+\two)\rangle_{\sigma}.
\end{equation}
Using \eq{eq:how-to-c-space}, \eq{eq:evolution-structure-factor},
\eq{eq:gamma-complex} and \eq{eq:structure-factor-randomized} and
\eq{eq:m-recover} it is obtained as,
\begin{eqnarray}
&& \langle\delta \phi^{\alpha}(r,\twt+\two) 
\delta \phi^{\alpha}(0,\twt+\two)\rangle_{\sigma}
 =   \int d^{d}k 
\left [  \frac{\Delta}{(2\pi)^{d}} \left (
1-2\rho^{2}+e^{-k^{2}\two/2} \right)
\right ] \frac{e^{-2k^{2}\twt}}{\gcycle(\twt+\two,\two)} \nonumber \\
& = & \mf^{2}(\twt,\two)
\exp \left (-\frac{r^{2}}{8\twt} \right)  
+  \tilde{m}^{2}(\twt,\two)
 \exp \left (-\frac{r^{2}}{8(\twt+\two/4)} \right).
\end{eqnarray}
In the last equation we used the parameters defined in 
\eq{eq:r-para-retrieval} and \eq{eq:tilde-m-para-retrieval}.
This result should be compared with the case of coarsening with un-biased
initial condition \eq{eq:domain-growth-random}.
For simplicity we assume that $\two$ has been taken very large so that
the second term can be neglected.
Then one finds that at short enough time scale 
the amplitude of the correlation function stays constant 
$\mf^{2} \simeq  1$ because of \eq{eq:gamma-complex-short}.
In this regime the behavior of the correlation function 
is essentially the same as in the usual case of  coarsening 
with un-biased initial condition \eq{eq:domain-growth-random}; this
contribution will therefore be associated to rejuvenation.
The range of correlation grows as,
\begin{equation}
\langle\delta \phi^{\alpha}(r,\twt+\two)
\delta \phi^{\alpha}(0,\twt+\two)\rangle_{\sigma} \simeq 
\exp \left [- \left (\frac{r}{L(\twt)}\right)^{2} \right] 
\qquad \twt \ll \trec.
\end{equation}
At larger  time scale compared with $\trec$, 
the amplitude $\mf^{2}$ vanishes because of \eq{eq:gamma-complex-long}
and the fluctuation disappears, i.e. the system is ordered again,
\begin{equation}
\langle\delta \phi^{\alpha}(r,\twt+\two)
\delta \phi^{\alpha}(0,\twt+\two)\rangle_{\sigma} \simeq 0
\qquad \twt \gg \trec.
\end{equation}

Finally let us consider the auto-correlation functions.
The simplest one which is useful is the correlation between 
the configuration at the beginning of the final coarsening with that 
at some later time. One easily obtains,
\begin{equation}
C(\twt+\two,\two) =
\frac{\gcycle(\twt/2+\two,\two)}{\sqrt{\gcycle(\twt+\two,\two)}}.
\end{equation}
This result should be compared with the case with zero bias \eq{eq:c}.
One can easily see that at the beginning $\twt \ll \trec$
where $\gcycle(\twt+\two,\two) \simeq \Gamma_{0}(\twt,0)$ 
\eq{eq:gamma-complex-short} holds,
the correlation function decreases as if starting from random 
initial conditions without bias,
\begin{equation}
C(\twt+\two,\two) \simeq C_{0}(\twt,0) \sim (L(\twt)/L_{0})^{-\lambda} \qquad \twt \ll \trec.
\label{eq:c-decay}
\end{equation} 
On the other hand it saturates at large enough time scales due to
$\gcycle(\twt+\two,\two) \simeq \rho^{2}$, see \eq{eq:gamma-complex-long}, as
\begin{equation}
C(\twt+\two,\two) = \rho \qquad \twt \gg \trec.
\label{eq:c-stop}
\end{equation} 
In section \ref{subsec.noise}, we expected this feature 
on general grounds and used it to estimate the recovery
time \eq{eq:recover-noise}.

\subsection{Double Coarsening in One Step Cycling}
\label{subsec.double-on}

Now we analyze the double coarsening process discussed in section
\ref{subsec.double} in the present specific model.
We first let the system coarsen towards the equilibrium state $\alpha$
for time $\twz$ starting from a random initial condition \eq{eq:random-ini-k}. 
Then we change the target state to $\beta$ for a time $\two$, given the configuration 
obtained above as the initial configuration.
And finally,  we switch back the coarsening towards the original 
equilibrium state $\alpha$ for time $\twt$. Note that the process 
we considered  in the previous
subsection can be regarded as the special case of $\twz=\infty$.

\subsubsection{The First and Second Coarsening}
\label{subsubsec.first-second-on}

The first stage is the usual coarsening from random initial conditions.
The structure-factor of the projection field 
with respect to $\alpha$ after time $\twz$ 
is given in \eq{eq:w-random} which reads as,
\begin{equation}
W^{\alpha}_{k}(\twz)=\frac{\Delta}{(2\pi)^{d}} 
\frac{e^{-2k^{2}\twz}}{\Gamma_{0}(\twz,0)}. \label{eq:w-start}
\end{equation}

As we discuss in appendix \ref{sec:appendix-noise-correlation}, the projection of this
configuration onto $\beta$ is random with zero mean  
\eq{eq:coarsening-noise-meanr-1} and short range correlation 
\eq{eq:coarsening-noise-meanr-2}.
Thus a new coarsening process begins in the second stage. 
After time $\two$, the correlation with
the configuration at time $\twz$ is,
\begin{equation}
\rho=C(\two,0) \sim (L(\two)/L_{0})^{-\lambda}.
\label{eq:rho-decay}
\end{equation}

In the following we analyze the time evolution of the coarsening in the second stage
(towards $\beta$) by projecting onto $\alpha$.
The basic information is the structure-factor of the projection field with 
respect to $\alpha$ which can be obtained using
\eq{eq:structure-factor-back-to-alpha} and \eq{eq:w-start}, 
\begin{eqnarray}
W^{\alpha}_{k}(\two+\twz) &=& 
  \rho^{2} W^{\alpha}_{k}(\twz)
+   \frac{\Delta}{ (2\pi)^{d}}  
\left [ 1 -  2 \rho^{2} \right.   \nonumber \\
& + &  \left. \frac{1}{\Gamma_{0}(\two,0)} \frac{\Delta}{(2\pi)^{d}} 
\int d^{d}k' \int d^{d}l'
e^{-(k-k')^{2}\two}e^{-(k+l')^{2}\two} W^{\alpha}_{k-k'+l}(\twz)
 \right ] \nonumber \\
& = &   \rho^{2} \frac{\Delta}{(2\pi)^{d}} 
\frac{e^{-2k^{2}\twz}}{\Gamma_{0}(\twz,0)}
 +   \frac{\Delta}{ (2\pi)^{d}} ( 1 -  2 \rho^{2}) \nonumber \\
& + & \rho^{2}
\frac{\Delta}{(2\pi\sqrt{A(\twz/\two)})^{d}} 
\frac{\exp \left (-2k^{2}\twz/A(\twz/\two) \right)}{\Gamma_{0}(\twz,0)}.
\label{eq:w-after-tw1}
\end{eqnarray}
where
\begin{equation}
A(y)\equiv 1+ 4y.
\label{eq:def-A}
\end{equation}

The spatial correlation function of the projection with respect 
to $\alpha$ is obtained immediately using \eq{eq:w-after-tw1} 
in \eq{eq:how-to-c-space} as,
\begin{eqnarray}
 C(r,\two+\twz,\two+\twz)
&=& \rho^{2}\exp \left (-\frac{r^{2}}{8\twz} \right)
+ (1 -  2 \rho^{2}) \Delta \delta^{d}(r) \nonumber \\
&+& 
\rho^{2} A^{-d/2}(\twz/\two)\exp \left (-\frac{r^{2}}{8\twz/A(\twz/\two)} \right). \label{eq:c-space-2nd}
\end{eqnarray}
This simple result immediately allows its physical interpretation. 
The first term in \eq{eq:c-space-2nd} can be written as
$\rho^{2} \exp(-(r/L(\twz))^{2})$ which implies the remnant of 
the spatial correlation established in the first stage. 
The amplitude $\rho$ decreases as \eq{eq:rho-decay} in the second stage. 
This can be interpreted as the correlation between the `ghost' domains 
discussed in section \ref{subsec.double} which is losing its amplitude. 
Thus no matter how large the noise becomes, 
the 'memory' of the spatial structure is conserved.
The second term \eq{eq:c-space-2nd}
represent the short-range noise induced by the coarsening of an unrelated
phase. The last term becomes the same as the first term 
in the limit $L(\two) \gg L(\twz)$ since $A \simeq 1$ as
one can see from \eq{eq:def-A}.
In the other limit, $L(\twz) \gg L(\two)$, it can be neglected.

\subsubsection{Inner- and Outer-Coarsening regimes  in the Third Stage}
\label{subsubsec.spatial-c-third}

We now restore the equilibrium state
$\alpha$ as the target state for time $\twt$ given the `noisy' spin-configuration 
at the end of the second stage as the input. Here we examine the scenario 
conjectured in section \ref{subsec.double} 
that the inner-coarsening transforms the `ghost domains' back into `real domains'.
We demonstrate that the inner-coarsening regime, the intermediate
plateau regime and the outer-coarsening regime show up explicitly
in various correlation functions and response functions.

\newcommand{\gone}{\Gamma_{\rm 1-step}}

Using \eq{eq:w-after-tw1} in \eq{eq:gamma-cal}, 
the $\Gamma$ factor in the third stage which we detnoe as
$\gone$ is obtained as,
\begin{eqnarray}
&& \gone(\twt+\two+\twz,\two+\twz)  =  \int d^{d}k e^{-2k^{2}\twt} W^{\alpha}_{k}(\two+\twz) \nonumber \\
& = & \rho^{2} \Gamma_{0}(\twt+\twz,\twz) 
+ (1 -  2 \rho^{2}) \Gamma_{0}(\twt,0)
+ \rho^{2} A^{-d/2}(\twz/\two)
\Gamma_{0} \left(\twt+\frac{\twz}{A(\twz/\two)},\twz \right).
\label{eq:gamma-after-tw2}
\end{eqnarray}
Here $\Gamma_{0}$ is the one obtained in the solution for
standard coarsening \eq{eq:random-solution} with un-biased 
random initial condition.
It is again useful to define the relative ratio of the contribution 
of the three terms in the last equation as,
\begin{eqnarray}
 \mb^{2}(\twt,\two,\twz) &= & \rho^{2} \Gamma_{0}(\twt+\twz,\twz)/
\gone(\twt+\two+\twz,\two+\twz),
\label{eq:mb-para}\\
 \mf^{2}(\twt,\two,\twz) &=& (1 -  2 \rho^{2}) \Gamma_{0}(\twt,0)/
\gone(\twt+\two+\twz,\two+\twz),
\label{eq:mf-para}\\
 \tilde{m}^{2}(\twt,\two,\twz) 
&=& \rho^{2} A^{-d/2}(\twz/\two)
\Gamma_{0} \left(\twt+\frac{\twz}{A(\twz/\two)},\twz \right)/
\gone(\twt+\two+\twz,\two+\twz).
\label{eq:tilde-m-para}
\end{eqnarray}
The structure-factor of the projected field to $\alpha$ is obtained using 
 \eq{eq:evolution-structure-factor} as,
\begin{equation}
W^{\alpha}_{k}(\twt+\two+\twz)=W^{\alpha}_{k}(\two+\twz)\frac{e^{-2k^{2}\twt}}{
\gone(\twt+\two+\twz,\two+\twz)}, 
\label{eq:w-after-tw2}
\end{equation}
where $W^{\alpha}_{k}(\two+\twz)$ is given in \eq{eq:w-after-tw1} and 
$\gone(\twt+\two+\twz,\two+\twz)$ is given in 
\eq{eq:gamma-after-tw2}.
Then the spatial correlation function of the projected field 
is obtained using \eq{eq:w-after-tw2}, \eq{eq:w-after-tw1} and 
\eq{eq:gamma-after-tw2} in \eq{eq:how-to-c-space} as,
\begin{eqnarray}
 C(r,\twt+\two+\twz,\twt+\two+\twz)  
& = & \mb^{2}(\twt,\two,\twz)
\exp \left (-\frac{r^{2}}{8(\twt+\twz)} \right) \nonumber \\
&+&  
\mf^{2}(\twt,\two,\twz) \exp \left (-\frac{r^{2}}{8\twt} \right) \nonumber \\
&+& \tilde{m}^{2}(\twt,\two,\twz)
\exp \left (-\frac{r^{2}}{8(\twt+\twz/A(\twz/\two))} \right). 
\label{eq:spatial-c-third-gene}
\end{eqnarray}
In the last equation we used the parameters introduced in \eq{eq:mb-para},
\eq{eq:mf-para} and \eq{eq:tilde-m-para}. Note that there is a sum rule,
\begin{equation}
\mb^{2}+\mf^{2}+\tilde{m}^{2}=1,
\label{eq:sum-rule}
\end{equation}
by the definition of the parameters.
Obviously the first term in \eq{eq:spatial-c-third-gene}
can be physically understood as the correlation due to the 
continuation of the `ghost domains' (memory). The amplitude 
$\mb$ can be regarded as the staggered magnetization
associated with the continuation of the `ghost domains'.
Similarly the second term can be interpreted as the correlation due to
re-start of coarsening within the ghost domains (rejuvenation), which we called
`inner-coarsening' in section \ref{subsec.double}. The role
of the last term depends on relative ratio of the
size of the domains in the first and second regime as we discuss later.

Now we discuss the change of the profile of the spatial correlation function
with increasing time $\twt$ in the third stage. We assume that the
duration of the second regime is large enough so that the bias 
has become very small $\rho\ll 1$. Then at the beginning for small $\twt$, 
we find the parameter defined in \eq{eq:mf-para} is  $\mf^{2} \simeq 1$ 
while the other two $\mb^{2}$ and $\tilde{m}^{2}$ defined in
\eq{eq:mb-para} and \eq{eq:tilde-m-para} are very small. 
Then the second term in \eq{eq:spatial-c-third-gene} which represent
the inner-coarsening (rejuvenation) is therefore dominant,
\begin{eqnarray}
 C(r,\twt+\two+\twz,\twt+\two+\twz)  & \simeq &
 \exp \left [ - \left (\frac{r}{L(\twt)} \right)^{2}\right] \nonumber \\
 {\mbox{inner-coarsening regime}} &&
\label{eq:spatial-c-inner-coarsening}
\end{eqnarray}
On the other hand, in the asymptotically large time scale
such that $L(\twt)\sim L(\twt+\twz)$, (i.e. $\twt \sim \twz$), the spatial 
correlation function becomes,
\begin{eqnarray}
 C(r,\twt+\two+\twz,\twt+\two+\twz)  & \simeq &
 \exp \left [ - \left (\frac{r}{L(\twt+\twz)} \right)^{2}\right], 
\nonumber\\
 \mbox{outer-coarsening regime} &&  
\label{eq:spatial-c-outer-coarsening}
\end{eqnarray}
Here we used the sum rule \eq{eq:sum-rule}.
This regime can be interpreted as the outer coarsening regime
we discussed in section \ref{subsec.double}. 

We have discussed  in section \ref{subsec.double} that the crossover 
between the inner-coarsening and outer-coarsening regime depends on 
the relative 
domain sizes of the first and second stages. 
We will indeed find below that it is the case in the present model.
In the following we consider the two limiting cases:
a) $L(\twz) \gg L(\two)$ and b) $L(\twz) \ll L(\two)$.

\subsubsection{Rejuvenation and memory :Case a)}
\label{sec.a}

We consider the case a) $L(\twz) \gg L(\two)$ : 
the duration of the first stage is much longer than the second. 
In this case the parameter $\tilde{m}^{2}$ is very small 
compared with $\mf^{2}$ since $A \gg 1$  as one can see in \eq{eq:def-A}.
Thus we neglect the last term in the spatial correlation function
given in \eq{eq:spatial-c-third-gene}.
Then the basic structure of the correlation function
\eq{eq:spatial-c-third-gene} can be naturally interpreted as
a sum of Gaussian packet due to the inner-coarsening (rejuvenation)
whose size is $L(\twt)$ (second term) and that due to the continuation of the 
`ghost domains' (memory) of size $L(\twt+\twz)$ (first term). 
In this case, the inner-coarsening regime 
\eq{eq:spatial-c-inner-coarsening} terminates due to the effect
of the bias as we discussed in section \ref{subsec.double}. 
Furthermore there is an intermediate regime which we called
`plateau regime' in section \ref{subsec.double} so that rejuvenation and memory effects
can be observed in a well separated manner. 

Let us consider the characteristic time scale at which the 
staggered magnetization of the ghost domain  
$\mb$ defined in  \eq{eq:mb-para} and $\mf$ defined in
\eq{eq:mf-para} become the same order. For short times 
such that $L(\twt) \ll L(\twz)$, we can assume 
$\Gamma_{0}(\twt+\twz,\twz) \sim 1$ in  \eq{eq:mb-para}.
Then the time scale $\trec(\rho)$ at which 
$\mb$ and $\mf$ become the same order is obtained using
\eq{eq:random-solution} as,
\begin{equation}
\rho \sim  (L(\trec(\rho)/L_{0})^{-\lambda} \label{eq:taueq-rho-2},
\end{equation}
which is the same as \eq{eq:taueq-rho} obtained in the limit
$L(\twz) \rightarrow \infty$. Note that 
the assumption $\Gamma_{0}(\twz+\twt,\twz) \sim 1$ is still satisfied
because the bias is $\rho \sim (L(\two)/L_{0})^{-\lambda}$ 
as given in \eq{eq:rho-decay} which implies 
\begin{equation}
L(\trec(\rho))\ll L(\twz).
\label{eq:taueq-twz}
\end{equation}

When the inner-coarsening regime ends, we are left with the
the `ghost domains' which have {\it almost} recovered 
their full staggered magnetization $\mb \simeq 1$.
The latter can be interpreted as the fact that the `real domain' 
are recovered.
However, the relation \eq{eq:taueq-twz} implies that the
relaxation time $\trec(\rho)$ is  not large enough to grow
the revived domain further. Thus there is an intermediate regime where:
\begin{eqnarray}
 C(r,\twt+\two+\twz,\twt+\two+\twz)  & \simeq &
 \exp \left [ - \left (\frac{r}{L(\twz)} \right)^{2}\right], 
\nonumber\\
 \mbox{plateau regime} &&  
L(\twz) \gg  L(\twt) \gg   L(\trec(\rho)).
\end{eqnarray}
This is the plateau regime in which the revived domain appears frozen in time.
Thus the memory (spatial structure of bias) conserved in the system is retrieved with its
original full amplitude recovered within this regime.
Much later in time, the plateau regime is followed by the asymptotic 
outer-coarsening regime \eq{eq:spatial-c-outer-coarsening}.

\subsubsection{Complete rejuvenation: Case b)}
\label{sec.b}

Next we discuss the case b) $L(\twz) \ll L(\two)$: the duration of the
first stage is much shorter than that of the second stage. 
In this case we find that the two contributions
defined in \eq{eq:mf-para} and \eq{eq:tilde-m-para} become essentially
equal $\mb^{2} \simeq \tilde{m}^{2}$  since
$A \simeq 1$ holds as one can see in \eq{eq:def-A}. 
Thus the basic structure of the correlation function
\eq{eq:spatial-c-third-gene} can be again naturally interpreted as
a sum of Gaussian packet due to the domain of inner-coarsening (rejuvenation)
whose size is $L(\twt)$ and that due to the continuation of the 
`ghost domain' (memory) of size $L(\twt+\twz)$.

However, the amplitude of the memory terms $\mb^{2} (\simeq \tilde{m}^{2})$
does not recover much and only saturate to a small value 
$\sim (L(\twz)/L(\two))^{\lambda} \ll 1$ as $L(\twt) \sim L(\twz)$ ($\twt \sim \twz$).
In the latter regime, the width of the memory and rejuvenation terms become of the same order,
\begin{equation}
L(\twt) \sim   L(\twz+\twt); \qquad  \twt \sim \twz
\end{equation}
Thus the correlation function crossovers very smoothly to the asymptotic outer-coarsening regime
\eq{eq:spatial-c-outer-coarsening} so that the memory cannot be retrieved.
The resultant behavior of the correlation function
is not very different from the case of $\tw=0$. In this sense the relaxation is
almost completely rejuvenated.

\subsubsection{U-turn in the Phase Space}
\label{subsubsec.u-turn}

The above result implies the time evolution of system in the phase space 
during the third stage is such that it makes an U-turn  to 
configuration before the second stage (inner-coarsening) 
and stay there for a while 
(plateau-regime) and finally make further excursion (outer-coarsening).
Such a feature can be elucidated by considering overlap $q$ between 
the configuration just after the first stage and the temporal 
configuration in the third stage.
It is readily obtained as,
\begin{eqnarray}
q(\twt) 
&=&\int\frac{d^{d}k}{(2\pi)^{d}}\int\frac{d^{d}l}{(2\pi)^{d}}
<<\phi_{k}(\twt+\two+\twz)>_{\sigma}\phi_{l}(\twz)>_{\rm ini}\nonumber \\
&=&\int d^{d}k  \frac{\rho e^{-k^{2}\twt}}{
\sqrt{\gone(\twt+\two+\twz,\two+\twz)}}W^{\alpha}_{k}(\twz) 
= \frac{\rho \gone((\twt+\twz+\twz)/2,\two)}{
\sqrt{\gone(\twt+\two+\twz,\two+\twz)}} \nonumber \\
&=& C_{0}(\twt+\twz,\twz) \mb(\twt,\two,\twz).
\label{eq:non-monotonic-c}
\end{eqnarray}
In the last equation we used the ratio $\mb$ defined in \eq{eq:mb-para}
and $C_{0}$ is the auto-correlation function of standard coarsening \eq{eq:c}.

As we found in the previous sections the ratio $\mb(\twt,\two,\twz)$ 
can be physically understood as the stuggard magnetization 
which increases with $\twt$ in the third stage.
On the other hand, the factor $C_{0}(\twt+\twz,\twz)$ 
which appears in \eq{eq:non-monotonic-c} 
describes de-correlation due to outer-coarsening.

The competing effects of the inner-coarsening and outer-coarsening
make the overlap $q(\twt)$ non-monotonic in time  $\twt$.
In the case a) $L(\twz) \gg L(\two)$ the behavior is the following.
It {\it increases} during the inner-coarsening because of the
increase of $\mb$ and almost saturate to $1$ at time scale $\twt$ 
at around the recovery time $\tau_{\rm rec}(\rho)$. 
It stays close to $1$ during plateau regime. Then in the outer-coarsening
regime, it {\it decreases} with time. This picture apparently becomes
invalid in the case of b)$L(\twz) \ll L(\two)$.

\subsection{Three-Stage Relaxation of Auto-Correlation Function 
after One-Step Cycling}
\label{subsubsec.c-third-2}

We now turn to more conventional observables with which 
the rejuvenation and memory effects can be see easily.
While the spatial correlation function of the projection field 
discussed above is convenient for theoretical discussions, it is obviously
impractical  in simulations and experiments of spin-glasses.
In this section, we consider auto-correlation functions  
and linear response functions in the next section. 
These quantities are invariant under changes of projections and 
can be measured directly in numerical simulations and experiments.
We will demonstrate that the characteristic three-stage relaxation
after one-step cycling:  inner-coarsening (rejuvenation), 
plateau and outer-coarsening (memory) regimes show up explicitly 
in these two-time quantities.

The auto-correlation function between two times in the third stage
is obtained using \eq{eq:gamma-after-tw2} in \eq{eq:how-to-c} as,
\begin{equation}
 C(\tau+\twtotal,\twtotal)  
 = 
\frac{\gone(\tau/2+\twtotal,\two+\twz)
}{\sqrt{\gone(\tau+\twtotal,\two+\twz)}\sqrt{\gone(\twtotal,\two+\twz)}}.
\label{eq:c-gene-third}
\end{equation}
with
\begin{equation}
\twtotal \equiv \twt+\two+\twz
\end{equation}
Here the explicit form of the $\Gamma$-factor 
is given in \eq{eq:gamma-after-tw2}. 
As we discussed in section \ref{subsec.dc-one-step}, the auto correlation
function can be related with the DC-magnetic susceptibilities
measured in experiments.

First, let us look at the simplest case  $\twt=0$ :
i.e. the correlation between the configuration just at the
beginning of the third stage and the  
configuration a time $\tau$ later.
Here we only consider the case a) $L(\twz) \gg L(\two)$
which allows clear separation between the inner- and outer-coarsening regimes.
Within the inner-coarsening regime $\tau \ll \trec$ the amplitude of the
ghost domains is small $\mb \ll  1$ and
$\mf \simeq 1$.  When the inner-coarsening ends at $\tau \sim \trec$, 
the amplitude of the ghost domains is almost
recovered:  $\mb \simeq  1$ and $\mf \ll 1$. 
Then one can see easily that the correlation function
have the following feature,
\begin{eqnarray}
C(\tau+\two+\twz,\two+\twz)= & C_{0}(\tau,0) 
& \qquad L(\tau) \ll L(\trec(\rho))  \\
& \rho C_{0}(\tau+\twz,\twz) 
& \qquad L(\tau) \gg L(\trec(\rho)).
\end{eqnarray}
Here $C_{0}$ is the auto-correlation function
in the standard coarsening given in  \eq{eq:non-monotonic-c}.
The result confirms the scaling property \eq{eq:c-third-step}
conjectured in section \ref{subsec.dc-one-step}.
It visualizes clearly the cross-over from
inner-coarsening, plateau regime and outer-coarsening.
Note that in the limit $L(\twz) \rightarrow \infty$, the last
relaxation does not occur $C_{0}(\tau+\twz,\twz)=1$ and  
we recover the result 
of the case in which we start from fully symmetry broken state 
with respect to phase $\alpha$. (see \eq{eq:c-decay} and \eq{eq:c-stop} )

\begin{figure}
\begin{center}
\leavevmode \epsfxsize=0.6\linewidth
\epsfbox{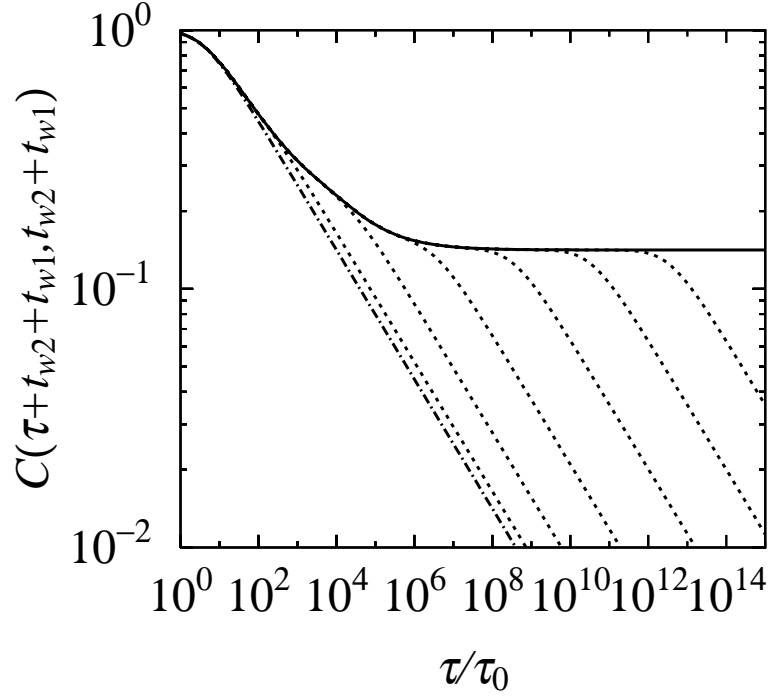}
\end{center}
\caption{The auto-correlation function
$C(\tau+\two+\twz,\two+\twz)$  in the
$O(n)$ Mattis model ($d=1$) under one-step cycling.
The dotted lines are curves with common $\two=10^{4}$ but varying
$\twz=10^{2},10^{4},10^{6},10^{8},10^{10},10^{12}$ (from left to right).
The solid line on the top 
is the curve with $\two=10^{4}$ but with $\twz=\infty$ (top).
The dash-dotted line is the reference curve with zero waiting time
of standard coarsening $C_{0}(\tau,0)$.(left most) } 
\label{on_c_1.fig}
\end{figure}

\begin{figure}
\begin{center}
\leavevmode \epsfxsize=0.6\linewidth
\epsfbox{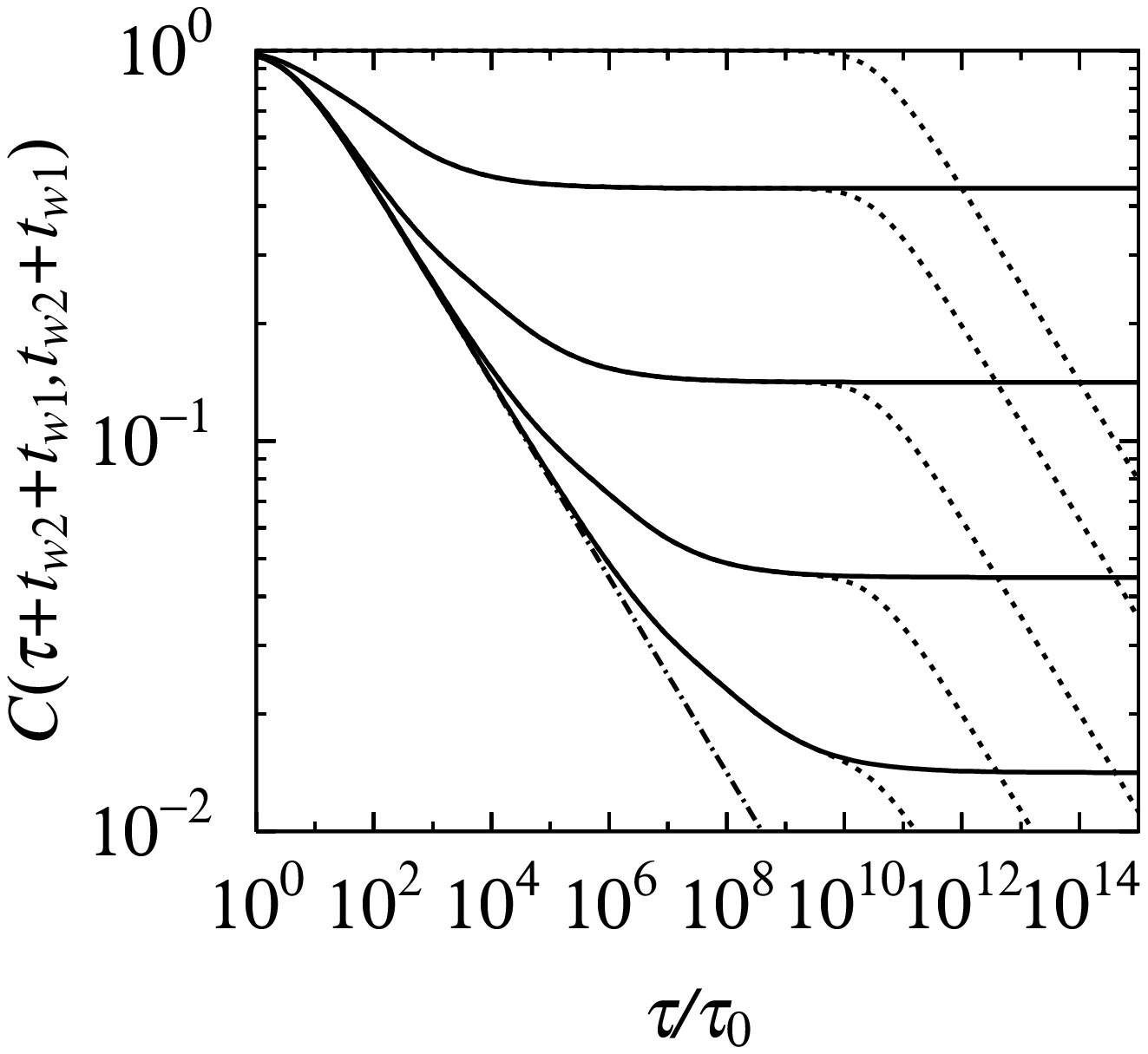}
\end{center}
\caption{The auto-correlation function 
$C(\tau+\two+\twz,\two+\twz)$ in the
$O(n)$ Mattis model ($d=1$) under one-step cycling.
The dotted lines are curves with common $\twz=10^{10}$
but varying $\two=0,10^{2},10^{4},10^{6},10^{8}$ (from top to bottom).
The solid lines are the reference curve of 
$\twz=\infty$ but varying $\two=10^{2},10^{4},10^{6},10^{8}$ (from top to bottom).
The dash-dotted line is the reference curve of zero waiting time
of standard coarsening $C_{0}(\tau,0)$.} 
\label{on_c_2.fig}
\end{figure}

We present some plots of the auto-correlation function 
$C(\tau+\two+\twz,\two+\twz)$ in Fig. \ref{on_c_1.fig}
and \ref{on_c_2.fig} for the case of $d=1$. The generic feature
is of course the same at any dimension but the dynamical exponent for
the decay depends on the dimension as $\lambda=d/2$.
Here the three-stage relaxation is clearly visible. Note that the
plateau is visible only for the cases in which $\twz \gg \two$,
which  is the condition to have sharp separation between
the inner-coarsening and outer-coarsening.

\begin{figure}
\begin{center}
\leavevmode \epsfxsize=0.7\linewidth
\epsfbox{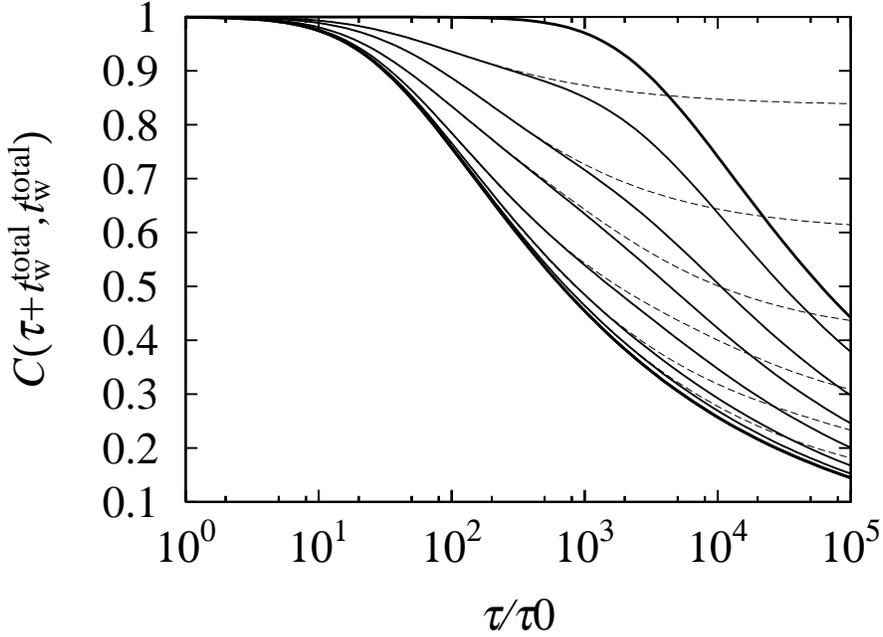}
\end{center}
\caption{The auto-correlation function 
$C(\tau+\twtotal,\twtotal)$ with 
$\twtotal=\twt+\two+\twz$ in the
$O(n)$ Mattis model ($d=1$) under one-step cycling
plotted against the time difference $\tau$ in the third stage.
In section \ref{sec:scenario}, we compare these curves with an
experimental data of the relaxation of thermo-remanent magnetization
after one-step temperature cycling.
The thin solid lines are curves with common $\twz=990$ and $\twt=10$ 
but varying $\two=10^{6},10^{5},10^{4},10^{3},10^{2},10^{1}$. 
(from left to right)
The dotted lines are the corresponding curves with $\twz=\infty$.
The top and bottom bold solid lines are the reference curves of 
standard coarsening $C_{0}(\tau+1000,1000$ and $C_{0}(\tau+10,10)$ respectively.} 
\label{on_c_3.fig}
\end{figure}

Second, let us consider more complicated cases with non-zero  $\twt > 0$.
The reason we analyzed it is to make comparison with 
a set of conventional TRM data \cite{saclay-exp-rev-1}
where such a protocol is used. 
\footnote{In any case, it may be impossible to realize strictly $\twt = 0$ 
in experiments.}
Later in section \ref{sec:scenario}, 
we compare the result with an experimental data.
In Fig. \ref{on_c_3.fig} we show a plot of 
the auto-correlation function
$C(\tau+\twtotal\twtotal)$ given in \eq{eq:c-gene-third}.
We also show curves of the special case where 
${\twz \rightarrow \infty}$. It can be seen that 
the initial decay does not depend on $\twz$ and decays 
as the case $\twz=\infty$.
This is the inner-coarsening regime which then 
crossovers to the plateau regime. Finally, the departure from the 
plateau due to outer-coarsening becomes visible 
at around $L(\tau) \sim L(\twt+\twz)$.

The plateau value is nothing but the staggered 
magnetization within the ghost domains at time $\twt+\two+\twz$.
The latter is naturally smaller for longer duration of the 
second stage $\two$. But note that it also depends on $\twt$ 
because the staggered magnetization tends to go back to the 
full value 1 in the third stage. (see \eq{eq:m-recover})
The last feature imply that rejuvenation is obscured in this protocol.
Nonetheless, for any $\two$, the presence of rejuvenation (inner-coarsening
regime) can be recognized by comparing with the curves with much larger
$\twz$ as demonstrated in the figure. 

\subsection{Short time Linear Response Functions in One-Step Cycling}
\label{sec.on-resp}

Within the spherical Mattis model, it is also possible to study
directly the linear response function exactly. We study it below for
the one-step cycling protocol and focus especially on its
behavior in the third stage. 
Here we will consider response functions $R(\Delta t + t,t)$ 
with fixed time separation $\Delta t$ as a function of increasing time $t$.
It is analogous to the relaxation of AC-susceptibilities $\chi''(\omega,t)$
of frequency $\omega=1/\Delta t$  at time $t$.
The advantage of studying $R(\Delta t + t,t)$ is that 
its analytical expression is  simpler than that of $\chi''(\omega,t)$.
Some details of the calculations are presented in appendix \ref{sec:response}.
We will find the anticipated crossover from inner-coarsening 
to outer-coarsening in the response  sketched in section 
\ref{subsec.ac-one-step}.

It should be noted however that mean field theories \cite{CK,FM,CKD96} 
suggests long time behavior of the response is likely to be quite 
different between spin-glass and usual coarsening systems 
as we noted in section \ref{subsec.dc-one-step}. 
Nonetheless, as far as short-time response is concerned the generic
behavior of the response function may be similar.

In the first stage the relaxation is the same as in the standard coarsening
and we naturally find new relaxation also in the 2nd stage,
\begin{eqnarray}
\tilde{R}_{\rm I}(\Delta t+t,t) & = & \tilde{R}_{0}(\Delta+t,t) \nonumber \\
\tilde{R}_{\rm II}(\Delta t+t,t) & = &
 \tilde{R}_{0}(\Delta t+(t-\twz),t-\twz) 
\end{eqnarray}
Here $\tilde{R}$ is the response function scaled by that in equilibrium and
$\tilde{R}_{0}(\Delta+t,t)$ is the rescaled response function
of the standard coarsening (see appendix \ref{sec:response} for the details) .

One finds richer behavior of the short-time response in 
the 3rd stage $ \twt+\two + \twz > t > \two+\twz$  
as obtained in  \eq{eq:rescaled-resp-1} and \eq{eq:rescaled-resp-2}.
we find the following.
At the beginning of the third stage which is the inner-coarsening regime,
we find new relaxation. On the other hand, in larger time scale which we 
called as plateau and outer-coarsening regime, 
we find continuation of the 1st stage.
\begin{equation}
\tilde{R}_{\rm III}(\Delta t+t,t)  \simeq   
 \tilde{R}_{0}(\Delta t+t-\two-\twz,t-\two-\twz)   
\qquad {\mbox{inner-coarsening regime}}
\end{equation}
\begin{equation}
\tilde{R}_{\rm III}(\Delta t+t,t)  \simeq   
 \tilde{R}_{0}(\Delta t+t-\twz,t-\twz)  
\qquad  \mbox{plateau/outer-coarsening regime} 
\end{equation}
In the case a) $L(\twz) \gg L(\two)$ which allows the plateau regime, the
inner-coarsening regime finishes before the significant relaxation due to the outer-coarsening
starts. 
The generic feature is consistent with the picture presented in section 
\ref{subsec.ac-one-step}. (see Fig.\ref{concept-one-step-ac.fig})

\subsection{Multiplicative Noise Effect and 2-Step U-turns in a 2-Step Cycling}
\label{subsec.u-turn-two-step-cycling}

One can extend the above calculations to two-step cycling
$\alpha \rightarrow \beta \rightarrow \gamma \rightarrow \beta
\rightarrow \alpha$, trying to mimic the multi-step cycling 
discussed in section \ref{subsec.multi}. Although the full 
calculation will be become too lengthy, we can readily have 
a glimpse of what happens in the spherical model.

Let us suppose that we are given a spin-configuration whose 
projection to a certain equilibrium state $\alpha$ is 
$\hat{\phi}_{k}^{\alpha}(0)$. We consider to give this
as initial condition for a one-step cycling $\beta (\twz) 
\rightarrow \gamma (\two) \rightarrow \beta (\twt)$ 
and monitor the time evolution of the
projection to $\alpha$. One finds that the expectation value
of the projection $\hat{\phi}_{k}^{\alpha}(\twt+\two+\twz)$
at time $\twt+\two+\twz$ as,
\begin{eqnarray}
&&<\hat{\phi}_{k}^{\alpha}(\twt+\two+\twz)>_{\sigma} =
\int \frac{d^{d}k_{1}}{(2\pi)^{d}}
\int \frac{d^{d}k_{2}}{(2\pi)^{d}}
\int \frac{d^{d}k_{3}}{(2\pi)^{d}}
\int \frac{d^{d}k_{4}}{(2\pi)^{d}}
<(\hat{\sigma}^{\alpha \beta})_{k_1}
(\hat{\sigma}^{\beta \gamma})_{k_2}
(\hat{\sigma}^{\gamma \beta})_{k_3}
(\hat{\sigma}^{\beta \alpha})_{k_4}>_{\sigma} \nonumber \\
&& \times \frac{e^{-(k-k_1)^{2}\twt}}{\sqrt{\gone(\twt+\two+\twz,\twz+\two)}}
\frac{e^{-(k-k_1-k_2)^{2}\two}}{\sqrt{G_{0}(\two,0)}}
\frac{e^{-(k-k_1-k_2-k_3)^{2}\twz}}{\sqrt{G_{0}(\twz,0)}}
\hat{\phi}_{k-k_1-k_2-k_3-k_4}^{\alpha}(0) \nonumber\\
&=& \mb(\twt,\two,\twz) C_{0}(\twt+\twz,0) \hat{\phi}_{k}^{\alpha}(0)
\label{eq:u-turn-2-step}
\end{eqnarray}
In the derivation of the last equation, we used \eq{eq:map-noise-correlation-3-states}, \eq{eq:random-solution} \eq{eq:c} and \eq{eq:mb-para}.
Thus we find that the expectation value of the projection field
is proportional to the initial one $\hat{\phi}_{k}^{\alpha}(0)$
with time-dependent prefactor.

A remarkable point is that the prefactor is product of $\mb(\twt,\two,\twz)$ 
and the auto-correlation function $C_{0}(\twt+\twz,0)$. At the beginning
of the last stage (2nd $\beta$-coarsening), we readily find that 
the prefactor is $C(\twz,0) \rho$ where $\rho$ is the reduction 
of the amplitude due to $\gamma$ coarsening. The latter means
multiplicative reduction of the amplitude of the projection onto
$\alpha$ due to the successive coarsening of two different phases
($\beta$ and $\gamma$) as we conjectured in section \ref{subsec.multi}.
The fact that noise effect is multiplicative, can be check explicitly
in more general multi-step coarsening as presented in appendix
\ref{sec:appendix-noise-correlation} (see \eq{eq:multiplicative-reduction}).

In the inner-coarsening regime of the last stage (2nd $\beta$-coarsening), 
we readily find that $\mb(\twt,\two,\twz)$ almost returns back to $1$ 
within the recovery time $\trec$ so that the noise due to $\gamma$-coarsening
onto $\beta$ is now removed. As far as time separation is wide enough
so that the plateau regime is allowed, the recovery time $\trec$ is
much shorter than $\twz$. Within the plateau regime 
$C_{0}(\twt+\twz,0)\simeq C_{0}(\twz,0)$. Remarkably, then the above 
formula \eq{eq:u-turn-2-step} implies that the noise onto $\alpha$ 
due to $\gamma$ coarsening is also 
cured thanks to the 2nd $\beta$-coarsening and the remnant noise is now
only that due to $\beta$ phase. But if one perform the 2nd $\beta$-coarsening
too long, the factor $C_{0}(\twt+\twz,0)$ will begin to decrease further
meaning that the noise due to $\beta$ coarsening will now begin to 
affect the projection to $\alpha$. 

If one stops the 2nd $\beta$-coarsening at around $\trec$ and then 
switch to coarsening of $\alpha$, the remnant noise due to $\beta$ 
phase will be removed. The above result demonstrates that the system 
can be returned back to the starting point in the phaser space 
by two-step U-turns. This is consistent with
our picture presented in section  \ref{subsec.multi} for the recovery
of memory in multi-step cycling.

\section{Two-dimensional Ising Mattis model}
\label{sec.ising}

While the $O(n)$ model in the spherical limit is analytically
tractable and contains the essential phenomenology of coarsening systems,
its drawback is that it does not contain topological defects 
like domain walls \cite{B94}. 
Thus it is desirable to study models which have clearly defined domain walls. 
In the present paper, we do not pursue more elaborate analytical calculations 
to take into account topological defects like the Ohta-Jasnow-Kawasaki approximation.
Instead, we directly study the Mattis model introduced in section \ref{subsec.mattis} 
with Ising spins on a two-dimensional square lattice by Monte Carlo simulations.

The algorithm is zero temperature Monte Carlo dynamics with 
multi-spin coding. 
The spin configurations $\{\sigma_{i}^{\alpha(\beta)}\}$ 
in the ground states are chosen to take $\pm 1$ randomly.

\subsection{Auto-Correlation Function after One-Step Cycling}

We simulated the one step cycling process 
$A \rightarrow B \rightarrow A$ described in section
\ref{subsec.double}, which was studied in the $O(n)$ model in section
\ref{subsec.double-on}. 
We examined the auto-correlation function between the configuration just at the
beginning of the third stage and the temporal 
configuration after time $\tau$ in the third stage, which
was studied in the $O(n)$ model in section \ref{subsubsec.c-third-2}.
The system size is $8192 \times 8192$
which is large enough to avoid finite size effects within 
the time window we have explored $\sim 10^{4}$(MCS).

\begin{figure}
\begin{center}
\leavevmode \epsfxsize=0.45\linewidth
\epsfbox{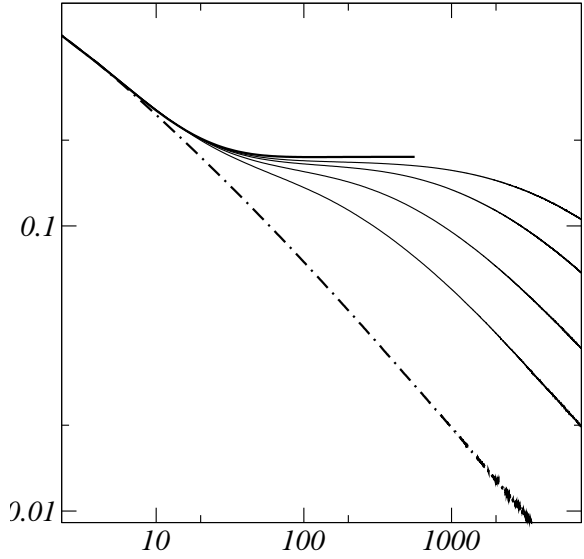}
\end{center}
\caption{The auto-correlation function
$C(\tau+\two+\twz,\two+\twz)$  in the
$d=2$ Ising Mattis model under one-step cycling plotted versus $\tau$ (MCS).
The dotted lines are 
$\two=20$ MCS and $\twz=100,300,1000,3000$ (MCS) from left to right.
The solid line on the top is the case $\twz=\infty$.
The  dash-dotted line is the
reference curve with zero-waiting time.} 
\label{tw0-tw1=20.fig}
\end{figure}

\begin{figure}
\begin{center}
\leavevmode \epsfxsize=0.45\linewidth
\epsfbox{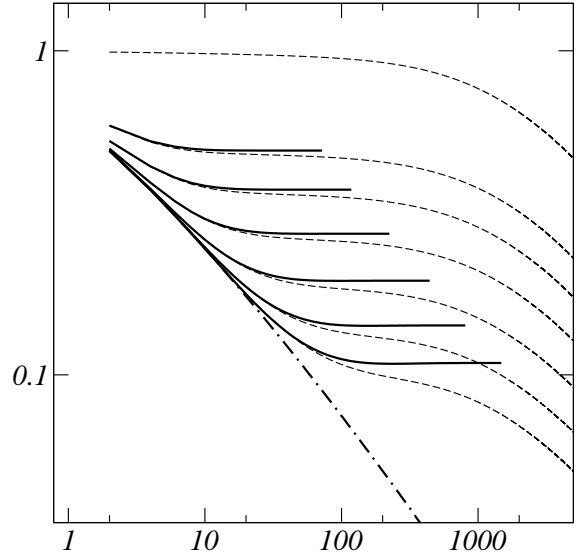}
\end{center}
\caption{The auto-correlation function
$C(\tau+\two+\twz,\two+\twz)$  in the
$d=2$ Ising Mattis model under one-step cycling plotted versus $\tau$ (MCS).
The dotted lines are 
$\twz=1000$ MCS $\two=0,2,4,8,16,30,50$ from the top to below.
The solid lines are with $\twz=\infty$. The dash-dotted line is the
reference curve with zero-waiting time.} 
\label{tw0=1000-inf.fig}
\end{figure}

In Fig. \ref{tw0-tw1=20.fig} and Fig. \ref{tw0=1000-inf.fig} we present
the result of the auto-correlation function. 
We included the results of simulations in which
the ground state of phase $\alpha$ is given to the
second stage as the input, i. e. the $\twz=\infty$ limit.
By comparing with the corresponding results within the
spherical model shown in Fig. \ref{on_c_1.fig} and Fig. \ref{on_c_2.fig},
one can see a very similar structure of the relaxation curves
in agreement with the conjecture presented in section \ref{subsec.dc-one-step}.
The initial decay implies inner-coarsening regime 
where the correlation decays as if the memory of the first stage was 
completely lost. The subsequent behavior implies however that
the memory of the first stage is not lost but its amplitude
is reduced from 1 to $\rho < 1$ at the beginning of the third stage.

\subsection{An Example of Multiple-Memory}

In section \ref{subsec.multi} we discussed coarsening under cycling
of multiple phases. Especially we argued that i) the noise effect
of multiple phases are multiplicative and that ii) the multiplicative
noise can be removed one by one by additional series of coarsening 
in the reversed order. We have verified the picture within the 
spherical model to a certain extent 
in section \ref{subsec.u-turn-two-step-cycling}.
Here we present a demonstration of the coarsening under cycling
of three independent target states in the 2-dimensional Ising Mattis model
where the two features i) and ii) appear explicitly.

\begin{figure}
\begin{center}
\leavevmode \epsfxsize=0.3\linewidth
\epsfbox{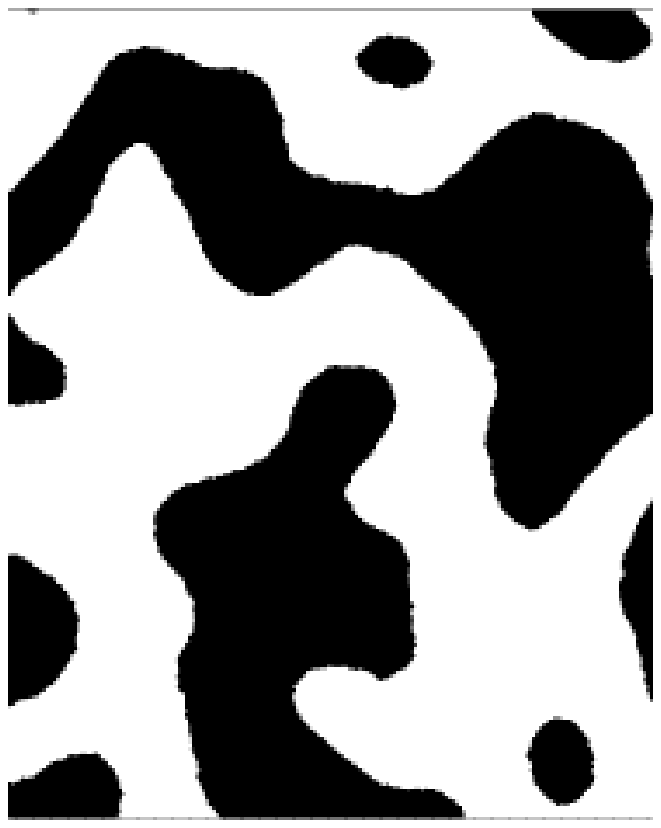}
\leavevmode \epsfxsize=0.3\linewidth
\epsfbox{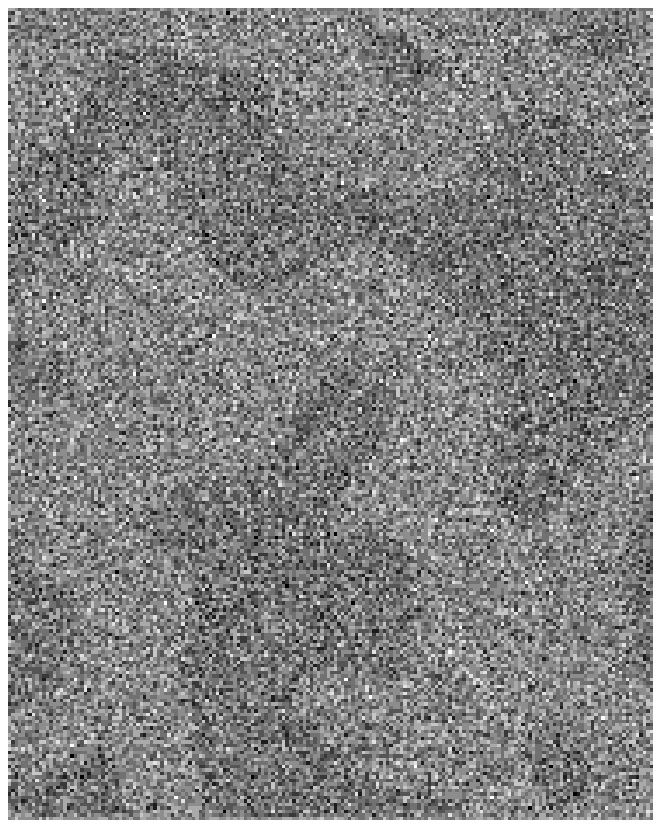}
\leavevmode \epsfxsize=0.3\linewidth
\epsfbox{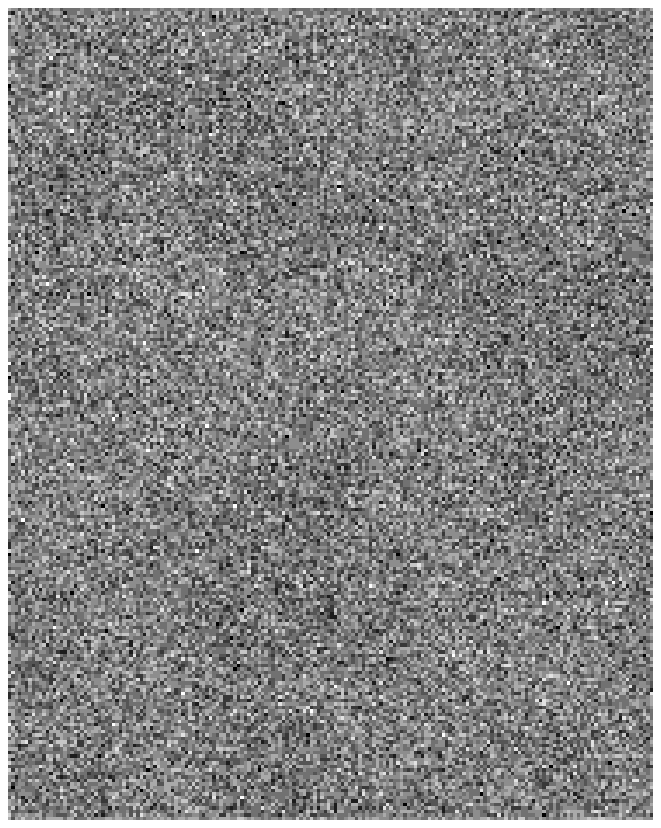}\\
\leavevmode \epsfxsize=0.3\linewidth
\epsfbox{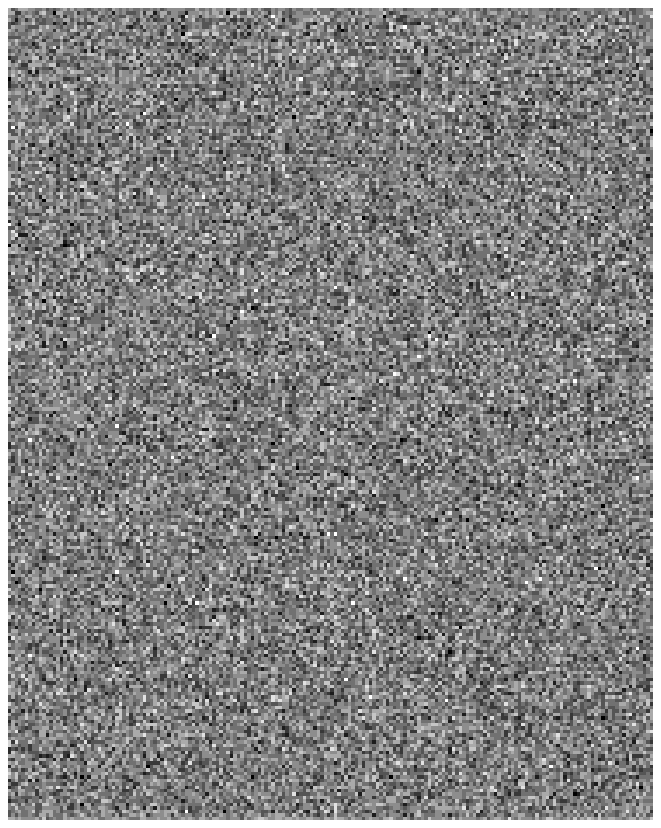}
\leavevmode \epsfxsize=0.3\linewidth
\epsfbox{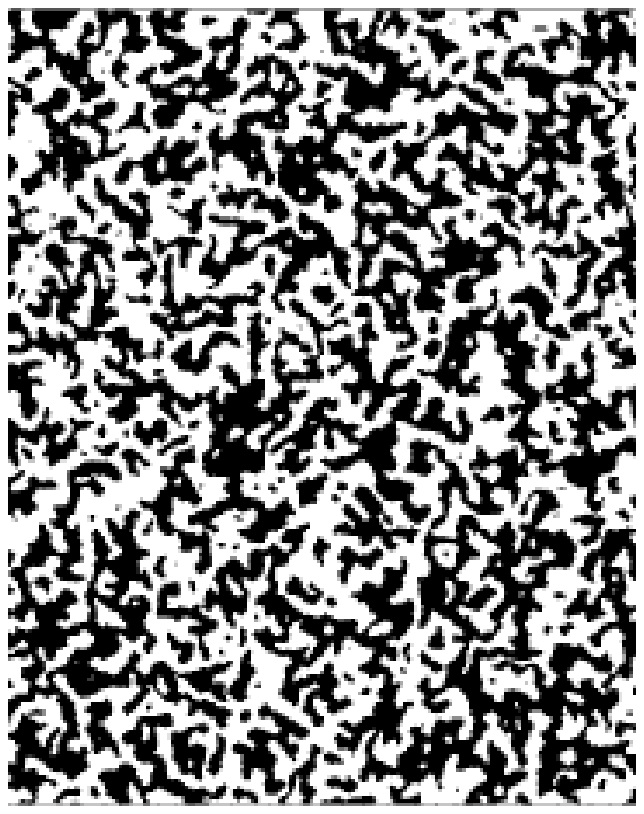}
\leavevmode \epsfxsize=0.3\linewidth
\epsfbox{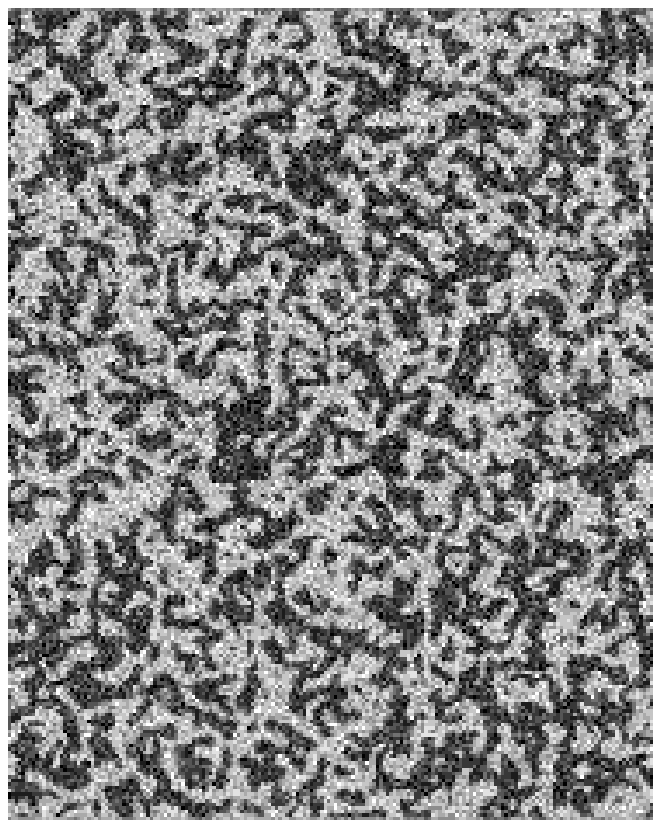}\\
\leavevmode \epsfxsize=0.3\linewidth
\epsfbox{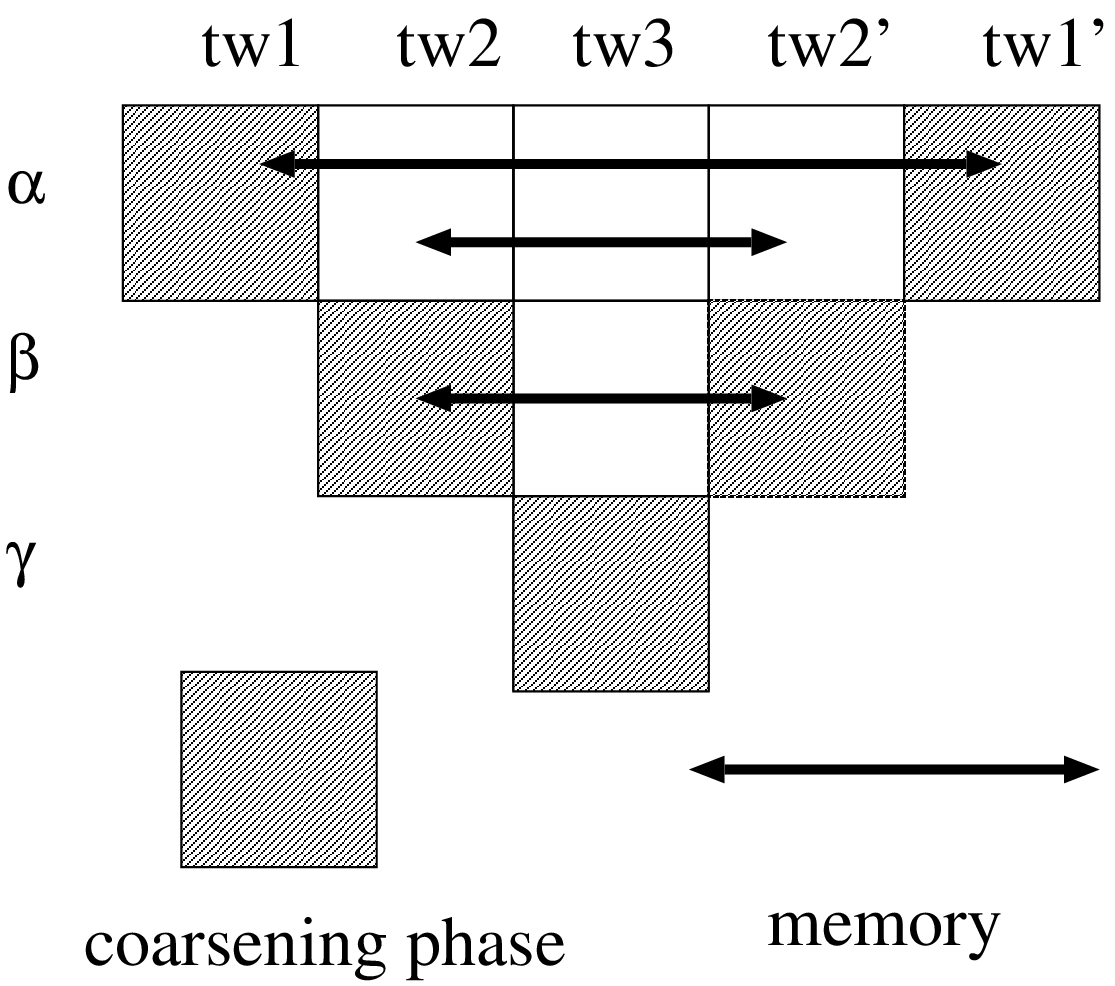}
\leavevmode \epsfxsize=0.3\linewidth
\epsfbox{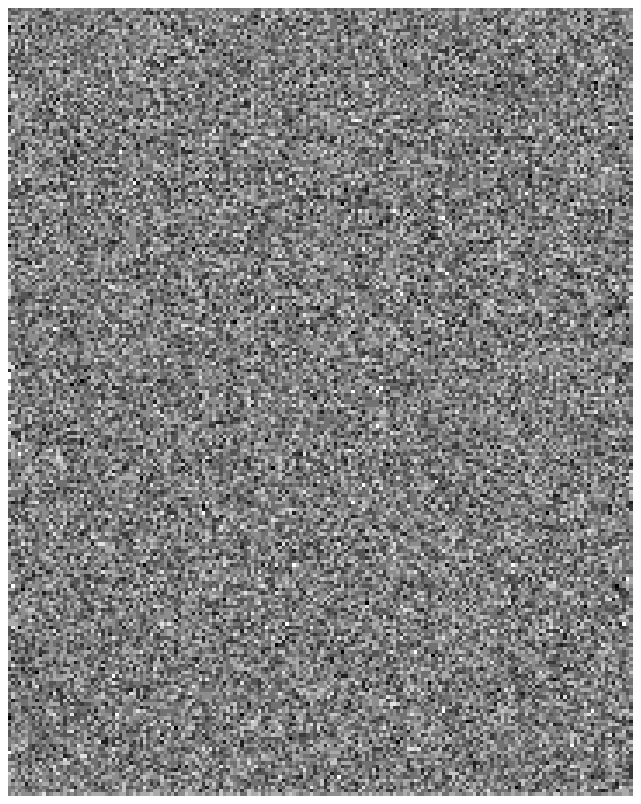}
\leavevmode \epsfxsize=0.3\linewidth
\epsfbox{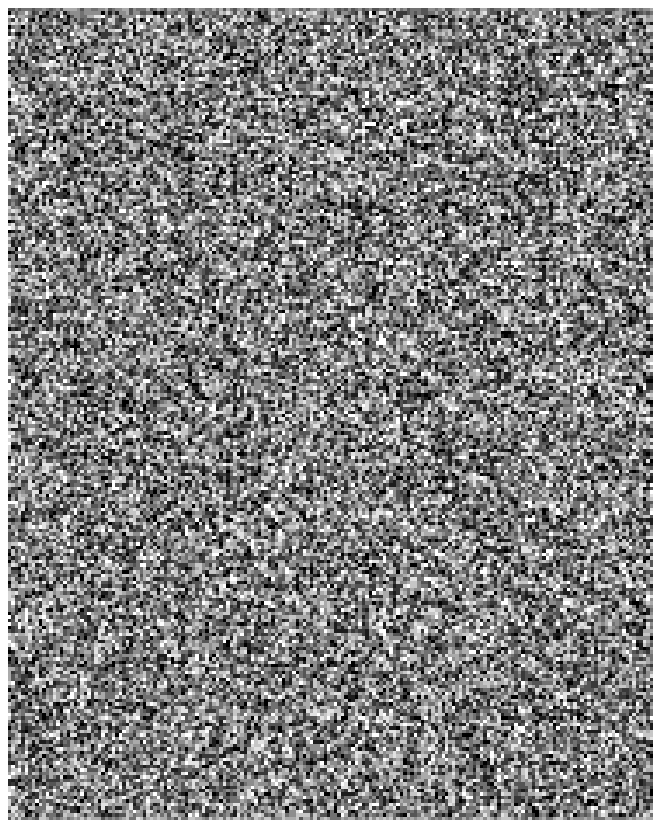}\\

\flushleft\leavevmode \epsfxsize=0.3\linewidth
\epsfbox{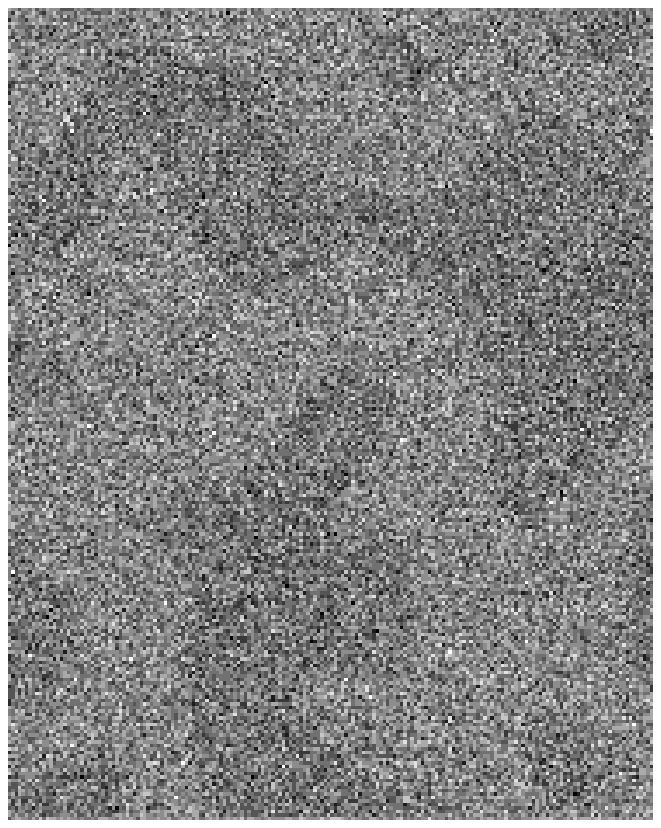}
\leavevmode \epsfxsize=0.3\linewidth
\epsfbox{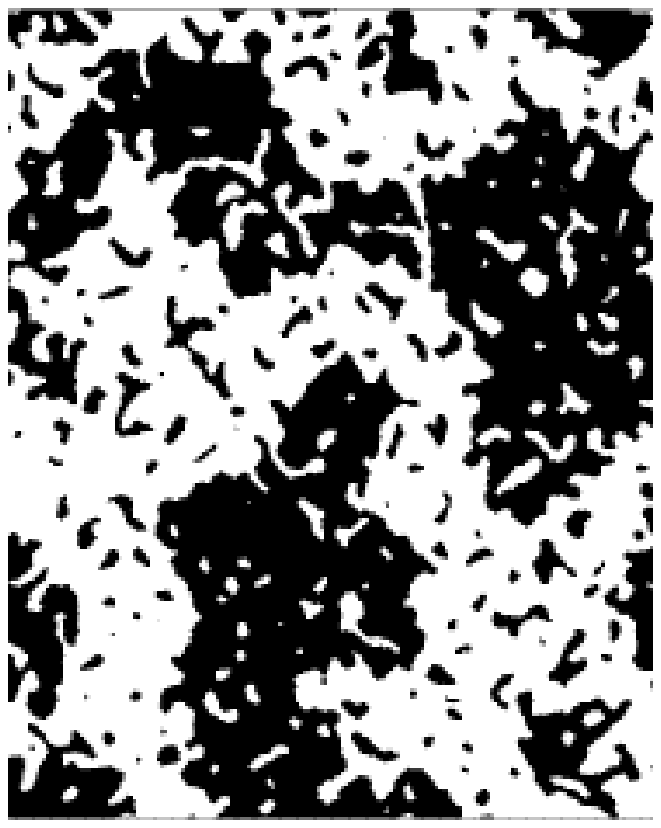}\\

\leavevmode \epsfxsize=0.3\linewidth
\epsfbox{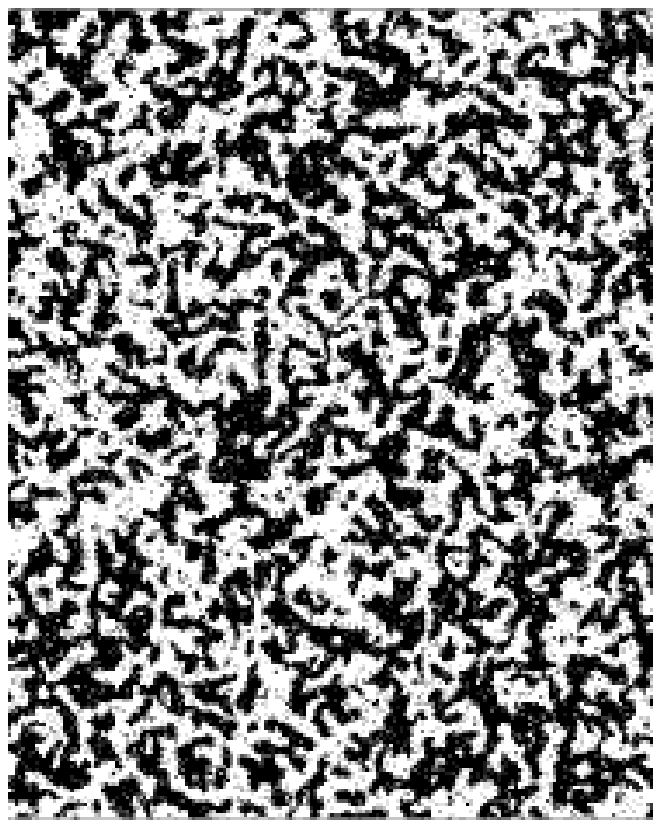} \hfill
\end{center}
\caption{The projection of the spin configuration
onto the ground state $A$, $B$ and $C$ (from top to bottom)
at the end of each stage in the in a 2-step cycling. 
The columns are in the chronological order.
Here $\twz=10000$ ,$\two=100$,$\twt=2$, $\two'=4$,$\twz'=200$ (MCS).
The system size is $1024 \times 1024$.
}
\label{image-2-step.fig}
\end{figure}

In Fig \ref{image-2-step.fig} we show the time evolution
of the projection to three completely different ground states
$A$, $B$ and $C$ during two-step coarsening 
$A (\twz) \rightarrow  B (\two) \rightarrow C (\twt) \rightarrow 
B (\two') \rightarrow C (\twz')$. We chose
$\twz=10000$, $\two=100$, $\twt=2$, $\two'=4$ and $\twz'=200$ (MCS).
The time schedule is decided 
according to the  principle explained in section \ref{subsec.multi}.
First, the first series $A \rightarrow B \rightarrow C$ is designed such that
the length scales of the three-phases are well-separated
$\twz \gg \two \gg \twt$. Second, the reversed series
$B \rightarrow A$ is designed such that multiplicative noise 
is removed one by one without accumulating additional noise:
$\two > \two' > \twt$ and $\twz > \twz' > \twt$.

One can find that noise on $A$ in the end of 3rd stage is very large.
The latter is due to the multiplicative effect of noise. 
It will take enormous time to remove the noise by a single stroke
which well exceeds $\twz$ so that recovery of memory is hopeless.
However, one can find that memory the spin-configuration at the end of
2nd and 1st stages are recovered by the 4th and 5th stages respectively
(as shown by arrows in the schematic picture.)
Thus this example demonstrates that multiple memories can be indeed 
stored and retrieved successively as argued in section \ref{subsec.double}.

\newpage

\newpage

\section{Comparison with Temperature-Cycling Experiments in Spin-Glasses}
\label{sec:scenario}

We now discuss on the temperature-cycling experiments in spin-glass
from the present point of view.
For reference, we consider the data of experiments 
on the $CdCr_{1.7}In_{0.3}S_4$ ($T_{c}=16.7K$) insulating spin glasses.
Because of the limitation of pages
we cannot discuss another rich set of experimental results 
obtained in  Cu:Mn metallic spin-glass  \cite{Nordblad,uppsala-1,uppsala-2} 
which show essentially same features.
For AC-susceptibility in one-step temperature cycling 
we refer to \cite{VBHL,LHMV} and \cite{JVHBN} for multi-step (continuous) 
temperature cycling. For DC-susceptibilities we refer to the measurements of 
thermo-remanent magnetization (TRM) on the same $CdCr_{1.7}In_{0.3}S_4$ system 
reported in \cite{saclay-exp-rev-1} which can also be found
in \cite{saclay-exp-rev-2}.  

\subsection{AC-susceptibility in One Step Temperature-Cycling Experiments}

\begin{figure}
\begin{center}
\leavevmode \epsfxsize=0.6\linewidth
\epsfbox{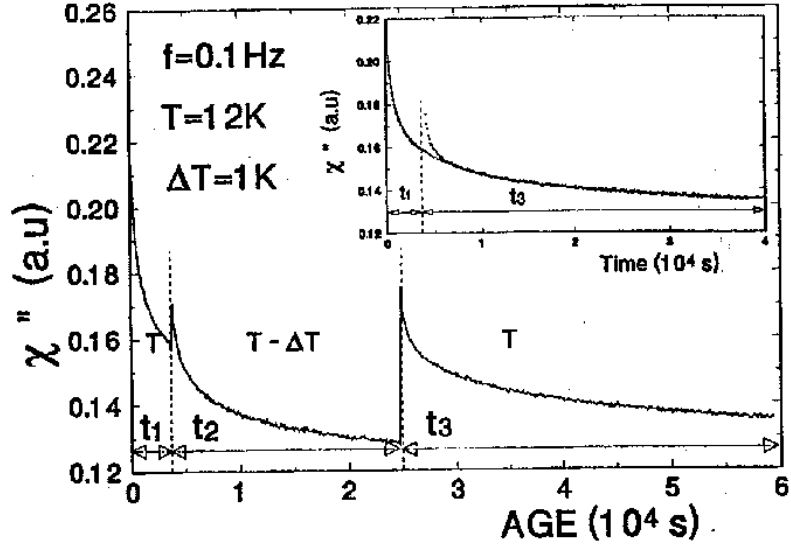}
\end{center}
\caption{Relaxation of $\chi''$ during an 
one-step negative temperature-cycling experiment 
$12K \rightarrow 11K \rightarrow 12K$ as a function of time. 
The inset shows points taken during the first and third stage 
at $T=12K$ plotted against the total time spend at $T=12K$.
The solid line is a reference relaxation curve of isothermal aging 
at $T=12K$.}  
\label{negative-1.fig}
\end{figure}

We consider first the measurements of the out-of-phase AC
susceptibility $\chi''(\omega,t)$ in a one-step temperature
cycling procedure, and interpret them according to the picture presented  
in section \ref{subsec.double} and section \ref{subsec.ac-one-step}.
In Fig. \ref{negative-1.fig} the data \cite{VBHL} of the relaxation of
$\chi''(\omega,t)$ at $\omega/2\pi=0.1 Hz$ is shown. 
Note that the schematic picture presented in Fig. 
\ref{concept-one-step-ac.fig}
agrees well with the general feature of the data.
The third regime can be naturally understood as containing both
the inner-coarsening regime and outer-coarsening regime in a well separated
manner. We expect that the duration of the inner-coarsening regime 
is given by \eq{eq:recover-noise-cycle}. This yields
$\trec \simeq 700$ (sec) at $12K$ using the microscopic time scale 
$\tau_{0}=10^{-13}$ (sec) and the duration of the second stage
$\two \sim 2 \times 10^{4}$ (sec) at $11K$. The order of magnitude of the latter is 
compatible with the duration of the transient 
relaxation $\sim 2500$ (sec) seen at the beginning of the third stage.

\begin{figure}
\begin{center}
\leavevmode \epsfxsize=0.6\linewidth
\epsfbox{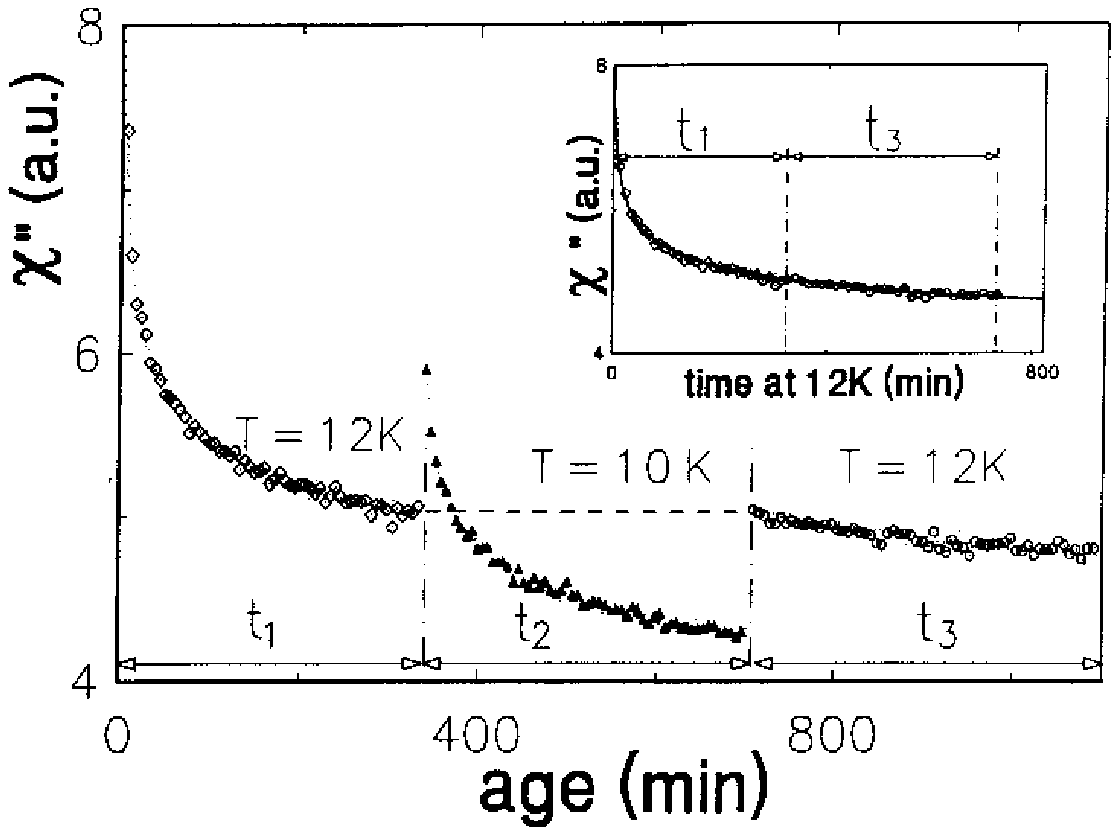}
\end{center}
\caption{
$\chi''$ relaxation during an one-step negative temperature-cycling experiment 
 $12K \rightarrow 10K \rightarrow 12K$ as a
function of time. The inset shows points taken during the first and
third stage at $T=12K$ plotted against the total time spend at $T=12K$.
The solid line is a reference relaxation curve of isothermal aging 
at  $T=12K$.} 
\label{negative-2.fig}
\end{figure}

In Fig. \ref{negative-2.fig} the data \cite{LHMV} of a similar experiment
but with a larger $\Delta T$ is shown. 
In this case the relaxation of the inner-coarsening regime is expected to be, 
from
\eq{eq:recover-noise-cycle}, $\trec \simeq 27$ (sec) using 
$\tau_{0}=10^{-13}$ (sec) and the duration of the second stage
$\two \sim 350$ (min). Note that this is of the same order of the period
of the AC measurement $2\pi/\omega \sim 10$ (sec)
so that condition \eq{eq:cond-inner-coarsening} needed to observe the
inner-coarsening regime is barely satisfied.
Because of the temperature dependence of the stiffness of the 
barrier \eq{eq:stiffness-temp}, $\trec$ would be even shorter and the
condition \eq{eq:cond-inner-coarsening} would be strongly violated.
This could explain why the transient relaxation seen in Fig.\ref{negative-1.fig}
is absent from the data, and that full memory can be expected.

\begin{figure}
\begin{center}
\leavevmode \epsfxsize=0.6\linewidth
\epsfbox{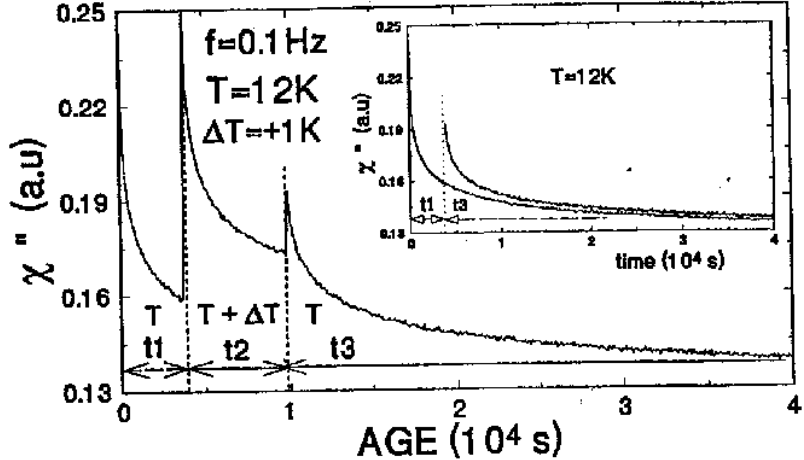}
\end{center}
\caption{
$\chi''$ relaxation during an one-step positive temperature-cycling experiment
 $12K \rightarrow 13K \rightarrow 12K$ as a
function of time. The inset shows points taken during the first and
third stage at $T=12K$ plotted against the total time spend at $T=12K$.
Note that the second curve can nearly be superimposed onto the first one by an
horizontal shift of $4000$ seconds.} 
\label{positive.fig}
\end{figure}

Now let us consider the data \cite{VBHL} corresponding to
a positive cycling in the same system,
which is shown in Fig. \ref{positive.fig}. In this case 
the duration of the second stage $\two\sim 6000$ (sec) at 
$13K$ amounts to effective time \eq{eq:enhanced-separation}
of $10^{5}$ sec at $12K$, which is much larger than the duration of the
first stage ($\twz\sim 4000$ (sec)). Thus in the third regime we expect 
that the inner-coarsening regime and outer-coarsening are not separated,
and that the obtained relaxation is very close to the one obtained in the 
first regime (no memory). 

To summarize we found that the present  picture is compatible with
the experimental data of the out-of-phase AC susceptibility during one-step
temperature cycling. In particular, the memory effect for negative cycling 
is explained by the fact that
the inner-coarsening regime ends after a time so short that it cannot be
observed. 

In the cases discussed above, we did not need to take into account 
the fact that the overlap length $\LOVLP$ become very large when $\Delta T
\to 0$. However it should be noted that the overlap length associated 
with any fluctuations of temperatures in experiments  should be large 
enough compared with the dynamical length scales explored. Otherwise
experimental temperature becomes meaningless.
Indeed, there are some data which suggest that $\LOVLP$  depends on
 $\Delta T$ \cite{uppsala-1}.
For example the data  of negative cycling between very close temperatures
$12 \rightarrow 11.7 K \rightarrow 12K$ of the insulating spin-glass
shown in \cite{LHMV} reveals that the relaxation during the 
second stage at $11.7$ K `helps' relaxation at $12K$ to a certain extent.

\subsection{DC-susceptibility in One Step Temperature-Cycling Experiments}

\begin{figure}
\begin{center}
\leavevmode \epsfxsize=0.6\linewidth
\epsfbox{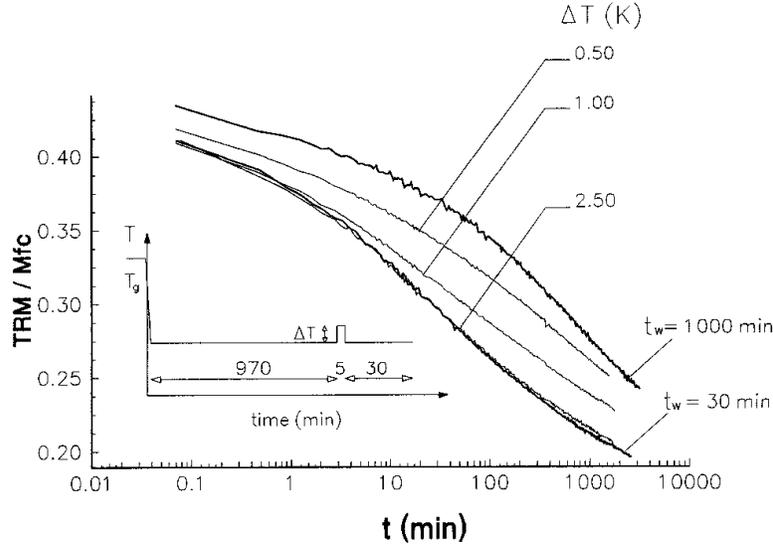}
\end{center}
\caption{Relaxation of the thermo-remanent magnetization (TRM)
at $12K$ after a one-step positive temperature-cycling. The procedure is
shown in the inset. The reference curves of $\tw=30$ (min) and 
$\tw=1000$ (min) of standard (isothermal) aging are also shown.}
\label{trm.fig}
\end{figure}

Let us now consider the relaxation of the
thermo-remanent magnetization (TRM), which is a DC susceptibility,
after similar one-step cycling.
A representative data set is shown in Fig. \ref{trm.fig}
which is taken from \cite{saclay-exp-rev-2}. Here a positive 
temperature cycling 
$12K \rightarrow 12K+\Delta T \rightarrow 12K$ is done under small
magnetic field which is then cut-off. 
Then the subsequent relaxation of the magnetization is measured. 
Here the duration of the second stage is fixed to $5$ minutes but
the amplitude of the shift $\Delta T$ is varied.

Now let us compare the experimental data with the correlation function
after a one-step cycling in the $O(n)$ Mattis model,
which we analyzed in section \ref{subsubsec.c-third-2}. 
Here we suppose that the generic features of the correlation function 
and TRM decay are similar. It should however be noted that
this is not obvious since the violation of the fluctuation 
dissipation theorem (FDT) in such non-stationary situations 
\cite{CK,CKD96,BCKM} does not allow an exact identification of the two quantities.
Second, we again disregard the possible finite value of $\LOVLP$ between the 
different equilibrium states at different temperatures and consider that 
they are completely uncorrelated.  

The activated dynamics \eq{eq:growth-law} implies that
for larger $\Delta T$, the coarsening in the 
second stage reaches larger length scales $L_{12K+\Delta T}(\two)$, 
so that more `noise' is added to the $12K$ domain structure. 
Thus we suppose that larger $\Delta T$ in the experiment 
corresponds to larger $\two$ in our model.
Indeed the experimental curves shown in Fig \ref{trm.fig} appear to be
very similar to the result in the $O(n)$ Mattis model shown in 
Fig. \ref{on_c_3.fig}.

In section \ref{subsubsec.c-third-2}, we found that the generic
features of the auto-correlation function can be understood as crossover 
from rejuvenation (inner coarsening) to memory (outer-coarsening).
This crossover is sharp only in the case a) where the second stage is
effectively much shorter than the first stage ($\two \ll \twz$).
On the other hand, in the case b) where the second stage is
effectively much longer than the first stage, rejuvenation is nearly complete.

Let us note that the curve in Fig.\ref{on_c_3.fig} with 
$\Delta T=2.5K$ belongs to the case b) because the effective
time \eq{eq:enhanced-separation} of $5$ minutes at $14.5K$ 
amounts to $8400$ min at 12K (with $\tau_{0} \sim 10^{-13}$ (sec)),
which is very large compared with $t_{\rm w1}=970$ min.
On the other hand, $\Delta T=1.5 K$  and $1K$ can be considered as cases of a) since
their effective time $430$ min and $97$ min at 12K are smaller than $t_{\rm w1}$.
It would be very interesting to look experimentally for the plateau regime
as seen in Fig.\ref{on_c_1.fig}, \ref{on_c_2.fig}, \ref{on_c_3.fig}
and Fig. \ref{tw0-tw1=20.fig}, \ref{tw0=1000-inf.fig}.

\subsection{Multiple Step Temperature-Cycling}

\begin{figure}
\begin{center}
\vspace*{1cm}
\leavevmode \epsfxsize=0.6\linewidth
\epsfbox{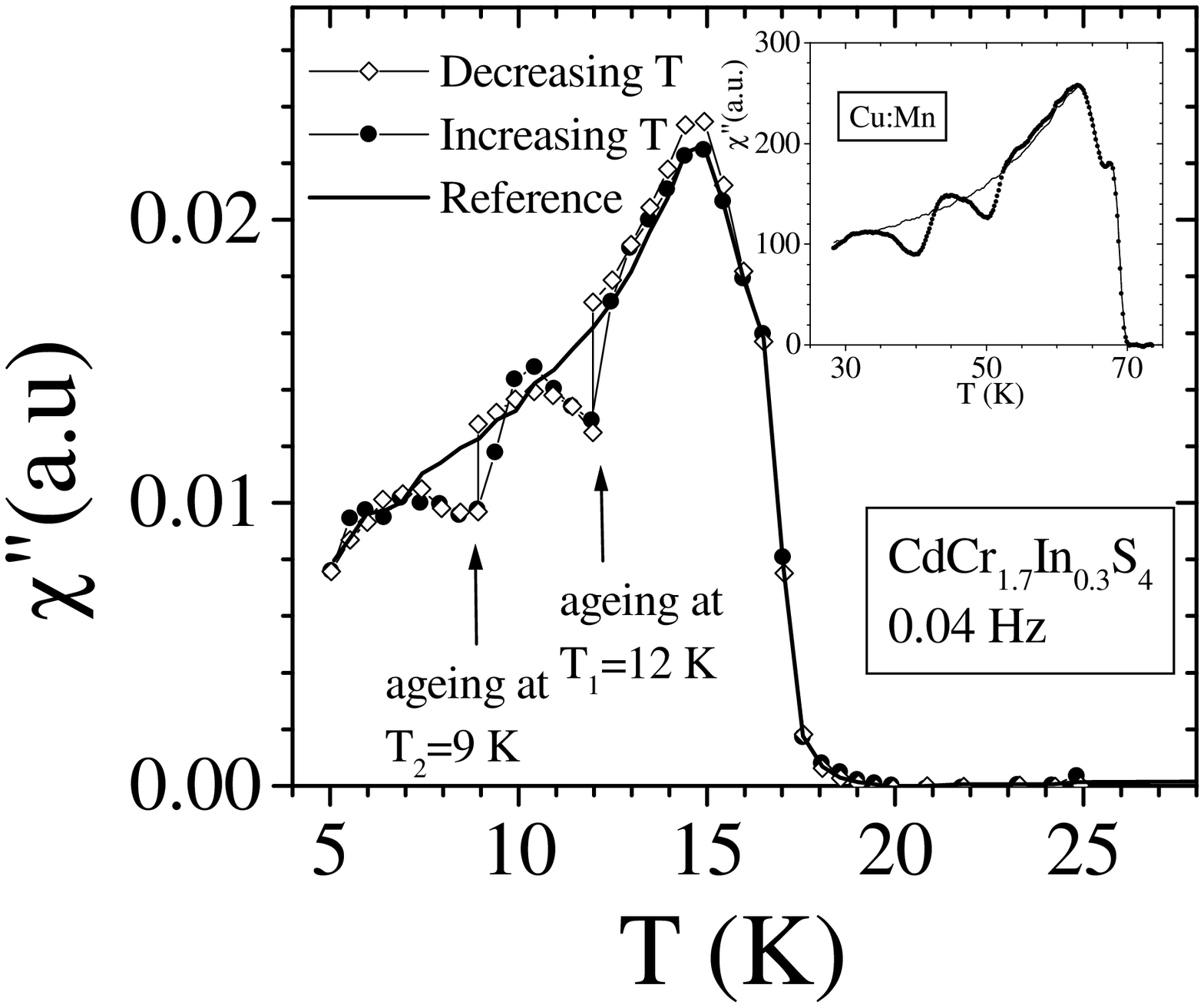}
\end{center}
\caption{
Out-of-phase susceptibility $\chi''$ of the
$CdCr_{1.7}In_{0.3}S_4$ spin glass.  The solid line is measured upon
heating the sample at a constant rate of $0.1K/min$ (reference
curve). The data during cooling (not shown) lies close to but 
slightly above the solid line.
Open diamonds: the measurement is done during cooling at this
same rate, except that the cooling procedure has been stopped at $12K$
during $7h$ and then $40h$ at $9K$
to allow for aging. Cooling then resumes down to $5K$:
$\chi''$ is not influenced and goes back to the reference curve
(chaos). Full circles: after this cooling procedure, the data is
taken while re-heating at the previous constant rate, exhibiting
memory of the aging stage both at $9K$ and $12K$.
The inset shows a similar ``double memory'' experiment performed on the
Cu:Mn metallic spin glass. }
\label{continuous.fig}
\end{figure}

Finally let us consider the multiple-step (continuous) version 
of the temperature cycling which has been studied 
in recent experiments \cite{Nordblad,uppsala-2,JVHBN,interference}
using the out-of-phase AC susceptibility $\chi''(\omega,t)$.
Here we discuss based on the scenario we developed 
in  section \ref{subsec.multi}.
In Fig. \ref{continuous.fig} an example of the 
data taken from \cite{JVHBN} is shown.

The basic schedule is a negative-cycling which consist of cooling 
and heating procedure.  In the cooling, 
the temperature is reduced successively by `micro-quenches' 
of small temperature step $\Delta T$, $T_{i}=T_{\rm start}-i \Delta T$ 
with ($i=1,2,\ldots,n$) where $T_{\rm start}$ is the starting 
temperature and $n$ is the number of steps. At every temperature $T_{i}$ 
a short period of time $t_{i}=\Delta t$ is spent. In this experiment, 
$\Delta T=0.5K$ and $\Delta t=5min$ \cite{typical-schedule}.
In the heating. the temperature is raised back successively 
by the same $\Delta T$ with a the same  period
$t'_{i}=\Delta t$ at each temperature.

Within the present scenario, coarsening of new phases are expected 
at every temperature in the cooling and the heating is the revered order 
series of coarsening.
Note that the above basic schedule satisfies 
the condition \eq{eq:retrieval-condition} to retrieve the memory at each 
temperature in the heating.
For instance \eq{eq:recover-noise-cycle} implies the noise 
due to $\Delta T=5$ min at $10$ K on the phase at $10+\Delta T=10.5$ K 
can be erased quickly within $\tau_{\rm rec} \sim 1$ (min). 
Indeed, it is found in the experiment that the reference curve 
of cooling and heating lie close to each other \cite{JVHBN}.
Concerning the frequency of AC $\omega/2\pi \sim 0.04$ (Hz),
we need to suppose that it is low enough so that the signal of the 
cleaning (inner-coasting) does not blur the observation of 
the memory.

In the experiment, an interesting modification is made for the cooling 
procedure: now two special temperatures ($9$ and $12$ K) are chosen 
to make relatively long stops $t_{\rm stop} \sim 40, 7 $ hours
in stead of the short period $\Delta t=5$ min. But all the rest of the 
basic schedule is unchanged. The dips in cooling curve mean relaxation 
due to the long aging. Then the fact that the cooling curve tend 
to return to the reference curve (without stops) after the long 
stops implies coarsening of the new phase within the
present scenario.

Let us the consider the heating curve. 
As the temperature comes up to the stop temperature 
$9$ K, the curve goes downward tracing the cooling curve. 
The latter is expected within the present scenario
since as soon as the noise due to coarsening at lower temperatures 
are erased, we are left with the large ghost domain 
at $9$ K with its full amplitude recovered.
The same phenomena happens also for $12$K.

The phases at the temperatures just above $9$ K and $12$ K are 
expected to be very noisy due to the long stops.
The latter belongs to the case b) 
discussed in section \ref{sec.one-step-third} and 
their memory may be hardly retrieved. Then the increase
of the heating curve after passing the stop temperatures
may be interpreted as the signal of inner-coarsening (rejuvenation).
Furthermore, the long stop at $9$ K of $40$ hours could have affected 
the memory at $12$K. However, its effective time turns out to be
only $4$ sec at $12$ K. 
If the stop temperatures are closer, the recovery 
of memory could have been affected more. Such a phenomena
is indeed observed in some experiments \cite{interference}.

Finally let us also note that the 'dips' are rounded in the 
experimental data, while
we would have expected sharper dips. In the present scenario, the 
rounding of the dips may be attributed to finiteness of the overlap length.

\section{Summary and Open problems}
\label{sec.discussion}

\subsection{What we have done in this paper}

The basis of our approach in the present study is the droplet picture.
The fundamental ingredients of the present paper are the tenets
of the droplet theory, namely: i) for a given temperature below 
the critical temperature, there are only two equilibrium states related 
by a global flip; ii) the relaxational dynamics is the coarsening 
of the domains of these equilibrium states. 
The growth of the domains is due to the thermal activation of droplets; 
iii) the underlying equilibrium states change chaotically
with weak external perturbations, such as a change of temperature. 

By a combination of scaling arguments, analytical calculations and numerical 
simulations, we have demonstrated {\it preservation and retrieval of memory}
is possible in a very dynamical way in the absence of any underlying static 
backbone structure. First, we showed that {\it preservation of memory} is 
possible by `ghost domains' (domain wall structure plus noise) 
 of {\it all} the phases the 
system goes through in the temporal spin-configuration.
Here coexistence of statistically independent 
domains only amount to mutual injection of uncorrelated short-range noise
which do not destroy large scale spatial structures of each other.
Thus in order to describe the spin-configuration, one needs to keep 
track of the projections to all phases. In this sense, there are indeed 
many `phases ' in the temporal spin-configuration, as already anticipated 
in the experimental studies \cite{saclay-exp-rev-1,saclay-exp-rev-2}. 
Then we studied in detail {\it retrieval of memory}: how the noise 
on a phase due to coexistence of the domain structures 
of other phases can be removed 
by additional conjugate coarsening. The analysis has been carried out 
explicitly in term of various spatial/temporal correlation 
functions and linear response functions. In particular, we have shown that 
when the time spent in the intermediate state is sufficiently small, 
rejuvenation effects and memory effects can be observed in a 
well separated manner. 
There the difficulty of the previous works \cite{FH88b,KH,uppsala-1} to 
explain the `memory effect', is resolved.  It is rather surprising 
to find that the simplest version of the droplet picture already 
allows coherent and rather rich phenomenology which can account 
for experiments at least qualitatively.

\subsection{Is the picture totally satisfactory ?}

The droplet picture itself is based on phenomenological scaling 
arguments and Migdal-Kadanoff type real-space renormalization-group 
calculations \cite{BM,FH88a}. It is by no means obvious that real 
spin-glasses can be described within this simple scenario. Many papers
have appeared recently, presenting relatively strong evidence against the
simplest version of the droplet theory \cite{Martin,ucstcrz,KA}. On the other
hand evidence for the validity of the basic droplet picture has also been 
provided \cite{MKRG,Middleton,HG,3DEA-KYT,4DEA-DF,4DEA-HYT}.
In the present paper we have focused on the simplest version of the droplet 
theory. Any modifications due to possible corrections to the simplest version
go beyond the scope of the present paper and we leave this issue for future studies.

The theoretical situation of temperature chaos is also
far from being settle neither in low dimensional systems nor in mean-field models.
Early numerical investigations at finite temperatures \cite{R94,N97} suggest
chaos with magnetic field and with the change of couplings. 
Temperature chaos has been found in 2-d Edwards-Anderson 
model \cite{HK} for very large length scales. Temperature chaos has also 
been suggested by some analytical studies of mean-field models \cite{KV,FN,Franzprivate}. 
However recent simulations of both the SK and the 3-d Edwards-Anderson models 
have concluded on the absence of chaos with temperature \cite{MaChaos} 
(as also suggested by another analytical calculation for the SK model \cite{PR}). 
But there is still a possibility that temperature-chaos was not observed in the latter
because the overlap length is larger than the system sizes studied.
Chaotic changes of equilibrium states have
also been looked for other related glassy systems. For instance,
a large scale transfer-matrix study of an elastic string pinned 
in random media \cite{FH91}, which describes for instance pinned domain walls,
has found  some evidence for the chaotic change of the free-energy landscape with 
temperature again for very large length scales. All studies therefore suggest
that `chaos', if it exists, only occurs on large length scales, whereas the
dynamical length is expected to be modest, even in experimental 
studies  \cite{JOW,BYHTTI}, due to its very slow growth with time.
 
\subsection{An alternative picture} 

Let us now briefly compare the present  scenario based on the droplet 
picture with another
phenomenology \cite{saclay-exp-rev-1,saclay-exp-rev-2}, based on a dynamical interpretation of the Parisi solution of the SK (mean-field) model, which 
suggests that the energy landscape is hierarchical. A concrete implementation
of this picture was proposed in \cite{BD}, in terms of a thermally activated
generalized random-energy model (GREM). To each level of the hierarchical
tree is associated a transition temperature, such that the dynamics at this 
level is stationary for higher temperatures, and aging for smaller temperatures.
A small decrease of temperature induces some rejuvenation by driving out of equilibrium 
a new level of the tree, while freezing out the dynamics at the upper levels, thereby
allowing the memory to be conserved. This model was recently studied further 
in \cite{SN}, where it is shown numerically that the rejuvenation/memory
effect is indeed already reproduced with two levels. 

A real space interpretation of this hierarchical 
tree proposed in \cite{BD,BBM},
and further developed in \cite{JP}, in terms of a multi-scale dynamics. 
Low levels of the tree corresponds to short 
wavelength modes, which are only frozen at low temperature, while large wavelengths
modes are frozen at a higher temperature and constitute the `backbone' where the 
memory is imprinted. The picture is particularly clear in the context of pinned 
domains walls \cite{JP}, and should apply to disordered ferromagnets where the
slow dynamics comes from the motion of these pinned domain walls \cite{pinned-domainwall,dipole-glass-2}. 

A closer look at the two different scenarios shows that they actually share
several important points. The fact that the dynamics is thermally activated 
is crucial in both pictures, as it provides a natural separation of time scales
for different length scales and/or temperatures. As was pointed out in 
\cite{JP}, for a given experimental time window, temperature acts as a microscope
by selecting the relevant dynamical length scales. The way memory is conserved
is therefore common in both approaches: it is stored in large length scales which
are to a large extent unaffected by the small length scale dynamics taking place
at lower temperatures. However it should be noted that in the droplet picture, 
there is no {\it static} backbone structure in the phase space and the memory is
entirely a dynamical effect. 
The mechanism leading to rejuvenation is therefore different
in the two pictures: in the droplet picture, it is due to a complete modification
of the free energy landscape with temperature, whereas in the hierarchical picture
it is due to the progressive freezing of smaller and smaller wavelengths \cite{JP}. This
picture is consistent with the simulations of \cite{MaChaos}: a broad peak in the
overlap probability at high temperature is resolved into sharper and sharper sub-peaks
as the temperature is lowered, with no overall shift. 

One should finally note that comparable rejuvenation/memory effects have now been seen
in very different materials, such as PMMA \cite{polymer-glass}, where
the droplet model is probably not the appropriate picture. In this respect, the 
idea of progressive freezing of smaller and smaller length scales is perhaps more
generic.

\subsection{Further developments}

In the present paper, we discussed experimental data only qualitatively. More
detailed and critical examinations of the scenario developed in the present
paper should be done. It would be very interesting to look for 
the plateau regime which separates the rejuvenation and memory both in  experiments 
and numerical simulations. 

A crucial difference between the two pictures is therefore the existence of an
overlap length $\LOVLP$ in the droplet picture. The role of this length 
(that we have neglected in the present paper to focus on the phenomena on
larger time scales) should show up when the temperature change
is very small (or equivalently shorter time scales are considered), 
and therefore be important to account for the possible cooling
rate effects \cite{RVHO87,JVHBN} in spin-glasses. 
We hope to come back to this important point in
the future. We have furthermore neglected the fact that the energy barriers 
are expected to be temperature dependent in the droplet picture
through the temperature dependence of the stiffness \eq{eq:stiffness-temp}.
This leads to a super-activated behavior of the time scales, as was 
seen in \cite{LOHOV},
that needs to be quantitatively accounted for \cite{usinprep,4DEA-HYT}.

If chaotic change of equilibrium states can be made by non-thermal perturbations like change of magnetic field \cite{R94} or pressure, they will also provide 
useful means to examine the present scenario. Temperature changes
may be used simultaneously to enhance the time separations.
If the present picture is correct, such a non-thermal perturbation 
at $T-\Delta T$ should induce the growth of domains on very short length scales, 
an effect that can be erased quickly by heating up the system. 
Therefore, complete memory should also be 
observed in a protocol where:
$T \to T-\Delta T \to T$ and the strength of the strength of non-thermal perturbation 
$P$ is changed as $P=0 \to P=\Delta P \to P=0$ simultaneously.

Another good testing ground for the two approaches may be systems with pinned-domain walls
which is intimately related with the problem of  elastic manifolds 
in random media \cite{HH85,MP,BF,BBM,M90,FH91,HF94} whose dynamics show aging effects 
\cite{KH93,FM,CKD96,Y96,B97,Y98,Y00}.
As we noted above a very natural scenario can be obtained by a GREM based approach \cite{JP}. 
On the other hand, it is also straightforward to construct a 
scenario based on a droplet picture \cite{FH91} for this problem in a way similar 
to our present work. A suitable toy model corresponding to the spherical Mattis model 
we studied here is the elastic manifold pinned by quenched random force field 
(Larkin model) subjected to cycling of the realization of the random force field.
Concerning this issue, it is interesting to note that recent experiments on disordered 
ferromagnets and ferro-electrics have revealed several similarities with spin-glasses \cite{pinned-domainwall,dipole-glass-2}.

\subsection{Some remarks on numerical simulations} 

A discouraging point of the numerical approach is that 
the separation of time scales cannot be made so dramatic
like in experiments of spin-glasses.
However, possible differences between experiments and
simulations due to such a difference of time scales should be amenable to
a quantitative analysis. 
One of the great advantages of the numerical approach is that one can 
directly obtain the size of the domains $L_{T}(t)$ at each time step 
using a spatial correlation functions between two real replicas.
\cite{Rieger-group} 
Then the scaling ansatz presented in the present paper can be tested 
very precisely since everything can be expressed in terms of the
size of the domains $L_{T}(t)$. Indeed this approach has been very useful to test some scaling ansatz by the droplet picture \cite{FH88b} 
for isothermal aging.\cite{3DEA-KYT,4DEA-HYT}
Furthermore, combined with an analysis on equilibrium properties 
around and below the critical temperature \cite{4DEA-DF},
this kind of approach allows one to quantitatively analyze possible 
crossovers between critical and low-temperature behaviors 
in the dynamical observables and in the domain growth law itself. \cite{4DEA-HYT}

It is important to note that due thermally activated processes
which presumably dominate relaxational dynamics 
in spin-glass like systems, the {\it length scales} explored 
in simulations are actually very small. This might be the reason why 
rejuvenation/memory effects are hard to observe numerically \cite{Heiko,KYT-cycle,picco}, because 
both the above scenarios rely on the existence of non overlapping 
(in length scales) dynamical processes. 

The fact that the growth law is so slow also implies that both experiments and simulations
will never be in the {\it ideal} regime of asymptotically large length scales
to test the scaling predictions of droplet picture. Whether one can observe 
some clear signatures of the droplet picture within a realistic time scale may then depend very much on details of the systems studied in experiments and simulations.
Whether the usual Edwards-Anderson model is the best model to describe
real spin-glass systems is now hotly discussed.\cite{chiral,picco}

We want to thank E. Vincent for the permission to use the curves of the
series of experiments in Saclay, useful discussions and encouragement. 
We want to thank S. Miyashita, M. Hammann, H. Chat\'e, 
S. Franz, F. Ricci, P. Nordblad, P. Jonoson and R. Matthieu  
for useful discussions.
H. Y. want to thank  H. Takayama, K. Hukushima, T. Komori and 
L. Bernardi for useful discussions in the collaborations with them 
which have motivated the present work.

\appendix

\section{Properties of the Transformation Field}
\label{sec:transformation-field}

Here we summarize some useful statistical properties
of the transformation field introduced in  \eq{eq:intro-transform-2}.
Due to \eq{eq:intro-transform-1}, the Fourier components 
of projection to different equilibrium states, say $\alpha$ and $\beta$,
are related as,
\begin{equation}
\hat{\phi}_{k}^{\alpha}=\int\frac{d^{d}k'}{(2\pi)^{d}}
(\hat{\sigma}^{\alpha \beta})_{k'}\hat{\phi}_{k-k'}^{\beta}, 
\label{eq:transform}
\end{equation}
where we defined
\begin{equation}
(\hat{\sigma}^{\alpha \beta})_{k}
=\int d^{d}x \sigma^{\alpha \beta}(x) e^{ikx}. 
\label{eq:trans-field}
\end{equation}

The average becomes zero due to \eq{eq:av-sig}, 
\begin{equation}
\langle(\hat{\sigma}^{\alpha \beta})_{k}\rangle_{\sigma}=0 \label{eq:map-noise-mean}.
\end{equation}
and the 2-body correlation function is obtained using \eq{eq:corr-sig} as,
\begin{equation}
\langle(\hat{\sigma}^{\alpha \beta})_{k}
(\hat{\sigma}^{\alpha \beta})_{l}\rangle_{\sigma}=
\int d^{d}xd^{d}x'
\langle\sigma^{\alpha \beta}_{x}\sigma^{\alpha \beta}_{x'}\rangle_{\sigma}
e^{ikx}e^{ikx'}
=
\Delta (2\pi)^{d} \delta^{d}(k+l)  \label{eq:map-noise-correlation}.
\end{equation}
The 4-body correlation function in the real space is,
\begin{eqnarray}
&& \langle\sigma^{\alpha \beta}(x_1)\sigma^{\alpha \beta}(x_2)\sigma^{\alpha \beta}(x_3)\sigma^{\alpha \beta}(x_4)\rangle_{\sigma} \nonumber \\
&= & \Delta^{2} \{ 
\delta^{d}(x_1-x_2)\delta^{d}(x_3-x_4)
+\delta^{d}(x_1-x_3)\delta^{d}(x_2-x_4)
+\delta^{d}(x_1-x_4)\delta^{d}(x_2-x_3) \nonumber \} \\
&-& 2 \Delta^{3} \delta^{d}(x_1-x_2)\delta^{d}(x_2-x_3)\delta^{d}(x_3-x_4),
\end{eqnarray}
which becomes in the Fourier space,
\begin{eqnarray}
&& \langle(\hat{\sigma}^{\alpha \beta})_{k_1}
(\hat{\sigma}^{\alpha \beta})_{k_2}
(\hat{\sigma}^{\alpha \beta})_{k_3}
(\hat{\sigma}^{\alpha \beta})_{k_4}
\rangle_{\sigma} \nonumber \\
&=&
\Delta^{2} \{ 
(2\pi)^{d} \delta^{d}(k_1+k_2)\times (2\pi)^{d} \delta^{d}(k_3+k_4) 
+(2\pi)^{d} \delta^{d}(k_1+k_3)\times (2\pi)^{d} \delta^{d}(k_2+k_4) 
+(2\pi)^{d} \delta^{d}(k_1+k_4)\times (2\pi)^{d} \delta^{d}(k_2+k_3) \} \nonumber \\
&-& 2 \Delta^{3}(2\pi)^{d} \delta^{d}(k_1+k_2+k_3+k_4).
 \label{eq:map-noise-correlation-4}
\end{eqnarray}

We will also need to consider projections to three different
states say $\alpha$, $\beta$ and $\gamma$. A useful correlation function is,
\begin{equation}
 \langle\sigma^{\alpha \beta}(x_1)\sigma^{\beta \gamma}(x_2)\sigma^{\gamma \beta}(x_3)\sigma^{\beta \alpha}(x_4)\rangle_{\sigma} 
=  \Delta^{2} \delta^{d}(x_1-x_4)\delta^{d}(x_2-x_3),
\end{equation}
which becomes in the Forier space,
\begin{equation}
 \langle(\hat{\sigma}^{\alpha \beta})_{k_1}
(\hat{\sigma}^{\beta \gamma})_{k_2}
(\hat{\sigma}^{\gamma \beta})_{k_3}
(\hat{\sigma}^{\beta \alpha})_{k_4}
\rangle_{\sigma} 
=\Delta^{2}
(2\pi)^{d} \delta^{d}(k_1+k_2)\times (2\pi)^{d} \delta^{d}(k_3+k_4) 
\label{eq:map-noise-correlation-3-states}
\end{equation}

Finally, correlation of the generalized type \eq{eq:corr-sig-general}
becomes in the Fourier space,
\begin{equation}
 \langle\hat{\sigma}^{\alpha_1 \alpha_2}_{k_1}\hat{\sigma}^{\alpha_2 \alpha_3}_{k_2}
\ldots \hat{\sigma}^{\alpha_{n-1} \alpha_{n}}_{k_{n-1}} \hat{\sigma}^{\alpha_n \alpha_1}_{k_n}\rangle_{\sigma}  \\
= (\Delta (2\pi)^{d})^{n} \delta^{d}(k_{1}+k_{2}+\ldots+k_{n}).
\label{eq:map-noise-correlation-general}
\end{equation}

\section{Properties of Noise Induced by Coarsening 
of Unrelated Phase}
\label{sec:appendix-noise-correlation}

Here we consider a projection field $\phi^{\alpha}(x)$ associated with
a ground state $\sigma^{\alpha}(x)$ and study the effect of 
coarsening with respect to an unrelated ground state $\sigma^{\beta}(x)$.

We assume that the initial projection field $\phi^{\alpha}(x,t')$
satisfies the normalization condition $(\phi^{\alpha}(x,t'))^{2}=1$  
at any $x$ which implies,
\begin{equation}
\hat{\phi}^{\alpha}_{k}(t')\hat{\phi}^{\alpha}_{l}(t')
=(2\pi)^{2d} \delta^{d}(k+l) W^{\alpha}_{k}(t'), \label{eq:norm-ini-1}
\end{equation}
where
\begin{equation}
\int d^{d}k W^{\alpha}_{k}(t')=1. \label{eq:norm-ini-2}
\end{equation}
due to \eq{eq:norm-prop-1} and \eq{eq:norm-prop-2}.
Given a spin configuration whose
projection to a ground state $\sigma^{\alpha}(x)$ is $\phi^{\alpha}(x,t')$
at time $t'$, we project it to the ground state $\sigma^{\beta}(x)$
to prepare the initial configuration. Using \eq{eq:transform}, the 
initial condition can be read as,
\begin{equation}
\hat{\phi}_{k}^{\beta}(t')=
\int \frac{d^{d}k'}{(2\pi)^{d}}
(\hat{\sigma}^{\alpha \beta})_{k'}
\hat{\phi}_{k-k'}^{\alpha}(t'). 
\end{equation}
Because of the random mapping through the transformation field
$(\hat{\sigma}^{\alpha \beta})_{k}$, the resultant 
$\hat{\phi}_{k}^{\beta}(t')$ should also be a random field.
Using \eq{eq:map-noise-mean}, we find that the mean  is zero,
\begin{equation}
\langle\hat{\phi}_{k}^{\beta}(t')\rangle_{\sigma}=
\int \frac{d^{d}k'}{(2\pi)^{d}}
\langle(\hat{\sigma}^{\alpha \beta})_{k'}\rangle_{\sigma}
\hat{\phi}_{k-k'}^{\alpha}(t')=0.
\end{equation}
Using \eq{eq:map-noise-correlation}, the correlation function becomes,
\begin{equation}
\langle\hat{\phi}_{k}^{\beta}(t')\hat{\phi}_{l}^{\beta}(t')\rangle_{\sigma}
=\int \frac{d^{d}k'}{(2\pi)^{d}}\int \frac{d^{d}l'}{(2\pi)^{d}}
\langle(\hat{\sigma}^{\alpha \beta})_{k'}(\hat{\sigma}^{\alpha \beta})_{l'}\rangle_{\sigma}
\phi^{\alpha}_{k-k'}\phi^{\alpha}_{l-l'}
=\Delta (2\pi)^{d}\delta^{d}(k+l).
\end{equation}
In the last equation, we used  \eq{eq:norm-ini-1} and \eq{eq:norm-ini-2}.
To summarize, the initial condition for the coarsening is a random initial 
condition which has only short-range correlation. The solution of the equation
of motion with such random initial condition is known and shown in
\eq{eq:random-solution}.

By transforming the solution of $\hat{\phi}_{k}^{\beta}(t)$ 
at time $t(>t')$ back to the ground state $\sigma^{\alpha}$ through
\eq{eq:transform}, we obtain the projection field as,
\begin{eqnarray}
\hat{\phi}_{k}^{\alpha}(t) &=&
\int \frac{d^{d}k'}{(2\pi)^{d}}(\hat{\sigma}^{\alpha \beta})_{k'}
\hat{\phi}_{k-k'}^{\beta}(t) \nonumber \\
&=& \int \frac{d^{d}k'}{(2\pi)^{d}}
(\hat{\sigma}^{\alpha \beta})_{k'}
\frac{e^{-(k-k')^{2}(t-t')}}{\sqrt{\Gamma_{0}(t,t')}}
\hat{\phi}_{k'}^{\beta}(t') \nonumber \\
&=& \int \frac{d^{d}k'}{(2\pi)^{d}}
(\hat{\sigma}^{\alpha \beta})_{k'}\frac{e^{-(k-k')^{2}(t-t')}}{\sqrt{\Gamma_{0}(t,t')}}
\int \frac{d^{d}k''}{(2\pi)^{d}}
(\hat{\sigma}^{\alpha \beta})_{k''}\hat{\phi}_{k-k'-k''}^{\alpha}(t').
\end{eqnarray}
In the following we examine the statistical property of the 
resultant projection field.

Using \eq{eq:map-noise-correlation},\eq{eq:random-solution} 
and \eq{eq:c} we obtain the expectation value of the resultant projection
field as,
\begin{eqnarray}
\langle\hat{\phi}_{k}^{\alpha}(t)\rangle_{\sigma} &=&
\int \frac{d^{d}k'}{(2\pi)^{d}}
\int \frac{d^{d}k''}{(2\pi)^{d}}
\langle(\hat{\sigma}^{\alpha \beta})_{k'}
(\hat{\sigma}^{\alpha \beta})_{k''}\rangle_{\sigma}
\frac{e^{-(k-k')^{2}t}}{\sqrt{\Gamma_{0}(t,t')}}
\hat{\phi}_{k-k'-k''}^{\alpha}(t') \nonumber \\
& = & \frac{\Delta}{(2\pi)^{d}}  \int d^{d}k'
\frac{e^{-(k-k')^{2}t}}{\sqrt{\Gamma_{0}(t,t')}}
\hat{\phi}_{k}^{\alpha}(t') \nonumber \\
& = & C_{0}(t-t',0)\hat{\phi}_{k}^{\alpha}(t').
\label{eq:coarsening-noise-meanr-1}
\end{eqnarray}
By taking inverse Fourier transform we obtain,
\begin{equation}
\langle\phi^{\alpha}(x,t)\rangle_{\sigma}=C_{0}(t-t',0)\phi^{\alpha}(x,t').
\label{eq:coarsening-noise-meanr-2}
\end{equation}
We find that profile of the field is maintained
on average with a reduced amplitude even after 
coarsening with respect to completely unrelated phase.
The amplitude decreases as 
the system de-correlates from the initial configuration by coarsening.
In section. We have discussed a special case with an initial condition 
in which symmetry fully-broken and found that the symmetry remains broken
at any time. The above result is the solution with  general initial 
conditions.

One can consider more generalized case where coarsening of 
multiple different equilibrium states are performed successively.
Suppose that coarsening of $n$ different equilibrium states, which are
unrelated with $\alpha$ and with each other, are performed in succession
between $(t_{n-1},t')$, $(t_{n-2},t_{n-1})$,\ldots, $(t_{1},t_{2})$
and $(t,t_{1})$. Then following similar calculation and using 
\eq{eq:map-noise-correlation-general} the projection field of $\alpha$ 
is obtained as,
\begin{equation}
\langle\phi^{\alpha}(x,t)\rangle_{\sigma}=C_{0}(t-t_{1},0)
C_{0}(t_{1}-t_{2},0)\ldots C_{0}(t_{n-1}-t',0)\phi^{\alpha}(x,t').
\label{eq:multiplicative-reduction}
\end{equation}
We find that profile of the field is maintained
with the amplitude reduced multiplicatively every time when 
a new unrelated phase is coarsened on top of it.

Next we consider the spatial correlation function of the 
resultant projection field. Using \eq{eq:map-noise-correlation-4},
\eq{eq:random-solution}, \eq{eq:c}, \eq{eq:norm-ini-1} and
\eq{eq:norm-ini-2} we obtain,
\begin{eqnarray}
\langle\hat{\phi}_{k}^{\alpha}(t)\hat{\phi}_{l}^{\alpha}(t)\rangle_{\sigma} &=&
\int \frac{d^{d}k'}{(2\pi)^{d}}
\int \frac{d^{d}k''}{(2\pi)^{d}}
\int \frac{d^{d}l'}{(2\pi)^{d}}
\int \frac{d^{d}l''}{(2\pi)^{d}}
\langle(\hat{\sigma}^{\alpha \beta})_{k'}
(\hat{\sigma}^{\alpha \beta})_{k''}
(\hat{\sigma}^{\alpha \beta})_{l'}
(\hat{\sigma}^{\alpha \beta})_{l''}
\rangle_{\sigma} \nonumber \\
&& \frac{e^{-(k-k')^{2}(t-t')}}{\sqrt{\Gamma_{0}(t,t')}}
\frac{e^{-(l-l')^{2}(t-t')}}{\sqrt{\Gamma_{0}(t,t')}}
\hat{\phi}_{k-k'-k''}^{\alpha}(t') 
\hat{\phi}_{l-l'-l''}^{\alpha}(t')  \nonumber \\
&= & C_{0}^{2}(t-t',0) 
\hat{\phi}_{k}^{\alpha}(t')\hat{\phi}_{l}^{\alpha}(t')
+   \Delta (2\pi)^{d} \delta^{d}(k+l)  \left[ 1 -  2 C_{0}^{2}(t-t',0)  \right. \nonumber \\
 &+&  \left. \frac{1}{\Gamma_{0}(t-t',0)} \frac{\Delta}{(2\pi)^{d}} 
\int d^{d}k'\int d^{d}l'
e^{-(k-k')^{2}(t-t')}e^{-(k+l')^{2}(t-t')} W^{\alpha}_{k-k'+l}(t')
 \right ].
\label{eq:coarsening-noise-corr-1}
\end{eqnarray}
The structure-factor \eq{eq:norm-prop-1} is obtained as,
\begin{eqnarray}
W^{\alpha}_{k}(t) &=&
  C_{0}^{2}(t-t',0) W^{\alpha}_{k}(t')
+   \frac{\Delta}{ (2\pi)^{d}}  
\left [ 1 -  2 C_{0}^{2}(t-t',0) \right.   \nonumber \\
& + &  \left. \frac{1}{\Gamma_{0}(t-t',0)} \frac{\Delta}{(2\pi)^{d}} 
\int d^{d}k'\int d^{d}l'
e^{-(k-k')^{2}(t-t')}e^{-(k+l')^{2}(t-t')} W^{\alpha}_{k-k'+l}(t')
 \right ].
\label{eq:structure-factor-back-to-alpha}
\end{eqnarray}
By taking inverse Fourier transform we obtain,
\begin{eqnarray}
&& \langle\phi^{\alpha}(x,t)\phi^{\alpha}(x',t)\rangle_{\sigma}
-\langle\phi^{\alpha}(x,t)\rangle_{\sigma}\langle\phi^{\alpha}(x',t)\rangle_{\sigma} \nonumber \\
&=& \Delta d^{d}(x-x') [1-2C_{0}^{2}(t-t',0)] \nonumber \\
&+& \frac{1}{\Gamma_{0}(t-t',0)} 
\left ( \frac{\Delta}{(2\pi)^{d}}\right)^{2}
\int d^{d}k\int d^{d'}k \int d^{d}l'
e^{-(k-k')^{2}(t-t')}e^{-(k+l')^{2}(t-t')} W^{\alpha}_{k-k'+l'}(t')
\label{eq:coarsening-noise-corr-2}
\end{eqnarray}
The above result implies that coarsening with respect 
an unrelated phase induce noise with certain spatial correlation.

\section{Response to Uniform Probing Field}
\label{sec:response}

Here we study the linear response of the $O(n)$ Mattis model to an uniform
external field. After describing the formal solutions, we obtained
the response function in the one-step cycling of equilibrium states discussed in
section \ref{sec.on-resp}.

\subsection{Formal Solution}

In the $O(n)$ model in the spherical limit, 
response of the field at wavelength $k$ is only due to perturbation
at the same wavelength. A pulse staggered field $\hat{h}_{k}(t')$ 
at time $t'$ induces a displacement of the projection field at time $t$ as,
\begin{equation}
\delta \phi_{k}(t)=R_{k}(t,t') \delta \hat{h}_{k}(t').
\end{equation}
The response function \cite{berthier}
associated with wave-vector $k$ reads as,
\begin{equation}
R_{k}(t,t')=\frac{e^{-k^{2}(t-t')}}{\Gamma(t,t')},
\end{equation}
where $\Gamma(t,t')$ is defined as \eq{eq:def-gamma}.

Now let us consider induced response  $\delta \psi(x,t)$ due to uniform
pulse field $\delta h_{\rm uni}(t')$ applied at time $(t>)t'$.
For simplicity, we assume system continues to do coarsening with respect
to the same equilibrium state between $t'$ and $t'$.
The relation  \eq{eq:uniform-staggered} implies that
the pulse of the uniform field $\delta h_{\rm uni}(t')$ induces
the pulse of the staggered  field at wavelength $k$ as,
\begin{equation}
\delta \hat{h}_{k}(t')= \int d^{d}y e^{-iky} \sigma(y) \delta h_{\rm uni}(t').
\end{equation}
Then using the relation \eq{eq:psi-phi},
the induced response of the real spin configuration becomes,
\begin{eqnarray}
\delta \psi(x,t) &=&  \sigma(x)\delta \phi(x,t)
= \sigma(x) \int \frac{d^{d}k}{(2\pi)^{d}}e^{ikx}\delta \hat{\phi}_{k}(t)
\nonumber \\
&=& \sigma(x) \int \frac{d^{d}k}{(2\pi)^{d}}e^{ikx}
R_{k}(t,t')  \int d^{d}y e^{-iky} \sigma(y) \delta h_{\rm uni}(t')
\end{eqnarray}
Finally the response function to uniform field is obtained using
\eq{eq:sig-sig} and \eq{eq:random-solution} as,
\begin{eqnarray}
R_{\rm uni}(t,t')  \equiv  
\frac{\langle\delta \psi(t)\rangle_{\sigma}}{\delta h_{\rm uni}(t')}
= \frac{\Delta}{(2\pi)^{d}} \int d^{d}k  R_{k}(t,t') 
= \frac{R^{\rm eq}_{\rm uni}(\tau)}{\sqrt{\Gamma(t,t')}}.
\label{eq:how-to-resp-1}
\end{eqnarray}
where
\begin{equation}
R^{\rm eq}_{\rm uni}(\tau)=\Gamma_{0}(\tau/2,0).
\label{eq:how-to-resp-2}
\end{equation}

It is convenient to define re-scaled response,
\begin{equation}
\tilde{R}(t,t')\equiv \frac{R_{\rm uni}(t,t')}{R_{\rm eq}(t-t')}=\frac{1}{\sqrt{\Gamma(t,t')}}.
\end{equation}
In the case of relaxation starting from random initial condition 
we obtain the re-scaled response using \eq{eq:how-to-resp-1}, 
\eq{eq:how-to-resp-2} and \eq{eq:random-solution} as,
\begin{equation}
\tilde{R}_{0}(t,t')\equiv 
1/\sqrt{\Gamma_{0}(t,t')}= 
\left (\frac{t}{t'} \right)^{-d/4}.
\label{eq:rescale-resp-random}
\end{equation}

\subsection{Response Function after One-Step Cycling}

Here we study the behavior of the response function
after one-step cycling of the equilibrium states discussed in section 
\ref{sec.on-resp}.
The response function in the third stage of the one-step cycling 
$\twt+\two + \twz > t (t') > \two+\twz$ is obtained formally 
using \eq{eq:how-to-resp-1}, \eq{eq:how-to-resp-2} as,
\begin{eqnarray}
 \tilde{R}_{\rm III}(t,t'))
&=& \frac{1}{\sqrt{\gone(t,t')}}
= \sqrt{\frac{\gone(t',\two+\twz)}{\gone(t,\two+\twz)}} 
\qquad \mbox{'third stage'}.
\label{eq:resp-third}
\end{eqnarray}
Here the $\gone$ factor is the one obtained in \eq{eq:gamma-after-tw2} 
which can be rewritten as,
\begin{equation}
\gone(t,\two+\twz)
 =  [\mb^{2}(t-s,\two,\twz)+\mf^{2}(t-s,\two,\twz)
+\tilde{m}^{2}(t-s,\two,\twz)]\gone(t,s)|_{s=\two+\twz}.
\end{equation}
(The same form holds for $\gone(t',\two+\twz)$.)
The three weights $\mb^{2}$, $\mf^{2}$ and $\tilde{m}^{2}$ 
describes 'memory' and 'rejuvenation' as we discussed in
in section \ref{subsubsec.spatial-c-third}. The sum of them equals $1$ by definition.

In the inner-coarsening regime, 
we found that $\mf^{2} \sim 1$ while others are negligible.
On the other hand, in the plateau and outer-coarsening regime,
we found that  $\mb^{2} \sim 1$  while others are negligible.
Here we are assuming the case of wide separation a) $\twz \gg \two$).
In the inner- and outer-coarsening regime, th re-scaled response function becomes,
\begin{eqnarray}
\tilde{R}_{\rm III}(t,t') & \simeq  & \tilde{R}_{0}(t-\two-\twz,t'-\two-\twz) 
\qquad {\mbox{inner-coarsening regime}} 
\label{eq:rescaled-resp-1} \\ 
\tilde{R}_{\rm III}(t,t') &  \simeq &
 \tilde{R}_{0}(t-\two,t'-\two) \qquad
  \mbox{plateau/outer-coarsening regime}.
\label{eq:rescaled-resp-2}
\end{eqnarray}

\end{document}